\documentclass[apj]{emulateapj}

\usepackage{amsmath}
\usepackage{hyperref} 

\slugcomment{To appear in the Astrophysical Journal}

\shorttitle{Characterizing the transition from diffuse atomic to dense
  molecular clouds in the Magellanic clouds} \shortauthors{Pineda, et
  al.}

\begin{document}

\title{Characterizing the transition from diffuse atomic to dense molecular
  clouds in the Magellanic clouds with [C\,{\sc ii}], [C\,{\sc i}], and
  CO }

\author{Jorge L. Pineda\altaffilmark{1}, 
William D. Langer\altaffilmark{1},
Paul  F. Goldsmith\altaffilmark{1}, 
Shinji Horiuchi\altaffilmark{2},
Thomas B. H. Kuiper\altaffilmark{1},
Erik Muller\altaffilmark{3},
Annie Hughes\altaffilmark{4},
J\"urgen Ott\altaffilmark{5},
Miguel A. Requena--Torres\altaffilmark{6},
Thangasamy Velusamy\altaffilmark{1},
and Tony Wong\altaffilmark{7}}
\affil{$^{1}$Jet Propulsion Laboratory, California Institute of
  Technology, 4800 Oak Grove Drive, Pasadena, CA 91109-8099, USA \\
$^{2}$CSIRO Astronomy \& Space Science/NASA Canberra Deep Space Communication Complex, PO Box 1035, Tuggeranong ACT 2901, Australia\\
$^{3}$National Astronomical Observatory of Japan, Chile Observatory, Tokyo, Mitaka, 181-8588, Japan\\
$^{4}$CNRS, IRAP, 9 Av. Colonel Roche, BP 44346, 31028, Toulouse Cedex 4, France; Universit\'e de Toulouse, UPS-OMP, IRAP, 31028, Toulouse Cedex 4, France\\
$^{5}$National Radio Astronomy Observatory, P.O. Box O, 1003 Lopezville Road, Socorro, NM 87801, USA\\
$^{6}$Space Telescope Science Institute, 3700 San Martin Dr., Baltimore, 21218 MD, USA\\
$^{7}$Department of Astronomy, University of Illinois, Urbana, IL 61801, USA
}
\email{Jorge.Pineda@jpl.nasa.gov}

\begin{abstract}

  We present and analyze deep {\it Herschel}/HIFI observations of the
  [C\,{\sc ii}] 158\,$\mu$m, [C\,{\sc i}] 609\,$\mu$m, and [C\,{\sc
    i}] 370\,$\mu$m lines towards 54 lines-of-sight (LOS) in the Large
  and Small Magellanic clouds.  These observations are used to
  determine the physical conditions of the line--emitting gas, which
  we use to study the transition from atomic to molecular gas and from
  C$^+$ to C$^0$ to CO in their low metallicity environments.  We
  trace gas with molecular fractions in the range $0.1<f({\rm
    H}_2)<1$, between those in the diffuse H$_2$ gas detected by UV
  absorption ($f({\rm H}_2)<0.2$) and well shielded regions in which
  hydrogen is essentially completely molecular. The C$^0$ and CO
  column densities are only measurable in regions with molecular
  fractions $f({\rm H}_2)>0.45$ in both the LMC and SMC.  Ionized
  carbon is the dominant gas--phase form of this element that is
  associated with molecular gas, with C$^{0}$ and CO representing a
  small fraction, implying that most (89\% in the LMC and 77\% in the
  SMC) of the molecular gas in our sample is CO--dark H$_2$.  The mean
  $X_{\rm CO}$ conversion factors in our LMC and SMC sample are larger
  than the value typically found in the Milky Way. When applying a
  correction based on the filling factor of the CO emission, we find
  that the values of $X_{\rm CO}$ in the LMC and SMC are closer to
  that in the Milky Way.  The observed [C\,{\sc ii}] intensity in our
  sample represents about 1\% of the total far--infrared intensity
  from the LOSs observed in both Magellanic Clouds.

%These LOSs represent different stages
%  in the H/H$_2$ and C$^+$/C$^0$/CO transition in interstellar clouds
%  and have been chosen to study the formation and evolution of
%  molecular clouds in metal poor galaxies. 
%We
%  determine the thermal pressure of the diffuse ISM to be $p_{\rm
%    th}/k_{\rm B}\simeq3.4\times$10$^{4}$\,K\,cm$^{-3}$ in the LMC and
%  1$\times$10$^{5}$\,K\,cm$^{-3}$ in the SMC. 

\end{abstract}


\section{Introduction}
\label{sec:introduction}

Understanding the life cycle of the interstellar medium in galaxies is
a primary goal in the study of galaxy evolution and star formation.
The formation of molecular clouds, their subsequent evolution to
initiate star formation, and the radiative and mechanical feedback of
stars into their progenitor molecular gas that terminates or reignites
star formation are fundamental aspects determining how galaxies evolve
over cosmic time.  
Most of the star formation at early cosmological times took place in
environments with reduced gas metallicity and dust--to--gas ratio.
With the implementation of ALMA it is now possible to study a large
number of high--redshift systems \citep[e.g.][]{Carilli2013}.  To
interpret these observations of poorly resolved high redshift galaxies
it is necessary to understand local templates of the low--metallicity
interstellar medium (ISM), that can be spatially and spectrally
resolved in a detail that is not achievable for distant galaxies.

The best local templates for studying the life cycle of the ISM and
star formation in low--metallicity environments are the Large and
Small Magellanic clouds (LMC; $Z$=0.5\,Z$_{\odot}$ and SMC,
$Z$=0.2\,Z$_{\odot}$; \citealt{Westerlund97}).  Due to their proximity
(LMC: $D=50$\,kpc; SMC: $D=60$\,kpc;
\citealt{Schaefer2008,Hilditch2005}), the Magellanic clouds provide a
unique opportunity to resolve individual clouds, allowing us to
conduct detailed studies of the different phases of the ISM in
low--metallicity environments using various gas and dust tracers.

The interstellar medium plays the critical role of fueling star
formation in galaxies. Obtaining a complete census of the mass of the
different phases of the interstellar medium is thus key for
understanding star formation and its effect in the evolution of
galaxies. Traditionally, the distribution of the ISM phases in
galactic disks has been studied in H$\alpha$ or radio continuum
emission to trace the ionized gas \citep[e.g.][]{Haffner2009}, in the
H\,{\sc i} 21 cm line which traces atomic gas
\citep[e.g.][]{Kalberla2009}, and in the CO line to trace shielded
molecular gas \citep[e.g.][]{Heyer2015}. The total dust mass of the
interstellar medium has been traced by modeling the spectral energy
distribution of dust continuum, but the conversion from dust mass to
hydrogen mass is often complicated by finding appropriate values for
the dust to gas ratio, dust emissivity and dust temperature,
parameters that can vary with environmental conditions
\citep{Roman-Duval2014}.  We lack, however, observations of very
important ISM phases constituting the transition between diffuse
atomic clouds and dense molecular clouds. This missing link between
diffuse atomic and dense molecular gas can be isolated and
characterized with observations of the [C\,{\sc ii}] 158\,$\mu$m line
that traces the diffuse ionized medium, warm and cold atomic clouds,
clouds in transition from atomic to molecular form, and dense and warm
photon dominated regions (PDRs).  In particular, the [C\,{\sc ii}]
line is a tracer of the CO--dark H$_2$ gas
\citep{Madden1997,Grenier2005,Wolfire2010, Langer2010,Langer2014a},
which is gas in which hydrogen is molecular but carbon is ionized and
thus not traced by CO but by [C\,{\sc ii}]. 
% This gas is the likely
%precursor of dense molecular gas that will eventually form stars.
 The
CO--dark H$_2$ gas represents $\sim$30\% of the molecular mass of the
Milky Way and this fraction increases with galactocentric distance,
which is an effect of the metallicity gradient of the Galaxy
\citep{Pineda2013}.

The reduced dust--to--gas ratio and lower abundance of species
responsible for gas cooling (e.g. C and O) in metal poor environments
impacts the relative distributions of different ISM phases and their
thermal balance.   In low dust--to--gas ratio environments, the
  dust column density required to form enough H$_2$ to self--shield
  efficiently against photo--destruction, enabling the transition from
  atomic to molecular gas, is achieved at larger gas column densities
  compared with larger dust--to--gas ratio environments
  \citep[e.g.][]{Gnedin2009, Krumholz2009,Sternberg2014}.  
Additionally, due to the reduced attenuation of FUV photons by dust in
low metallicity gas, the CO molecule is more readily
photo--dissociated pushing the C$^+$/C$^0$/CO transition to higher
molecular hydrogen column densities compared to higher metallicity
systems \citep{vanDishBlack88}. This effect results in an enhanced
CO--to--H$_2$ conversion factor in the Magellanic Clouds
\citep{Cohen88,Rubio91,Israel97,Israel2000,Bolatto2013}.  Observations
of the distribution of [C\,{\sc ii}], [C\,{\sc i}], and CO line
emission in the Magellanic clouds are thus important tools for
studying how metallicity affects the properties of the different
phases of the interstellar medium.

The [C\,{\sc ii}] line has been imaged in the entire Large Magellanic
cloud with the BICE balloon by \citet{Mochizuki1994}, with
15\arcmin\ angular resolution and 175 km\,s$^{-1}$ velocity
resolution. These observations showed an enhanced [C\,{\sc ii}]/CO
ratio compared with the Milky Way and that [C\,{\sc ii}] constitutes
about 1.32\% of the far--infrared luminosity of the LMC
\citep{Rubin2009}.  A handful of star forming regions in the LMC and
SMC have been studied with the {\it Kuiper Airborne Observatory} both
with low \citep{Poglitsch95,Israel96,Israel2011} and with high
\citep{Boreiko91} velocity resolution observations. The 30\,Doradus
region in the LMC has been studied in detail with the PACS instrument
on {\it Herschel} \citep{Chevance2016}.  High velocity resolution
[C\,{\sc ii}] images of H\,{\sc ii} regions in the LMC and SMC are
starting to become available using SOFIA
\citep{Okada2015,Requena-Torres2016}.

The [C\,{\sc ii}] 158\,$\mu$m, [C\,{\sc i}] 609\,$\mu$m and
370\,$\mu$m fine structure, and $^{12}$CO and $^{13}$CO rotational,
transitions are diagnostics of the physical conditions of PDRs
\citep[e.g.][]{Kim2006,Minamidani2007,Minamidani2011,Pineda2008,Pineda2012,Okada2015,Lee2016,Chevance2016}.
When compared with predictions from PDR models, the [C\,{\sc ii}] line
constrains the strength of the FUV field and volume density.  The
excitation of the [C\,{\sc i}] 609\,$\mu$m and 370\,$\mu$m lines can
give independent estimates of the kinetic temperature and H$_2$ volume
density \citep{Stutzki1997}.  The ionized gas component can be
characterized with the [N\,{\sc ii}] 122 and 205\,$\mu$m fine structure
line emission (\citealt{Goldsmith2015}, \citealt{Langer2016}) as well
as with hydrogen recombination lines and free--free continuum emission
in the cm wavelength regime.  Characterizing the physical conditions
of the gas is key for our understanding to the origin of the [C\,{\sc
  ii}] emission in the Magellanic clouds and will have implications in
the interpretation of observations of distant galaxies.

In this paper we present deep observations of the [C\,{\sc ii}]
158\,$\mu$m, [C\,{\sc i}] 609\,$\mu$m, [C\,{\sc i}] 370\,$\mu$m, and
$^{12}$CO $J=7\to6$ lines towards 54 LMC and SMC lines-of-sight (LOS).
These data were obtained using the HIFI \citep{deGraauw2010}
instrument onboard the {\it Herschel Space Observatory}\footnote{{\it
    Herschel} is an ESA space observatory with science instruments
  provided by European-led Principal Investigator consortia and with
  important participation from NASA.}  \citep{Pilbratt2010}. We
complement this data set with observations of the $J=1\to0$ and
$J=3\to2$ transitions of $^{12}$CO and $^{13}$CO from the Mopra and
APEX telescopes, respectively.  We base our target selection on maps
of H\,{\sc i}, 160\,$\mu$m dust continuum emission, and CO emission as
well as on results from the FUSE survey of H$_2$ absorption towards
the Magellanic Clouds \citep{Cartledge2005}. The targets are
distributed throughout the LMC and SMC in order to study spatial
variations of the properties of their ISM. By studying clouds with
different physical conditions we aim to determine the key factors that
characterize the evolution of the interstellar matter in the
Magellanic clouds.

This paper is organized as follows. In Section~\ref{sec:observations}
we describe our sample selection and observations. We discuss the
determination of the contributions of the different phases of the
interstellar medium to the observed [C\,{\sc ii}] emission in
Section~\ref{sec:origin-c-sc} and we estimate in
Section~\ref{sec:physical-parameters} the physical parameters of
different ISM components. In Section~\ref{sec:discussion}, we study
the H$^0$ to H$_2$ as well as the C$^+$/C$^0$/CO transitions in the
LMC and SMC. We also discuss the CO--to--H$_2$ conversion factor, the
relationship between [C\,{\sc ii}] and far--infrared dust emission in
our sample, and we compare our single dish observations with pencil
beam FUV and optical studies to characterize the substructure of the
ISM in the Magellanic clouds.  We present our conclusions in
Section~\ref{sec:conclusions}.

\begin{figure*}[t]
\centering
\includegraphics[width=0.75\textwidth,angle=0]{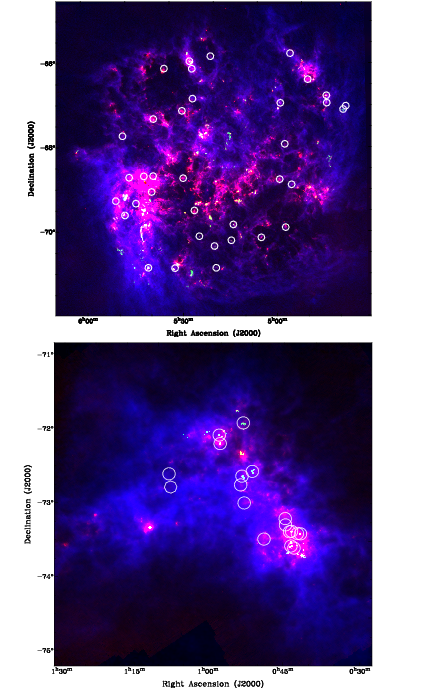}
\caption{ Images showing H$^{0}$ column density map (blue;
  \citealt{Kim2003}, \citealt{Stanimirovic1999}), {\it Herschel}
  160\,$\mu$m continuum emission (red; \citealt{Meixner2013}), and
  MAGMA CO line emission (green; \citealt{Wong2011} and Muller et
  al. 2017 in preparation) in the Large ({\it top}) and Small ({\it
    bottom}) Magellanic clouds.  The white circles denote the
  positions studied in this paper.  The size of the circles does not
  denote the beam size employed in any of the observations presented
  here. }
\label{fig:locations}
\end{figure*}

\begin{figure*}[t]
\centering
\includegraphics[width=\textwidth,angle=0]{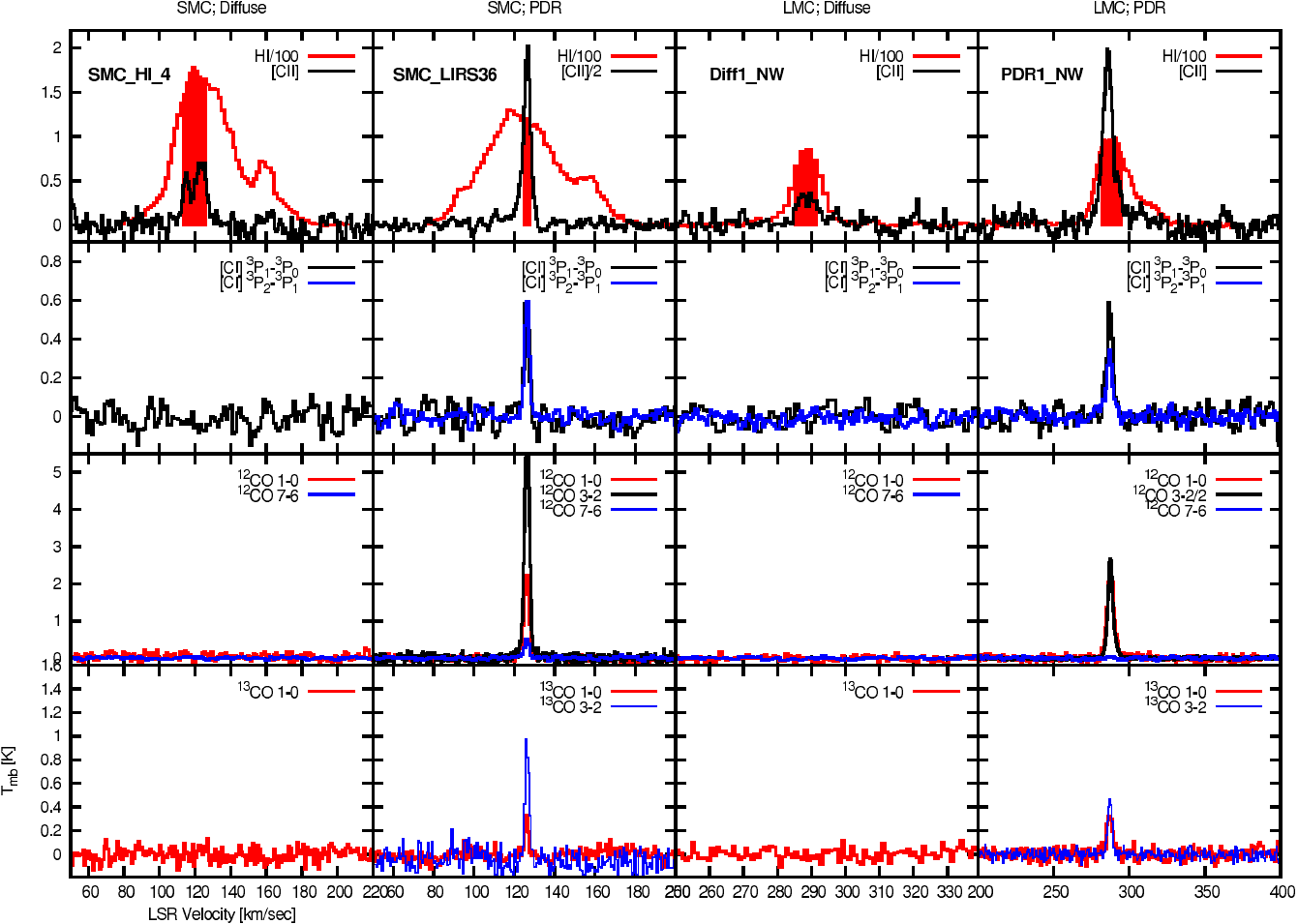}
\caption{Sample spectra of ionic, atomic, and molecular species in the
  Large and Small Magellanic Clouds. We show examples of diffuse
  regions as well as of dense photon dominated regions.  The
    shaded region in the H\,{\sc i} spectra represents the velocity
    range where we assume that the H\,{\sc i} emission is in the form of CNM (see
    Section~\ref{sec:atomic-gas}). }
\label{fig:spectra}
\end{figure*}

\begin{figure*}[t]
\centering
\includegraphics[width=\textwidth,angle=0]{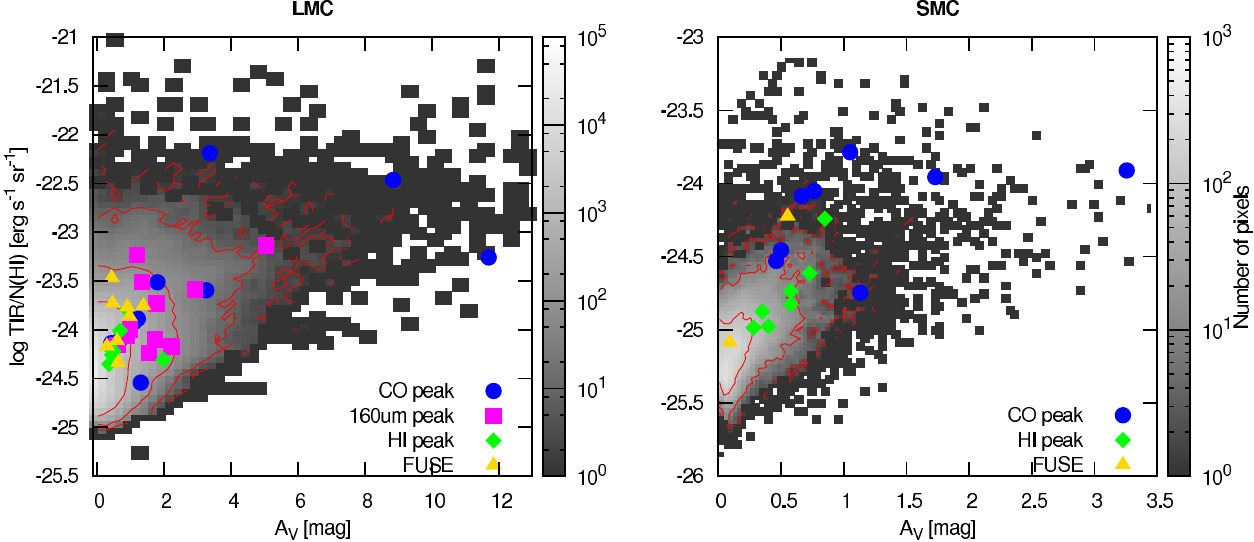}
\caption{The ratio of the total far--infrared intensity
  (Section~\ref{sec:relat-betw-c}) to the H$^0$ column density
  (Section~\ref{sec:atomic-gas}) as a function of the visual
  extinction (Section~\ref{sec:visu-extinct-determ}) for the entire
  LMC and SMC. We also include data points that correspond to those in
  our sample in the LMC and SMC.  The gray--scale represents the
  number of pixels at a given TIR/$N$(H$^0$) and $A_{\rm V}$ bin, with
  contour lines representing 2, 10, 100, and 1000 pixels. }
\label{fig:context}
\end{figure*}

\section{Sample Selection and Observations}
\label{sec:observations}

\subsection{Sample Selection}

%\label{sec:sample-selection}
% In this section I want to justify and describe why we selected the different LOS in our survey. 

To characterize the transition from diffuse atomic to dense molecular
clouds in the Magellanic clouds we need to study many different clouds
having different physical conditions in their low metallicity
environments. Therefore our work results in statistical properties of
the sample of clouds rather than full understanding of an individual
region. This strategy also allows us to search for locations that
could be followed up with current or future observatories (e.g. SOFIA,
STO2).

% to study the substructure of the ISM.

Our sample consists of 36 LOSs in the LMC and 18 in the SMC. We show
the locations used in our analysis in Figure~\ref{fig:locations}.
 The locations of our LOSs were selected to be as uniformly
  distributed as possible over the LMC and SMC, and they do not
  necessarily represent the brightest emission peaks in these
  galaxies.   We present sample spectra representing diffuse and
dense photon dominated regions (PDRs) in Figure~\ref{fig:spectra}.
The full spectral line data set used in this paper including
  images of the spectra in each LOS is available as a
  \href{http://www.cosmos.esa.int/web/herschel/user-provided-data-products}{{\it
      Herschel} User Provided Data Product}\footnote{{\tt
      http://www.cosmos.esa.int/web/herschel/user-\\provided-data-products}
  }.
%\hyperref{http://www.cosmos.esa.int/web/herschel/user-provided-data-products}
In Table\,1 and 2, we show the observational parameters in our sample,
including H$_2$ column densities and molecular fractions derived in
our analysis and discussed in Section~\ref{sec:h-h_2-transition}.  By
fitting Gaussians to the observed [C\,{\sc ii}], [C\,{\sc i}], and CO
spectra we identified 49 velocity components in the LMC and 28
velocity components in the SMC.  We show the integrated intensities of
the spectral lines detected in our survey in Table\,3 for the LMC and in
Table\,4 for the SMC. 

Our selection of positions to sample was initially based on H\,{\sc i}
(LMC; \citealt{Kim2003}, SMC; \citealt{Stanimirovic1999}) and CO (LMC;
\citealt{Wong2011}, SMC; \citealt{Muller2010,Rubio93}) maps of the LMC
and SMC. In the case of the LMC, we also used {\it Spitzer}
160\,$\mu$m continuum emission \citep{Meixner06}\footnote{The
  160\,$\mu$m continuum map, used as a proxy for higher column
  density, warmer gas, was only available for the LMC when the sample
  was originally selected.}. We included 7 H\,{\sc i} peaks in both
the LMC and SMC that have faint or no associated CO line emission or,
in the case of the LMC, 160\,$\mu$m continuum. These peaks represent
atomic hydrogen--dominated LOSs.  We also included 12 H\,{\sc i} peaks
in the LMC that are associated with 160\,$\mu$m continuum emission but
are undetected in the CO maps.   This sample is likely tracing clouds
that are in transition from diffuse molecular to dense molecular
clouds.  We also included 10 CO peaks in the LMC and 7 in the SMC
representing regions that have enough column density to show CO
emission.  They might still have a large fraction of H$_2$ gas traced
by [C\,{\sc ii}] if they are clumpy and the volume filling factor of
CO cores is low.  Finally, our sample also includes 8 lines--of--sight
in the LMC and 3 in the SMC studied in UV absorption with FUSE by
\citet{Cartledge2005}.  With these sources we have {\it a priori}
knowledge of the H$_2$ column density, and thus they can be used to
compare with our determination of the H$_2$ column densities from
spectral line data.

We studied whether the physical conditions derived in our sample are
representative of the average conditions in the LMC and SMC. For that
purpose, we compared the ratio of the total far--infrared intensity
(TIR; Section~\ref{sec:relat-betw-c}) to the H$^0$ column density
(Section~\ref{sec:atomic-gas}) as a function of the visual extinction
in our sample with that derived in maps over the entire LMC and SMC.
In Figure~\ref{fig:context}, we show the TIR/$N$(H$^0$) ratio as a
function of $A_{\rm V}$ for both the entire LMC and SMC and for the
locations of our sample. The color scale in Figure~\ref{fig:context}
represents the number of pixels at a given TIR/$N$(H$^0$) and $A_{\rm
  V}$ bin, with contour lines representing 2, 10, 100, and 1000
pixels.  The TIR/$N$(H$^0$) ratio is an approximate measure of the FUV
radiation field per unit hydrogen atom, and therefore it is sensitive
to how closely a LOS is associated with star formation, while the
$A_{\rm V}$ is a measure of the total column density. Thus, a diffuse
LOS away from star formation would be in the lower left corner of the
plot, while a warm and dense photon dominated region close to newly
formed stars would be in the upper right corner of the plot.  The TIR
and $A_{\rm V}$ maps used here were smoothed to 60\arcsec\ to match
the resolution of the H\,{\sc i} data.  The bulk of the LOSs have
TIR/$N$(H$^0$) $\simeq 10^{-25}-10^{-23}$\,erg\,s$^{-1}$\,sr$^{-1}$
and $A_{\rm V}\lesssim 2$\,mag in the LMC, and $\simeq
10^{-25.5}-10^{-24.5}$\,erg\,s$^{-1}$\,sr$^{-1}$ and $\lesssim
0.5$\,mag in the SMC.  In the LMC, a large fraction (90\%) of our
sample have TIR/$N$(H$^0$) and $A_{\rm V}$ values that are similar to
those of the bulk of the pixels in this galaxy.  This correspondence
suggests that the properties derived in our sample are representative
of the average properties in the LMC. There are four LOSs that have
larger TIR/$N$(H$^0$) and $A_{\rm V}$ values, corresponding to CO
peaks associated with dense photon dominated regions, which represent
a small volume fraction in the LMC.  In the SMC, half the LOSs have
conditions that are similar to the bulk of the pixels in this galaxy
while the other half tend to have larger values of TIR/$N$(H$^0$) and
$A_{\rm V}$. However, most of the [C\,{\sc ii}] emission is detected
in the latter half of the sample. Thus, the conditions derived in our
sample in the SMC might represent those of more active regions
compared with the bulk of the SMC.

\begin{deluxetable*}{lcccccccc} 
\tabletypesize{\scriptsize} \centering \tablecolumns{9} \small
\tablewidth{0pt}
\tablecaption{Derived Parameters for LMC and SMC Sample}
\tablenum{1}
\tablehead{\colhead{LOS} & \colhead{R.A} & \colhead{Decl.} &   \colhead{$N({\rm H^0})^1$} & \colhead{$N({\rm H^+})^2$} &  \colhead{$N({\rm H_2})^3$}  & \colhead{$f(\rm H_2)^4$} &\colhead{$A_{\rm V}^5$} & \colhead{Total far--IR$^6$} \\ 
\colhead{} & \colhead{} & \colhead{} &   \colhead{log} & \colhead{log} &  \colhead{log}  & \colhead{} &\colhead{} & \colhead{log} \\ 
\colhead{} & \colhead{J2000} & \colhead{J2000} &   \colhead{[cm$^{-2}$]} & \colhead{[cm$^{-2}$]} &  \colhead{[cm$^{-2}$]} &\colhead{} &\colhead{[mag]}  & \colhead{[erg\,s$^{-1}$cm$^{-2}$sr$^{-1}$]}}\\
\startdata
\cutinhead{Large Magellanic Cloud}
%#\sidehead{H\,{\sc i} peaks} %#\sidehead{{H\,{\sc i} Peaks}
  Diff1\_NW$^{8}$  &  5:31:59.2  &  -66:22:52.3  &  21.23$\pm$0.02  &  20.51$\pm$0.01  &  --  &  --  &  0.91$\pm$0.36  &  -2.61$\pm$0.04 \\ %	
   Diff2\_SE$^{8}$  &  4:59:35.5  &  -70:11:04.6  &  21.39$\pm$0.01  &  20.48$\pm$0.01  &  $20.9^{+0.2}_{-0.1}$  &  $0.4^{+0.1}_{-0.1}$  &  0.70$\pm$0.28  &  -2.70$\pm$0.04 \\ %
Diff3\_RIDGE$^{8}$  &  5:31:50.6  &  -71:12:41.6  &  21.53$\pm$0.01  &  19.61$\pm$0.10  &  --  &  --  &  2.00$\pm$0.80  &  -2.82$\pm$0.06 \\ %
   Diff4\_NE$^{8}$  &  5:01:47.7  &  -65:59:05.2  &  21.39$\pm$0.01  &  19.85$\pm$0.06  &  --  &  --  &  0.38$\pm$0.15  &  -2.91$\pm$0.07 \\ %
   Diff5\_SE$^{8}$  &  4:58:54.0  &  -69:08:29.9  &  21.47$\pm$0.01  &  19.67$\pm$0.09  &  --  &  --  &  0.42$\pm$0.17  &  -2.78$\pm$0.05 \\ %
   Diff6\_NW$^{8}$  &  5:43:34.9  &  -67:56:08.2  &  21.35$\pm$0.01  &  20.06$\pm$0.03  &  --  &  --  &  0.46$\pm$0.18  &  -2.95$\pm$0.08 \\ %
   Diff7\_NW$^{8}$  &  5:25:17.3  &  -67:08:03.6  &  21.48$\pm$0.01  &  19.91$\pm$0.05  &  20.9$\pm$0.1  &  $0.4^{+0.0^7}_{-0.0^7}$  &  0.50$\pm$0.20  &  -2.68$\pm$0.04 \\ %
   LMC10\_NE$^{9}$  &  4:51:51.1  &  -67:05:45.0  &  21.58$\pm$0.01  &  20.84$\pm$0.16  &  --  &  --  &  2.23$\pm$0.89  &  -2.63$\pm$0.04 \\ %
LMC11\_Ridge$^{9}$  &  5:25:33.8  &  -69:50:16.6  &  21.36$\pm$0.01  &  19.92$\pm$0.05  &  $21.3^{+0.3}_{-0.4}$  &  $0.6^{+0.2}_{-0.2}$  &  1.21$\pm$0.48  &  -1.77$\pm$0.01 \\ %
   LMC12\_SE$^{9}$  &  5:02:13.7  &  -69:02:16.4  &  21.40$\pm$0.01  &  20.25$\pm$0.02  &  $21.5^{+0.2}_{-0.5}$  &  $0.7^{+0.1}_{-0.2}$  &  2.94$\pm$1.18  &  -1.91$\pm$0.01 \\ %
  LMC\_1\_NW$^{9}$  &  5:28:1.9  &  -67:25:14.0  &  21.43$\pm$0.01  &  21.14$\pm$0.00$^7$  &  $21.6^{+0.0^7}_{-0.1}$  &  $0.8^{+0.0^7}_{-0.0^7}$  &  5.05$\pm$2.02  &  -1.72$\pm$0.00$^7$ \\ %
    LMC2\_NW$^{9}$  &  5:25:16.3  &  -66:24:40.8  &  21.58$\pm$0.01  &  20.32$\pm$0.02  &  21.0$\pm$0.1  &  $0.3^{+0.1}_{-0.1}$  &  0.91$\pm$0.36  &  -2.44$\pm$0.02 \\ %
    LMC3\_NW$^{9}$  &  5:20:44.8  &  -66:06:58.2  &  21.25$\pm$0.02  &  19.74$\pm$0.07  &  $20.5^{+0.3}_{-0.2}$  &  $0.3^{+0.1}_{-0.1}$  &  0.63$\pm$0.25  &  -2.85$\pm$0.06 \\ %
 LMC4\_RIDGE$^{9}$  &  5:28:22.5  &  -69:28:22.5  &  21.49$\pm$0.01  &  20.53$\pm$0.01  &  21.2$\pm$0.1  &  $0.5^{+0.0^7}_{-0.0^7}$  &  1.79$\pm$0.72  &  -2.18$\pm$0.01 \\ % 
    LMC5\_SE$^{9}$  &  5:06:23.1  &  -70:28:08.7  &  21.26$\pm$0.02  &  20.47$\pm$0.01  &  21.2$\pm$0.1  &  $0.6^{+0.0^7}_{-0.1}$  &  1.35$\pm$0.54  &  -2.08$\pm$0.01 \\ %
 LMC7\_RIDGE$^{9}$  &  5:45:6.9  &  -69:50:42.6  &  21.74$\pm$0.01  &  20.49$\pm$0.01  &  21.2$\pm$0.1  &  $0.4^{+0.0^7}_{-0.0^7}$  &  1.00$\pm$0.40  &  -2.20$\pm$0.01 \\ %
 LMC8\_RIDGE$^{9}$  &  5:47:11.8  &  -69:28:35.1  &  21.81$\pm$0.00$^7$  &  20.52$\pm$0.01  &  --  &  --  &  1.75$\pm$0.70  &  -2.20$\pm$0.01 \\ % 
    LMC9\_NE$^{9}$  &  5:03:20.8  &  -67:11:44.2  &  21.61$\pm$0.01  &  20.17$\pm$0.03  &  $20.5^{+0.3}_{-0.2}$  &  $0.1^{+0.1}_{-0.1}$  &  1.56$\pm$0.62  &  -2.63$\pm$0.04 \\ %	LMC9\_NE 
  NT127$^{10}$  &  5:24:19.8  &  -70:27:48.7  &  21.05$\pm$0.03  &  19.49$\pm$0.13  &  $21.1^{+0.4}_{-0.7}$  &  $0.7^{+0.2}_{-0.2}$  &  1.23$\pm$0.49  &  -2.74$\pm$0.05 \\ %	NT127 
NT2\_NE$^{10}$  &  4:47:36.8  &  -67:12:13.7  &  21.35$\pm$0.01  &   &  --  &  --  &  1.32$\pm$0.53  &  -3.01$\pm$0.09 \\ %	NT2\_NE 
   NT74$^{10}$  &  5:14:33.4  &  -70:10:51.9  &  21.33$\pm$0.01  &  20.87$\pm$0.15  &  $21.4^{+0.6}_{-0.3}$  &  $0.7^{+0.3}_{-0.1}$  &  1.11$\pm$0.45  &  -2.49$\pm$0.03 \\ %	NT74 
   NT77$^{10}$  &  5:15:9.4  &  -70:35:42.0  &  21.00$\pm$0.03  &  20.66$\pm$0.24  &  $21.3^{+0.2}_{-0.3}$  &  $0.8^{+0.1}_{-0.1}$  &  1.82$\pm$0.73  &  -1.56$\pm$0.00$^7$ \\ %	NT77 
   NT97$^{10}$  &  5:19:27.8  &  -71:13:52.4  &  21.03$\pm$0.03  &  19.52$\pm$0.12  &  --  &  --  &  0.44$\pm$0.18  &  -3.04$\pm$0.10 \\ %	NT97 
   NT99$^{10}$  &  5:19:57.6  &  -70:42:21.7  &  21.17$\pm$0.02  &  20.87$\pm$0.15  &  --  &  --  &  0.68$\pm$0.27  &  -2.83$\pm$0.06 \\ %	NT99 
   PDR1\_NW$^{10}$  &  5:25:46.9  &  -66:13:41.6  &  21.71$\pm$0.01  &  20.40$\pm$0.01  &  $21.7^{+0.3}_{-0.6}$  &  $0.6^{+0.2}_{-0.2}$  &  3.27$\pm$1.31  &  -1.93$\pm$0.01 \\ %	PDR1\_NW 
   PDR2\_NW$^{10}$  &  5:35:22.4  &  -67:35:00.5  &  21.37$\pm$0.01  &  21.78$\pm$0.00$^7$  &  22.2$\pm$0.1  &  $0.9^{+0.0^7}_{-0.0^7}$  &  3.38$\pm$1.35  &  -0.51$\pm$0.00$^7$ \\ %	PDR2\_NW 
   PDR3\_NE$^{10}$  &  4:52:8.3  &  -66:55:13.7  &  21.50$\pm$0.01  &  21.47$\pm$0.00$^7$  &  22.5$\pm$0.1  &  $1.0^{+0.0^7}_{-0.0^7}$  &  8.84$\pm$3.54  &  -0.33$\pm$0.00$^7$ \\ %	PDR3\_NE 
PDR4\_RIDGE$^{10}$  &  5:39:48.7  &  -71:09:27.4  &  21.65$\pm$0.01  &  20.43$\pm$0.01  &  $21.5^{+0.2}_{-0.4}$  &  $0.6^{+0.1}_{-0.2}$  &  11.70$\pm$4.68  &  -1.42$\pm$0.00$^7$ \\ %	PDR4\_RIDGE 
 SK-66D35$^{11}$  &  4:57:4.5  &  -66:34:38.0  &  21.49$\pm$0.01  &  20.64$\pm$0.00$^7$  &  $21.1^{+0.1}_{-0.2}$  &  $0.6^{+0.3}_{-0.1}$  &  1.40$\pm$0.56  &  -2.29$\pm$0.02 \\ %	SK-66D35 
  SK-67D2$^{11}$  &  4:47:4.4  &  -67:06:53.0  &  21.31$\pm$0.01  &  20.13$\pm$0.03  &  --  &  --  &  0.31$\pm$0.12  &  -2.61$\pm$0.04 \\ %	SK-67D2 
SK-68D129$^{11}$  &  5:36:26.8  &  -68:57:32.0  &  21.55$\pm$0.01  &  20.66$\pm$0.00$^7$  &  --  &  --  &  0.98$\pm$0.39  &  -2.34$\pm$0.02 \\ %	SK-68D129 
SK-68D140$^{11}$  &  5:38:57.3  &  -68:56:53.0  &  21.69$\pm$0.01  &  21.20$\pm$0.00$^7$  &  21.0$\pm$0.1  &  $0.3^{+0.1}_{-0.1}$  &  0.45$\pm$0.18  &  -1.63$\pm$0.00$^7$ \\ %	SK-68D140 
SK-68D155$^{11}$  &  5:42:54.9  &  -68:56:54.0  &  21.76$\pm$0.01  &  20.81$\pm$0.00$^7$  &  21.4$\pm$0.0  &  $0.5^{+0.0^7}_{-0.0^7}$  &  0.92$\pm$0.37  &  -1.77$\pm$0.01 \\ %	SK-68D155 
 SK-68D26$^{11}$  &  5:01:32.2  &  -68:10:43.0  &  21.24$\pm$0.02  &  19.91$\pm$0.05  &  --  &  --  &  0.61$\pm$0.24  &  -2.86$\pm$0.06 \\ %	SK-68D26 
SK-69D228$^{11}$  &  5:37:9.2  &  -69:20:20.0  &  21.52$\pm$0.01  &  20.85$\pm$0.00$^7$  &  20.9$\pm$0.1  &  $0.3^{+0.1}_{-0.1}$  &  0.48$\pm$0.19  &  -2.17$\pm$0.01 \\ %	SK-69D228 
SK-69D279$^{11}$  &  5:41:44.7  &  -69:35:15.0  &  21.62$\pm$0.01  &  20.41$\pm$0.01  &  --  &  --  &  0.64$\pm$0.26  &  -2.61$\pm$0.04 \\ 
\tablenotetext{1}{H$^0$ column density derived from the  integrated intensity of the H\,{\sc i} 21\,cm line over the full velocity range (see Section~\ref{sec:atomic-gas}). }
\tablenotetext{2}{H$^+$ column density derived from H$\alpha$ observations as described in Section~\ref{sec:ionized-gas}. }
\tablenotetext{3}{H$_2$ column density derived from [C\,{\sc ii}], [C\,{\sc i}], and CO observations as described in Section~\ref{sec:h-h_2-transition}. }
\tablenotetext{4}{Molecular fraction, $f({\rm H}_2)=2N({\rm H}_2)/(N({\rm H^0})+2N({\rm H}_2))$ (Section~\ref{sec:h-h_2-transition}).}
\tablenotetext{5}{Visual extinction derived from dust continuum maps as described in Section~\ref{sec:visu-extinct-determ}. }
\tablenotetext{6}{Total far--infrared intensity derived from Spitzer 24\,$\mu$m  and {\it Herschel} 100\,$\mu$m maps as described in Section~\ref{sec:relat-betw-c}. }
\tablenotetext{7}{The uncertainty is below 0.005 or 0.05. }
\tablenotetext{8}{H\,{\sc i} Peak (See Section~\ref{sec:observations}).}
\tablenotetext{9}{160\,$\mu$m Peak  (See Section~\ref{sec:observations}). }
\tablenotetext{10}{CO Peak  (See Section~\ref{sec:observations}). }
\tablenotetext{11}{FUSE LOSs  (See Section~\ref{sec:observations}). }

\enddata
\end{deluxetable*}

\begin{deluxetable*}{lcccccccc} 
\tabletypesize{\scriptsize} \centering \tablecolumns{9} \small
\tablewidth{0pt}
\tablecaption{Derived Parameters for LMC and SMC Sample}
\tablenum{2}
\tablehead{\colhead{LOS} & \colhead{R.A} & \colhead{Decl.} &   \colhead{$N({\rm H^0})^1$} & \colhead{$N({\rm H^+})^2$} &  \colhead{$N({\rm H_2})^3$}  & \colhead{$f(\rm H_2)^4$} &\colhead{$A_{\rm V}^5$} & \colhead{Total far--IR$^6$} \\ 
\colhead{} & \colhead{} & \colhead{} &   \colhead{log} & \colhead{log} &  \colhead{log}  & \colhead{} &\colhead{} & \colhead{log} \\ 
\colhead{} & \colhead{J2000} & \colhead{J2000} &   \colhead{[cm$^{-2}$]} & \colhead{[cm$^{-2}$]} &  \colhead{[cm$^{-2}$]} &\colhead{} &\colhead{[mag]}  & \colhead{[erg\,s$^{-1}$cm$^{-2}$sr$^{-1}$]}}\\ 
\startdata
\cutinhead{Small Magellanic Cloud}
AzV18$^{11}$  &  0:47:13.1  &  -73:06:25.0  &  22.10$\pm$0.00$^7$  &  20.54$\pm$0.00$^7$  &  21.9$\pm$0.1  &  $0.5^{+0.1}_{-0.0^7}$  &  0.55$\pm$0.22  &  -1.88$\pm$0.01 \\ %	AzV18 
AzV456$^{11}$  &  1:10:55.8  &  -72:42:55.0  &  21.64$\pm$0.00$^7$  &  20.52$\pm$0.01  &  --  &  --  &  0.09$\pm$0.04  &  -3.15$\pm$0.12 \\ %	AzV456 
AzV462$^{11}$  &  1:11:25.9  &  -72:32:21.0  &  21.62$\pm$0.00$^7$  &  19.77$\pm$0.07  &  --  &  --  &  0.00$\pm$0.00$^7$  &  -3.54$\pm$0.30 \\ %	AzV462 
SMC\_HI\_1$^{8}$  &  0:58:40.5  &  -72:34:52.4  &  21.99$\pm$0.00$^7$  &  20.10$\pm$0.03  &  --  &  --  &  0.28$\pm$0.11  &  -3.04$\pm$0.10 \\ %	SMC\_HI\_1 
SMC\_HI\_2$^{8}$  &  0:57:35.7  &  -72:48:56.3  &  21.99$\pm$0.00$^7$  &  20.11$\pm$0.03  &  --  &  --  &  0.40$\pm$0.16  &  -2.98$\pm$0.08 \\ %	SMC\_HI\_2 
SMC\_HI\_3$^{8}$  &  0:53:2.0  &  -73:15:27.9  &  22.02$\pm$0.00$^7$  &  20.27$\pm$0.02  &  --  &  --  &  0.58$\pm$0.23  &  -2.82$\pm$0.06 \\ %	SMC\_HI\_3 
SMC\_HI\_4$^{8}$  &  0:48:41.5  &  -73:06:08.3  &  22.13$\pm$0.00$^7$  &  20.36$\pm$0.01  &  21.8$\pm$0.1  &  $0.5^{+0.0^7}_{-0.0^7}$  &  0.85$\pm$0.34  &  -2.11$\pm$0.01 \\ %	SMC\_HI\_4 
SMC\_HI\_5$^{8}$  &  0:47:20.4  &  -73:18:28.4  &  22.05$\pm$0.00$^7$  &  20.63$\pm$0.01  &  --  &  --  &  0.73$\pm$0.29  &  -2.73$\pm$0.05 \\ %	SMC\_HI\_5 
SMC\_HI\_6$^{8}$  &  0:49:53.2  &  -72:56:15.6  &  22.05$\pm$0.00$^7$  &  20.48$\pm$0.01  &  --  &  --  &  0.58$\pm$0.23  &  -2.76$\pm$0.05 \\ %	SMC\_HI\_6 
SMC\_HI\_7$^{8}$  &  0:49:38.1  &  -73:01:14.2  &  22.06$\pm$0.00$^7$  &  20.16$\pm$0.03  &  --  &  --  &  0.35$\pm$0.14  &  -2.80$\pm$0.06 \\ %	SMC\_HI\_7 
SMC\_B2\_6$^{10}$  &  0:47:57.2  &  -73:17:16.4  &  22.00$\pm$0.00$^7$  &  20.98$\pm$0.00$^7$  &  22.2$\pm$0.1  &  $0.7^{+0.0^7}_{-0.0^7}$  &  0.48$\pm$0.19  &  -1.79$\pm$0.01 \\ %	SMC\_B2\_6 
SMC\_LIRS36$^{10}$  &  0:46:40.3  &  -73:06:10.5  &  22.04$\pm$0.00$^7$  &  20.96$\pm$0.00$^7$  &  22.2$\pm$0.2  &  $0.7^{+0.1}_{-0.1}$  &  0.60$\pm$0.24  &  -1.53$\pm$0.00$^7$ \\ %	SMC\_LIRS36 
SMC\_LIRS49$^{10}$  &  0:48:21.1  &  -73:05:29.0  &  22.12$\pm$0.00$^7$  &  20.53$\pm$0.00$^7$  &  $22.5^{+0.1}_{-0.2}$  &  $0.8^{+0.0^7}_{-0.1}$  &  0.90$\pm$0.36  &  -1.69$\pm$0.00$^7$ \\ %	SMC\_LIRS49 
SMC\_NE\_1a$^{10}$  &  0:59:43.8  &  -71:44:47.0  &  21.60$\pm$0.00$^7$  &  20.23$\pm$0.02  &  $21.5^{+0.1}_{-0.2}$  &  $0.6^{+0.1}_{-0.1}$  &  0.41$\pm$0.16  &  -2.38$\pm$0.02 \\ %	SMC\_NE\_1a 
SMC\_NE\_3c$^{10}$  &  1:03:30.0  &  -71:57:00.0  &  21.78$\pm$0.00$^7$  &  20.48$\pm$0.01  &  $21.5^{+0.3}_{-0.8}$  &  $0.5^{+0.2}_{-0.4}$  &  0.46$\pm$0.19  &  -2.70$\pm$0.04 \\ %	SMC\_NE\_3c 
SMC\_NE\_3g$^{10}$  &  1:03:9.9  &  -72:03:46.9  &  21.79$\pm$0.00$^7$  &  20.84$\pm$0.00$^7$  &  $21.4^{+0.1}_{-0.2}$  &  $0.5^{+0.1}_{-0.1}$  &  0.82$\pm$0.33  &  -2.29$\pm$0.02 \\ %	SMC\_NE\_3g 
SMC\_NE\_4a\_hi$^{10}$  &  0:57:0.0  &  -72:22:40.0  &  21.86$\pm$0.00$^7$  &  20.21$\pm$0.02  &  $22.1^{+0.6}_{-1.9}$  &  $0.8^{+0.3}_{-0.2}$  &  1.13$\pm$0.45  &  -2.77$\pm$0.05 \\ %	SMC\_NE\_4a\_hi 
SMC\_NE\_4c\_low$^{10}$  &  0:58:40.0  &  -72:27:40.0  &  21.93$\pm$0.00$^7$  &  20.39$\pm$0.02  &  22.2$\pm$0.7  &  $0.8^{+0.4}_{-0.2}$  &  0.67$\pm$0.27  &  -2.63$\pm$0.04 \\ %	SMC\_NE\_4c\_low 
\tablenotetext{1}{H$^0$ column density derived from the  integrated intensity of the H\,{\sc i} 21\,cm line over the full velocity range (see Section~\ref{sec:atomic-gas}). }
\tablenotetext{2}{H$^+$ column density derived from H$\alpha$ observations as described in Section~\ref{sec:ionized-gas}. }
\tablenotetext{3}{H$_2$ column density derived from [C\,{\sc ii}], [C\,{\sc i}], and CO observations as described in Section~\ref{sec:h-h_2-transition}. }
\tablenotetext{4}{Molecular fraction, $f({\rm H}_2)=2N({\rm H}_2)/(N({\rm H^0})+2N({\rm H}_2))$ (Section~\ref{sec:h-h_2-transition}).}
\tablenotetext{5}{Visual extinction derived from dust continuum maps as described in Section~\ref{sec:visu-extinct-determ}. }
\tablenotetext{6}{Total far--infrared intensity derived from Spitzer 24\,$\mu$m  and {\it Herschel} 100\,$\mu$m maps as described in Section~\ref{sec:relat-betw-c}. }
\tablenotetext{7}{The uncertainty is below 0.005 or 0.05. }
\tablenotetext{8}{H\,{\sc i} Peak (See Section~\ref{sec:observations}).}
\tablenotetext{9}{160$\mu$m Peak  (See Section~\ref{sec:observations}). }
\tablenotetext{10}{CO Peak  (See Section~\ref{sec:observations}). }
\tablenotetext{11}{FUSE LOSs  (See Section~\ref{sec:observations}). }

\enddata
\end{deluxetable*}

\begin{deluxetable*}{lcccccccccc} 
\tabletypesize{\scriptsize} \centering \tablecolumns{11} \small
\tablewidth{0pt}
\tablecaption{Integrated Intensities of Spectral Line  for LMC  Sample}
\tablenum{3}
\tablehead{\colhead{LOS} & \colhead{$V_{\rm LSR}$} & \colhead{$I^{1}_{\rm [HI];CNM}$} & \colhead{$I_{\rm [CII]}$} &   \colhead{$I_{\rm [CI](1-0)}$} &  \colhead{$I_{\rm [CI](2-1)}$} &    \colhead{$I_{\rm CO(1-0)}$} &  \colhead{$I_{\rm CO(3-2)}$}  &  \colhead{$I_{\rm CO(7-6)}$} & \colhead{$I_{\rm ^{13}CO(1-0)}$} &  \colhead{$I_{\rm ^{13}CO(3-2)}$}\\
\colhead{} &\colhead{[km\,s$^{-1}$]} &\multicolumn{9}{c}{[K\,km\,s$^{-1}$]} }\\ %& \colhead{[K\,km\,s$^{-1}$]} &  \colhead{[K\,km\,s$^{-1}$]} & \colhead{[K\,km\,s$^{-1}$]} &  \colhead{[K\,km\,s$^{-1}$]} & \colhead{[K\,km\,s$^{-1}$]} & \colhead{[K\,km\,s$^{-1}$]} &\colhead{[K\,km\,s$^{-1}$]}  & \colhead{[K\,km\,s$^{-1}$]}}\\
\startdata
\cutinhead{Large Magellanic Cloud} 
Diff1\_NW\_1 & 288.8 & 510$\pm$8 & 2.4$\pm$0.3 & -- & -- & -- & -- & -- & -- & --  \\
Diff2\_SE\_1 & 233.3 & 169$\pm$6 & 2.8$\pm$0.6 & 0.9$\pm$0.3 & -- & -- & -- & -- & -- & \\
Diff3\_RIDGE\_1 & -- &  -- & -- & -- & -- & 0.7$\pm$0.2 & -- & -- & -- & \\
Diff5\_SE\_1 & 245.0 & 405$\pm$7 & 0.9$\pm$0.2 & -- & -- & -- & -- & -- & -- & --  \\
Diff6\_NW\_1 & -- &  -- & -- & -- & -- & 1.0$\pm$0.1 & 0.3$\pm$0.1 & -- & -- & -- \\
Diff7\_NW\_1 & 288.1 & 253$\pm$6 & 2.7$\pm$0.2 & -- & -- & -- & -- & -- & -- & --  \\
LMC10\_NE\_1 & 275.4 & 496$\pm$7 & 1.8$\pm$0.2 & 0.3$\pm$0.1 & -- & 1.7$\pm$0.2 & -- & -- & -- & \\
LMC11\_Ridge\_1 & 250.3 & 341$\pm$7 & 7.9$\pm$0.1 & 0.6$\pm$0.1 & 0.4$\pm$0.1 & 4.1$\pm$0.2 & 3.4$\pm$0.1 & -- & -- & 0.3$\pm$0.1 \\
LMC12\_SE\_1 & 265.7 & 224$\pm$6 & 6.4$\pm$0.2 & 0.1$\pm$0.1 & -- & 1.9$\pm$0.2 & 2.8$\pm$0.1 & -- & -- & 0.2$\pm$0.1 \\
LMC12\_SE\_2 & 270.6 & 288$\pm$6 & 5.3$\pm$0.2 & 1.1$\pm$0.1 & 0.9$\pm$0.1 & 4.7$\pm$0.2 & 7.4$\pm$0.1 & -- & 0.4$\pm$0.1 & 0.5$\pm$0.1 \\
LMC\_1\_NW\_1 & 293.1 & 288$\pm$7 & 6.6$\pm$0.0$^2$ & 0.6$\pm$0.1 & 0.4$\pm$0.0$^2$ & 1.8$\pm$0.3 & 1.3$\pm$0.2 & 1.5$\pm$0.1 & -- & 0.3$\pm$0.0$^2$ \\
LMC\_1\_NW\_2 & 287.6 & 341$\pm$7 & 6.1$\pm$0.0$^2$ & 0.9$\pm$0.1 & 0.6$\pm$0.1 & 2.3$\pm$0.3 & 2.5$\pm$0.3 & 0.4$\pm$0.1 & -- & 0.3$\pm$0.1 \\
LMC\_1\_NW\_3 & 288.0 & 890$\pm$12 & 10.1$\pm$0.0$^2$ & -- & -- & -- & -- & -- & -- & --  \\
LMC2\_NW\_1 & 290.5 & 716$\pm$8 & 4.7$\pm$0.4 & 0.5$\pm$0.1 & 0.4$\pm$0.1 & 2.3$\pm$0.2 & -- & -- & -- & \\
LMC3\_NW\_1 & 289.7 & 257$\pm$6 & 1.6$\pm$0.3 & -- & -- & -- & -- & -- & -- & --  \\
LMC4\_RIDGE\_1 & 246.6 & 618$\pm$9 & 5.7$\pm$0.3 & -- & -- & 0.6$\pm$0.2 & -- & -- & -- & \\
LMC5\_SE\_1 & 219.0 & 75$\pm$4 & 9.7$\pm$0.1 & 0.7$\pm$0.1 & 0.8$\pm$0.2 & 2.5$\pm$0.2 & 3.2$\pm$0.1 & -- & -- & 0.4$\pm$0.1 \\
LMC7\_RIDGE\_1 & 232.7 & 745$\pm$9 & 6.2$\pm$0.5 & -- & -- & 1.9$\pm$0.2 & 0.5$\pm$0.2 & -- & -- & -- \\
LMC8\_RIDGE\_1 & 238.5 & 688$\pm$9 & 3.7$\pm$0.4 & -- & -- & -- & -- & -- & -- & --  \\
LMC9\_NE\_1 & 274.2 & 310$\pm$7 & 2.0$\pm$0.3 & -- & -- & 3.3$\pm$0.3 & 1.3$\pm$0.1 & -- & -- & -- \\
NT127\_1 & 235.0 & 73$\pm$4 & 0.6$\pm$0.1 & 1.1$\pm$0.1 & 0.2$\pm$0.0$^2$ & 3.7$\pm$0.2 & 4.2$\pm$0.1 & -- & -- & 0.3$\pm$0.1 \\
NT2\_NE\_1 & -- &  -- & -- & 1.2$\pm$0.2 & -- & 8.2$\pm$0.2 & -- & -- & 0.9$\pm$0.1 & \\
NT74\_1 & 236.1 & 310$\pm$8 & 1.6$\pm$0.4 & 2.6$\pm$0.1 & 0.5$\pm$0.1 & 17.1$\pm$0.3 & 11.0$\pm$0.1 & -- & 2.5$\pm$0.2 & 0.6$\pm$0.1 \\
NT77\_1 & 217.4 & 59$\pm$4 & 8.3$\pm$0.1 & 1.0$\pm$0.2 & 1.0$\pm$0.1 & 8.1$\pm$0.2 & 15.5$\pm$0.1 & 1.1$\pm$0.1 & 1.5$\pm$0.1 & 2.7$\pm$0.1 \\
NT97\_1 & -- &  -- & -- & 0.4$\pm$0.1 & -- & 3.8$\pm$0.1 & -- & -- & 0.7$\pm$0.1 & \\
NT99\_1 & 227.0 & 270$\pm$7 & 1.3$\pm$0.3 & -- & -- & 5.0$\pm$0.1 & -- & -- & 0.6$\pm$0.1 & \\
PDR1\_NW\_1 & 285.8 & 775$\pm$9 & 13.7$\pm$0.3 & 3.1$\pm$0.2 & 1.7$\pm$0.1 & 13.6$\pm$0.2 & 23.0$\pm$0.1 & -- & 1.6$\pm$0.1 & 1.6$\pm$0.1 \\
PDR1\_NW\_2 & 293.3 & 287$\pm$6 & 1.0$\pm$0.2 & -- & -- & -- & -- & -- & -- & --  \\
PDR2\_NW\_1 & 291.4 & 184$\pm$7 & 3.1$\pm$0.2 & -- & 0.4$\pm$0.0$^2$ & 2.1$\pm$0.5 & 3.9$\pm$0.1 & 1.5$\pm$0.1 & -- & \\
PDR2\_NW\_2 & 284.3 & 144$\pm$7 & 47.0$\pm$0.3 & 0.8$\pm$0.1 & 1.8$\pm$0.1 & 7.1$\pm$0.6 & 11.8$\pm$0.1 & 5.9$\pm$0.2 & -- & 1.0$\pm$0.2 \\
PDR2\_NW\_3 & 279.6 & 203$\pm$8 & 21.0$\pm$0.2 & 1.2$\pm$0.1 & 0.5$\pm$0.0$^2$ & 4.2$\pm$0.3 & 2.8$\pm$0.1 & -- & -- & -- \\
PDR2\_NW\_4 & 301.4 & 173$\pm$11 & 13.6$\pm$0.5 & -- & -- & -- & -- & -- & -- & --  \\
PDR3\_NE\_1 & 275.8 & 737$\pm$9 & 42.7$\pm$1.4 & 0.9$\pm$0.4 & 0.4$\pm$0.1 & 3.7$\pm$0.5 & 14.7$\pm$0.3 & 5.0$\pm$0.2 & 0.3$\pm$0.3 & 2.6$\pm$0.1 \\
PDR3\_NE\_2 & 271.4 & 391$\pm$7 & 34.1$\pm$0.2 & 2.5$\pm$0.4 & 4.0$\pm$0.1 & 15.1$\pm$0.5 & 32.0$\pm$0.3 & 7.6$\pm$0.2 & 2.1$\pm$0.3 & 5.3$\pm$0.0$^2$ \\
PDR3\_NE\_3 & 276.6 & 1370$\pm$13 & 36.6$\pm$1.4 & -- & -- & -- & -- & -- & -- & --  \\
PDR4\_RIDGE\_1 & 228.0 & 602$\pm$9 & 9.3$\pm$0.1 & 4.2$\pm$0.2 & 2.2$\pm$0.1 & 33.9$\pm$0.3 & 34.2$\pm$0.1 & 2.4$\pm$0.1 & 5.4$\pm$0.2 & 4.0$\pm$0.1 \\
SK-66D35\_1 & -- &  -- & -- & 1.0$\pm$0.3 & 0.3$\pm$0.1 & 0.7$\pm$0.1 & 2.0$\pm$0.1 & 0.5$\pm$0.1 & -- & 0.3$\pm$0.0$^2$ \\
SK-66D35\_2 & 279.5 & 584$\pm$9 & 5.9$\pm$0.4 & -- & -- & -- & 0.4$\pm$0.1 & -- & -- & -- \\
SK-67D2\_1 & -- &  -- & -- & -- & -- & 0.7$\pm$0.1 & 0.3$\pm$0.0$^2$ & -- & -- & -- \\
SK-68D129\_1 & 259.0 & 708$\pm$10 & 2.2$\pm$0.5 & -- & -- & 0.6$\pm$0.2 & -- & -- & -- & \\
SK-68D140\_1 & 250.0 & 635$\pm$9 & 7.0$\pm$0.6 & -- & -- & -- & -- & -- & -- & --  \\
SK-68D140\_2 & 273.9 & 809$\pm$14 & 6.2$\pm$1.5 & -- & -- & -- & -- & -- & -- & --  \\
SK-68D140\_3 & 262.0 & 590$\pm$9 & 5.4$\pm$1.2 & -- & -- & -- & -- & -- & -- & --  \\
SK-68D155\_1 & 246.1 & 383$\pm$8 & 8.0$\pm$0.5 & -- & -- & -- & -- & -- & -- & --  \\
SK-68D155\_2 & 256.6 & 374$\pm$7 & 1.8$\pm$0.4 & -- & -- & -- & -- & -- & -- & --  \\
SK-68D26\_1 & 251.3 & 476$\pm$9 & 2.2$\pm$0.5 & -- & -- & -- & -- & -- & -- & --  \\
SK-69D228\_1 & 247.0 & 314$\pm$9 & 6.1$\pm$0.3 & -- & -- & -- & -- & -- & -- & --  \\
\enddata
\tablenotetext{1}{H\,{\sc i} intensity integrated over a
  velocity range defined by the FWHM of the observed [C\,{\sc ii}] line (Section~\ref{sec:atomic-gas}). }
\tablenotetext{2}{Error below 0.05. }
\end{deluxetable*}

\begin{deluxetable*}{lcccccccccc} 
\tabletypesize{\footnotesize}

 \centering \tablecolumns{11} \small
\tablewidth{0pt}
\tablecaption{Integrated Intensities of Spectral Line  for  SMC Sample}
\tablenum{4}
\tablehead{\colhead{LOS} & \colhead{$V_{\rm LSR}$} & \colhead{$I^{1}_{\rm [HI];CNM}$} & \colhead{$I_{\rm [CII]}$} &   \colhead{$I_{\rm [CI](1-0)}$} &  \colhead{$I_{\rm [CI](2-1)}$} &    \colhead{$I_{\rm CO(1-0)}$} &  \colhead{$I_{\rm CO(3-2)}$}  &  \colhead{$I_{\rm CO(7-6)}$} & \colhead{$I_{\rm ^{13}CO(1-0)}$} &  \colhead{$I_{\rm ^{13}CO(3-2)}$}\\
\colhead{} &\colhead{[km\,s$^{-1}$]} &\multicolumn{9}{c}{[K\,km\,s$^{-1}$]} }\\ %& \colhead{[K\,km\,s$^{-1}$]} &  \colhead{[K\,km\,s$^{-1}$]} & \colhead{[K\,km\,s$^{-1}$]} &  \colhead{[K\,km\,s$^{-1}$]} & \colhead{[K\,km\,s$^{-1}$]} & \colhead{[K\,km\,s$^{-1}$]} &\colhead{[K\,km\,s$^{-1}$]}  & \colhead{[K\,km\,s$^{-1}$]}}\\
\startdata
\cutinhead{Small Magellanic Cloud}
AzV18\_1 & 122.0 & 1709$\pm$11 & 5.9$\pm$0.6 & -- & -- & --  & -- & -- & -- & -- \\
AzV18\_2 & 136.7 & 764$\pm$8 & 3.4$\pm$0.5 & -- & -- & --  & -- & -- & -- & -- \\
SMC\_B2\_6\_1 & 120.9 & 731$\pm$4 & 15.6$\pm$0.4 & 0.9$\pm$0.1 & 0.8$\pm$0.1 & 5.5$\pm$0.2 & 5.6$\pm$0.1 & 0.8$\pm$0.1 & 0.8$\pm$0.1 & 0.7$\pm$0.1 \\
SMC\_B2\_6\_2 & 125.3 & 1273$\pm$5 & 5.3$\pm$0.4 & -- & -- & --  & -- & -- & -- & -- \\
SMC\_HI\_2\_1 & -- &  -- & -- & 0.7$\pm$0.2 & -- & -- & -- & -- & -- & \\
SMC\_HI\_3\_1 & 145.4 & 2347$\pm$8 & 4.4$\pm$0.8 & 0.6$\pm$0.1 & -- & -- & -- & 0.4$\pm$0.1 & -- & \\
SMC\_HI\_4\_1 & 123.3 & 1103$\pm$4 & 5.5$\pm$0.4 & -- & -- & --  & -- & -- & -- & -- \\
SMC\_HI\_4\_2 & 114.8 & 508$\pm$3 & 2.3$\pm$0.3 & -- & -- & --  & -- & -- & -- & -- \\
SMC\_HI\_6\_1 & 149.7 & 431$\pm$4 & 1.3$\pm$0.3 & -- & -- & --  & -- & -- & -- & -- \\
SMC\_LIRS36\_1 & 126.4 & 383$\pm$3 & 16.4$\pm$0.1 & 2.0$\pm$0.1 & 1.8$\pm$0.1 & 7.6$\pm$0.2 & 18.9$\pm$0.1 & 1.5$\pm$0.1 & 0.7$\pm$0.1 & 2.8$\pm$0.2 \\
SMC\_LIRS36\_2 & -- &  -- & -- & -- & -- & 0.5$\pm$0.1 & -- & -- & -- & \\
SMC\_LIRS49\_1 & 110.4 & 448$\pm$3 & 5.6$\pm$0.0$^2$ & -- & -- & -- & 0.8$\pm$0.1 & -- & -- & -- \\
SMC\_LIRS49\_2 & 115.0 & 502$\pm$3 & 14.3$\pm$0.0$^2$ & 2.5$\pm$0.2 & 1.3$\pm$0.1 & 8.6$\pm$0.2 & 18.1$\pm$0.1 & 0.6$\pm$0.1 & 0.6$\pm$0.1 & 2.0$\pm$0.2 \\
SMC\_LIRS49\_3 & 125.9 & 1173$\pm$5 & 5.2$\pm$0.0$^2$ & -- & -- & 1.3$\pm$0.2 & 2.4$\pm$0.1 & -- & -- & 0.3$\pm$0.1 \\
SMC\_LIRS49\_4 & 137.2 & 191$\pm$2 & 1.4$\pm$0.0$^2$ & -- & -- & --  & -- & -- & -- & -- \\
SMC\_NE\_1a\_1 & 148.9 & 496$\pm$4 & 4.5$\pm$0.1 & 0.5$\pm$0.1 & 0.5$\pm$0.1 & 4.0$\pm$0.1 & 4.3$\pm$0.1 & 0.6$\pm$0.1 & 0.4$\pm$0.0$^2$ & \\
SMC\_NE\_3c\_1 & 175.1 & 170$\pm$2 & 1.1$\pm$0.1 & 1.2$\pm$0.1 & 0.7$\pm$0.1 & 4.2$\pm$0.2 & 4.1$\pm$0.1 & -- & 0.7$\pm$0.0$^2$ & 0.4$\pm$0.1 \\
SMC\_NE\_3g\_1 & 169.1 & 690$\pm$4 & 4.0$\pm$0.1 & 0.6$\pm$0.2 & 0.6$\pm$0.1 & 4.3$\pm$0.1 & 3.2$\pm$0.1 & 0.5$\pm$0.1 & 0.4$\pm$0.0$^2$ & 0.5$\pm$0.1 \\
SMC\_NE\_4a\_hi\_1 & -- &  -- & -- & -- & -- & -- & 0.7$\pm$0.1 & -- & -- & 0.2$\pm$0.1 \\
SMC\_NE\_4a\_hi\_2 & 153.9 & 131$\pm$2 & 1.6$\pm$0.2 & 0.5$\pm$0.1 & -- & 3.1$\pm$0.1 & 2.1$\pm$0.1 & -- & 0.2$\pm$0.0$^2$ & \\
SMC\_NE\_4a\_hi\_3 & 122.2 & 551$\pm$4 & 1.3$\pm$0.3 & -- & -- & --  & -- & -- & -- & -- \\
SMC\_NE\_4c\_low\_1 & 121.6 & 780$\pm$4 & 2.1$\pm$0.4 & 0.7$\pm$0.1 & -- & 5.8$\pm$0.1 & 1.7$\pm$0.1 & 0.2$\pm$0.1 & 0.7$\pm$0.0$^2$ & \\
\enddata
\tablenotetext{1}{H\,{\sc i} intensity integrated over a
  velocity range defined by the FWHM of the observed [C\,{\sc ii}] line (Section~\ref{sec:atomic-gas}). }
\tablenotetext{2}{Error below 0.05. }
\end{deluxetable*}
% We present a selection of GOT\,C+ lines--of--sight in [C\,{\sc ii}],
% H\,{\sc i}, and $^{12}$CO emission in Figure~\ref{fig:spectra}.

\subsection{[C\,{\sc ii}] and [C\,{\sc i}] observations}
\label{sec:c-sc-ii}

We surveyed the Magellanic Clouds in the [C\,{\sc i}]
$^3$P$_1$--$^3$P$_0$, [C\,{\sc i}] $^3{\rm P}_2-^3$P$_1$, and [C\,{\sc
  ii}] $^2$P$_{3/2}-^2$P$_{1/2}$ fine--structure lines at
492.1607\,GHz, 809.3420\,GHz, and 1900.5469\,GHz (rest frequency),
respectively, with the HIFI \citep{deGraauw2010} instrument aboard the
{\it Herschel Space Observatory} \citep{Pilbratt2010}.  These
observations are part of the {\it Herschel} Open Time 1 Project {\tt
  OT1\_jpineda\_1}. The $^{12}$CO $J=7\to6$ (806.6518\,GHz) rotational
line was observed simultaneously with the [C\,{\sc i}]
$^3$P$_2$--$^3$P$_1$ line. There are 9 LOS where only the $^{12}$CO
$J=7\to6$ was observed due to an error in the frequency
configuration. These LOS correspond to diffuse LOS where the [C\,{\sc
  i}] $^3$P$_1$--$^3$P$_0$ line was not detected and therefore we do
not expect a detection of the usually weaker [C\,{\sc i}]
$^3$P$_2$--$^3$P$_1$ line.

The [C\,{\sc ii}] 1.9\,THz observations were carried out with the HIFI
Band 7b receiver, which is employs Hot Electron Bolometer (HEB)
mixers, in the LoadChop with reference observing mode.  The HEB bands
in HIFI show prominent electrical standing waves that are produced
between the HEB mixing element and the first low noise amplifier. The
standing wave shape is not a standard sinusoid and is difficult to
remove from the resulting spectrum using standard fitting methods
\citep{Higgins2009}.  To remove these standing waves we used a
procedure available in {\tt HIPE} \citep{Ott2006} version 12, which
uses a library of standing wave shapes to find the best fit to correct
the observed spectrum (see \citealt{Higgins2011} for a detailed
description of this method).  In {\tt HIPE} we also removed residual
standing waves by fitting a single sinusoidal function using the {\tt
  FitHIFIFringe()} procedure. After all standing waves are removed, we
exported our data to the \href{http://www.iram.fr/IRAMFR/GILDAS}{\tt
  CLASS90}\footnote{{\tt http://www.iram.fr/IRAMFR/GILDAS}} data
analysis software, which we used to combine the two polarization, fit
the data with polynomial baselines (typically of order 3), and smooth
in velocity.

The angular resolution of the [C\,{\sc ii}] observations is 12\arcsec.
We divided the data by a factor of 0.61 to transform the data from an
antenna temperature to a main-beam temperature scale\footnote{The beam
  efficiencies of all HIFI bands are presented in {\tt
    http://herschel.esac.esa.int/twiki/pub/Public/\\
    HifiCalibrationWeb/HifiBeamReleaseNote\_Sep2014.pdf}}.  The data
were produced by the wide band spectrometer (WBS), which has a channel
width of 1\,MHz (0.16 km s$^{-1}$ at 1.9 THz).  We later smoothed the
data to a resolution of 0.8 km s$^{-1}$.  For this resolution the
average rms noise of our data is\footnote{ For the typical [C\,{\sc
    ii}] FWHM line width of about 3\,km s$^{-1}$, this sensitivity
  limit corresponds to $1.1\times10^{-6}$ erg s$^{-1}$ cm$^{-2}$
  sr$^{-1}$. The integrated intensity in units of K\,km\,s$^{-1}$ can
  be converted that in units of erg s$^{-1}$ cm$^{-2}$ sr$^{-1}$ using
  $I$[K\,km\,s$^{-1}$]=$1.43\times10^{5}$$I$[erg s$^{-1}$ cm$^{-2}$
  sr$^{-1}$] \citep{Goldsmith2012}. } 0.1\,K.

We observed the [C\,{\sc i}] $^3$P$_1$--$^3$P$_0$ line using HIFI Band
1, while the [C\,{\sc i}] $^3$P$_2$--$^3$P$_1$ and $^{12}$CO $J=7\to6$
lines were observed simultaneously using HIFI Band 3. The angular
resolutions are 44\arcsec\ and 26.5\arcsec\ for Bands 1 and 3,
respectively. We applied main-beam efficiencies of 0.651 and 0.645,
respectively, to convert these data from an antenna temperature to a
main-beam temperature scale.  The typical rms noise for the [C\,{\sc
  i}] $^3$P$_1$--$^3$P$_0$ line was 0.035\,K in a 0.91 km\,s$^{-1}$
channel width. For the [C\,{\sc i}] $^3$P$_2$--$^3$P$_1$ and $^{12}$CO
$J=7\to6$ lines the typical rms noise was 0.02\,K in a
1.1\,km\,s$^{-1}$ channel width.

\subsection{CO observations}\label{sec:mopra-observations}

To complement the {\it Herschel} data, we observed the $J = 1 \to 0$
transitions of $^{12}$CO, $^{13}$CO, and C$^{18}$O with the ATNF
Mopra\footnote{The Mopra radio telescope is part of the Australia
  Telescope which is funded by the Commonwealth of Australia for
  operation as a National Facility managed by CSIRO.}  Telescope
(project {\tt M580}).  The Mopra 22m telescope has an angular
resolution of 33\arcsec\ at 115\,GHz.  Typical system temperatures
were 600, 300, and 250\,K for $^{12}$CO, $^{13}$CO, and C$^{18}$O,
respectively.  To convert from antenna to main--beam temperature
scale, we used a main-beam efficiency of 0.42 \citep{Ladd05}. All
lines were observed simultaneously with the MOPS spectrometer in zoom
mode. The spectra were smoothed in velocity to 0.87\,km s$^{-1}$ for
$^{12}$CO and to 0.91\,km\,s$^{-1}$ for $^{13}$CO. The typical rms
noise is 0.06\,K for $^{12}$CO and 0.05\,K for $^{13}$CO. The
C$^{18}$O line was detected only in a small number of LOSs, and we
will not use these observations in the present analysis.  We checked
pointing accuracy every 60 minutes using a nearby SiO maser.

We used the APEX\footnote{This publication is based in part on data
  acquired with the Atacama Pathfinder Experiment (APEX). APEX is a
  collaboration between the Max-Planck-Institut f\"ur Radioastronomie,
  the European Southern Observatory, and the Onsala Space Observatory}
12m telescope \citep{Guesten2006} to observe the $^{12}$CO $J=3\to2$
and $^{13}$CO $J=3\to2$ lines in 25 LOS in the LMC and SMC that have
been detected in our low--$J$ observations (project {\tt M0036\_93}).
The angular resolution of the APEX telescope is 17.5\arcsec\ at
345\,GHz. We converted from antenna to main--beam temperature scale
using an efficiency of 0.69 for 345\,GHz.  The typical rms noise in
the $^{12}$CO $J=3\to2$ data is 0.08\,K for a 0.33\,km\,s$^{-1}$
channel width and in the $^{13}$CO $J=3\to2$ data is 0.05\,K for a
0.33\,km\,s$^{-1}$ channel width.

\subsection{H\,{\sc i}, H$\alpha$,  H66$\alpha$, and Dust Continuum data}
\label{sec:hi-radio-dust}

We used the H\,{\sc i} 21\,cm maps of the entire LMC and SMC presented
by \citet{Kim2003} and \citet{Stanimirovic1999} (see also
\citealt{Staveley-Smith2003}), respectively. These maps of the
Magellanic clouds were made by combining interferometric (ATCA) and
single dish (Parkes) observations.  The map of the LMC has an angular
resolution of 60\arcsec, corresponding to a spatial resolution of 15
pc, while the SMC map has an angular resolution of 98\arcsec\
corresponding to a spatial scale of 29\,pc.
% Both maps have an angular resolution of 1\arcmin, corresponding to a
% spatial resolution of 15\,pc in the LMC and 29\,pc in the SMC.
The rms noise in the LMC map is 2.4\,K in a 1.65\,km\,s$^{-1}$ channel
width and in the SMC map is 1.3\,K over a 1.65\,km\,s$^{-1}$ channel
width.

% To compare with the
%{\it Herschel} spectral line observations, we assume that the
%distribution of the atomic gas is smooth on spatial scales between
%3\,pc and 40\,pc, the spatial resolution of our observations.
%This assumption
%is confirmed in the Milky Way (CHECK THOR).

To estimate the contribution to the [C\,{\sc ii}] emission from
ionized gas, we used the H$\alpha$ map from Southern Halpha Sky Survey
Atlas survey \citep[SHASSA;][]{Gaustad2001}. The SHASSA maps cover the
entire LMC and SMC with an angular resolution of 48\arcsec\ and a
sensitivity level of 0.5 Rayleigh
(R=$10^{6}$/4$\pi$\,photons\,cm$^{-2}$\, s$^{-1}$\,sr$^{-1}$).  We
also used the NASA 70--m Deep Space Network telescope (DSS--43) to
observe the H66$\alpha$ hydrogen radio recombination line at
22.364\,GHz in one of our sources with the aim to test the effect of
dust extinction on the H$\alpha$ observations. The angular resolution
of the DSS--43 at 22.364\,GHz is 48\arcsec. We converted the data from
an antenna temperature to a main beam temperature scale using a
main--beam efficiency of 0.50. The H66$\alpha$ spectrum has a rms
noise of 5.4\,mK in a 3.5\,km\,s$^{-1}$ channel width.

We also compared our {\it Herschel} spectral line observations with
dust continuum maps of the LMC and SMC taken using the PACS and SPIRE
instruments on {\it Herschel}.  These maps are part of the HERITAGE
survey \citep{Meixner2013} which provides images at 100\,$\mu$m,
160\,$\mu$m (PACS), 250\,$\mu$m, and 350\,$\mu$m, and 500\,$\mu$m
(SPIRE).  The angular resolution of these maps are 7.7\arcsec\
(100\,$\mu$m), 12\arcsec\ (160\,$\mu$m), 18\arcsec\ (250\,$\mu$m),
25\arcsec\ (350\,$\mu$m), and 40\arcsec\ (500\,$\mu$m).  The
foreground dust emission originating from the Milky Way was subtracted
using a linear baseline \citep{Meixner2013}.  We also used the {\it
  Spitzer} 24\,$\mu$m MIPS maps observed as part of the SAGE survey
\citep{Meixner06} and has an angular resolution of 6\arcsec.

%#$ and was
%$convolved to a 12\arcsec\ resolution for comparison with the [C\,{\sc
%  ii}] observations.

\begin{figure}[t]
\centering
\includegraphics[width=0.49\textwidth,angle=0]{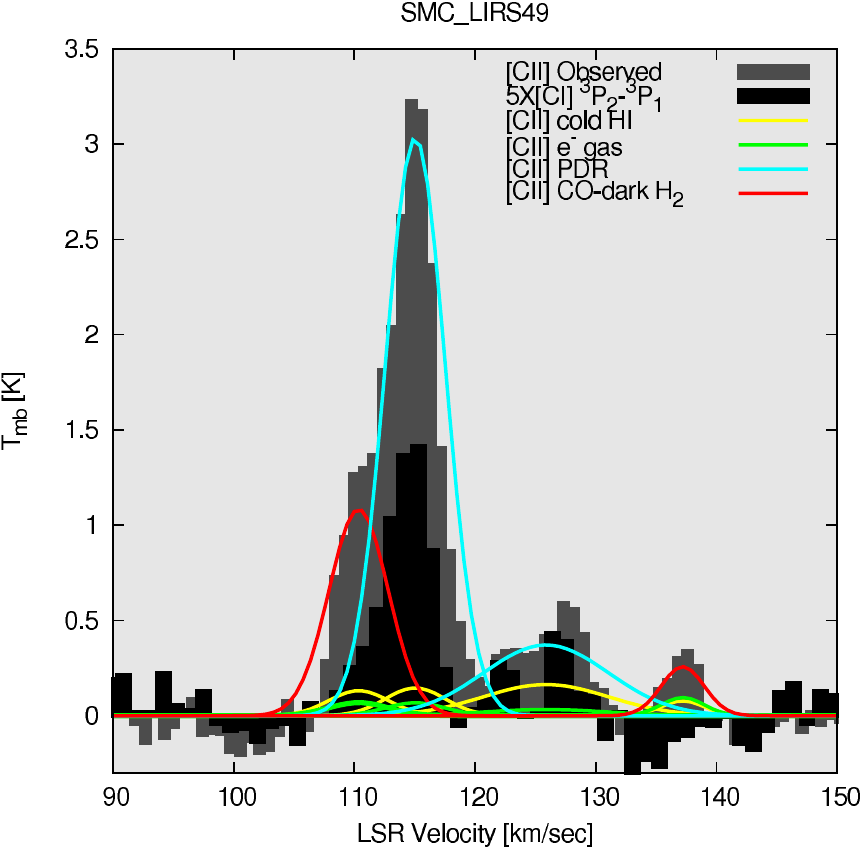}
\caption{Sample [C\,{\sc ii}] and [C\,{\sc i}] $^3$P$_2-^3$P$_1$
  spectrum illustrating the decomposition between [C\,{\sc ii}]
  emission associated with PDRs, CO--dark H$_2$ gas, cold H\,{\sc i},
  and ionized gas. See Section~\ref{sec:origin-c-sc} for details on
  the decomposition of the [C\,{\sc ii}] emission.}
\label{fig:decomposition_demo}
\end{figure}

\begin{figure*}[t]
\centering
\includegraphics[width=0.8\textwidth,angle=0]{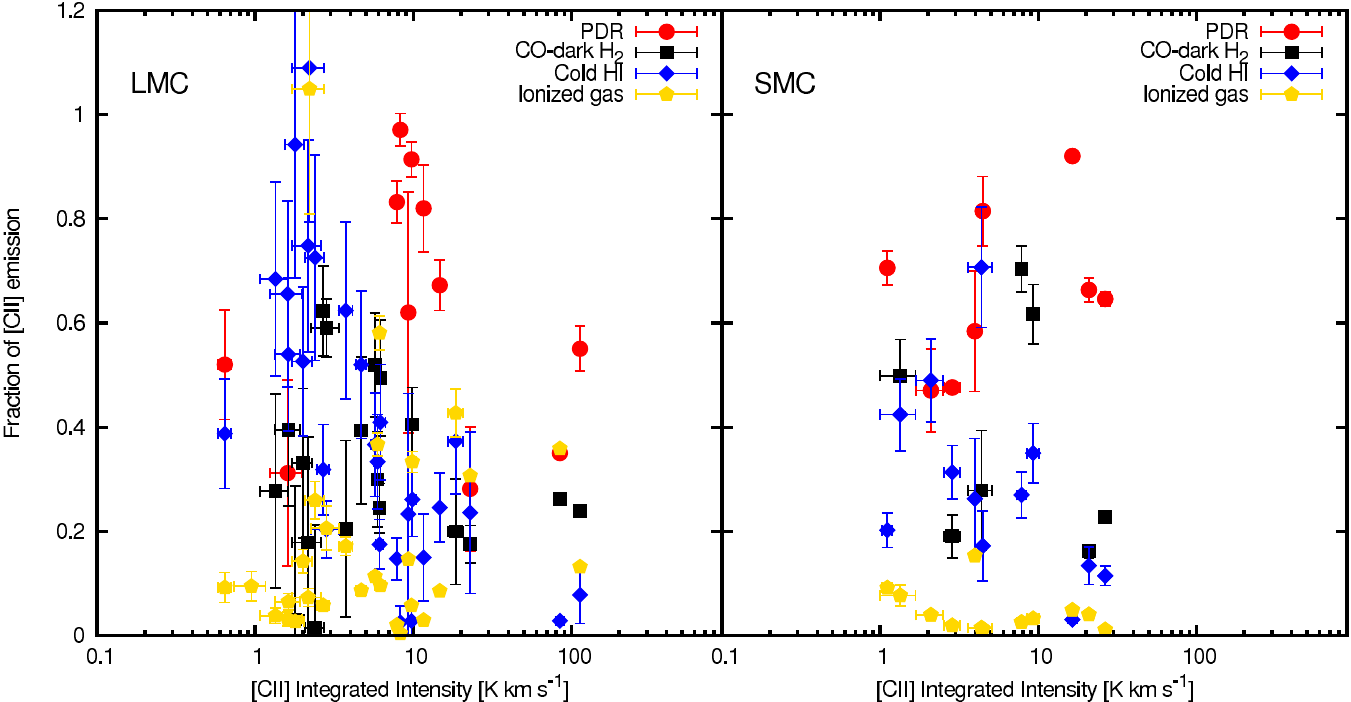}
\caption{The fraction of the [C\,{\sc ii}] emission that we estimate
  arises from ionized gas, cold atomic gas, CO--dark H$_2$, and
  photon dominated regions (PDRs) as a function of the observed
  [C\,{\sc ii}] emission in the LMC and SMC.}
\label{fig:percentage_lmc}
\end{figure*}

\subsection{Angular Resolution}
% Define problem

  The analysis presented in this paper uses a multi--wavelength data
  set from different telescopes. Therefore, the angular resolution of
  the observations is not uniform. While maps can be convolved to a
  uniform resolution, this is not the case for pointed [C\,{\sc ii}],
  [C\,{\sc i}], and CO observations, with angular resolutions ranging
  from 12\arcsec\ for [C\,{\sc ii}] and 60\arcsec\ for H\,{\sc i}.
  Therefore, absolute intensities and line ratios used in our analysis
  can be affected by beam dilution effects.

  We studied the effect of beam dilution in our observations by
  smoothing the 12\arcsec\ angular resolution 160\,$\mu$m HERITAGE
  dust continuum map, which is assumed to be a proxy for the
  distribution of gas in the LMC and SMC. We then compared the
  resulting intensities at different angular resolutions.  In the case
  of ionized gas traced by H$\alpha$ observations, we used the {\it
    Spitzer} 24\,$\mu$m images of the Magellanic clouds which is a
  proxy for hot dust emission associated with ionized gas. Details of
  this study are presented in Appendix~\ref{sec:beam-dilut-corr}.  The
  effects of beam dilution in our results will be discussed as they
  are presented throughout the paper. In general, we find that beam
  dilution has a minor effect in our results for most of our sample,
  with the exception of a handful of point--like, unresolved PDR
  sources. Note that the data used in our analysis is not corrected
  for beam dilution effects, due to the uncertainty on whether the
  24\,$\mu$m or 160\,$\mu$m are good tracers of the ionized and
  neutral gas components, respectively.

\section{The Origin of the [C\,{\sc ii}] emission in the Magellanic Clouds}
\label{sec:origin-c-sc}

The observed [C\,{\sc ii}] emission arises from gas associated with
hydrogen in the form of H$^0$, H$_2$, and H$^+$. In the latter
gas component, the collisions are mainly with electrons due to their
higher speeds, so we refer to it as $\mathrm{e^-}$ gas. Thus, the
observed [C\,{\sc ii}] emission is given by,
\begin{equation}
  I_{\rm [C\,II]}=I^{\rm H^0}_{\rm [C\,II]}+I^{\rm H_2}_{\rm [C\,II]}+I^{\rm e^-}_{\rm [C\,II]}.
\end{equation}

For optically thin emission, the [C\,{\sc ii}] intensity (in units of
K\,km\,s$^{-1}$) is related to the C$^+$ column density, $N_{\rm C^+}$
(cm$^{-2}$), and volume density of the collisional partner (assumed to
be uniform along the line of sight), $n$ (e$^-$, H$^0$, or H$_2$;
cm$^{-3}$), as \citep[see e.g.][]{Goldsmith2012}

\begin{multline}
\label{eq:6}
%\begin{aligned}
I_{\rm [CII] } = \\ N_{{\rm C}^+} \left [3.05\times10^{15} \left (1+0.5
\left (1+\frac{A_{ul}}{R_{ul}n} \right )e^{\frac{91.21}{T_{\rm kin}}} \right
)\right ]^{-1},
%\end{aligned}
%\end{equation}
\end{multline}
where $A_{ul}=2.3\times10^{-6}$\,s$^{-1}$ is the Einstein spontaneous
decay rate and $R_{ul}$ is the collisional de--excitation rate
coefficient at a kinetic temperature $T_{\rm kin}$, with $u$ and $l$
denoting the upper and lower energy levels. Values of $R_{ul}$ for
collisions with H$^0$, H$_2$, and $e^-$ as a function of the kinetic
temperature are available from \citet{Barinovs2005},
\citet{Wiesenfeld2014}, and \citet{Wilson2002}, respectively.  For
example, at $T_{\rm kin}=100$\,K, $R_{ul}=7.58\times10^{-10}$ and
$5.12\times10^{-10}$\,cm$^{3}$\,s$^{-1}$ for collisions with atomic
and molecular hydrogen\footnote{Assuming an ortho--to--para H$_2$
  ratio, OPR=1. See discussion in \citet{Wiesenfeld2014}.},
respectively, while at $T_{\rm kin}=8000$\,K,
$R_{ul}=5.2\times10^{-8}$\,cm$^{3}$\,s$^{-1}$ for collisions with
electrons.

\citet{Dufour1982} studied the carbon abundance in H\,{\sc ii} regions
in the Magellanic clouds deriving 12+log(C/H)=7.9 for the LMC and
12+log(C/H)=7.16 for the SMC. Relative to the gas--phase carbon
fractional abundance of [C/H]=1.4$\times10^{-4}$ determined in the
Milky Way \citep{Sofia1997}, the carbon abundances in the LMC and SMC
are 1.8 and 10 times lower, respectively. We used the
fractional carbon abundances derived by \citet{Dufour1982} in the
analysis presented here.

In the following we study the relative contribution to the observed
[C\,{\sc ii}] emission from ionized, atomic, and molecular gas. In
Figure~\ref{fig:decomposition_demo} we show an example of the
decomposition of the [C\,{\sc ii}] emission originating from different
ISM components.  The derived [C\,{\sc ii}] emission fractions as a
function of the  observed [C\,{\sc ii}] intensity for the entire
sample are summarized in Figure~\ref{fig:percentage_lmc}.

\subsection{Ionized gas}
\label{sec:ionized-gas}

We estimated the [C\,{\sc ii}] intensity originating from ionized gas
($\mathrm{e^-}$ gas) using the Southern Halpha Sky Survey Atlas survey
\citep[SHASSA;][]{Gaustad2001} H$\alpha$ emission maps of the LMC and
SMC.  A commonly defined quantity used to relate the properties of the
ionized gas and the observed H$\alpha$ emission is the emission
measure (EM), defined as the integral of the electron volume density
squared along the line of sight,
\begin{equation}\label{eq:1}
EM=\int n^2_e dl.
\end{equation}

Assuming that the electron density is constant along the line of
sight, this equation can be simplified to
\begin{equation}\label{eq:8}
EM=\left<n_e\right>N_e\simeq\left<n_e\right>N_{\rm H^+}.
\end{equation}

The emission measure is related to the intensity of the H$\alpha$
line, in units of Rayleigh (R), as (e.g. \citealt{Reynolds1991})
\begin{equation}\label{eq:7}
\left ( \frac{ EM}{{\rm pc\, cm}^{-6}} \right) = 2.75 \left (
  \frac{T_{\rm kin}}{10^4\,{\rm K}} \right)^{0.9} \left ( \frac{I_{\rm H\alpha}}{\rm R}
  \right).
\end{equation}

The critical density, $n_{\rm cr}\equiv A_{\rm ul}/R_{\rm ul}$, is the
density at which the collisional de--excitation rate is equal to the
effective spontaneous decay rate, and for collisions of C$^+$ with
electrons at $T_{\rm kin}=8000$\,K is equal to 44\,cm$^{-3}$.  For
electron densities much smaller than the critical density $n_e \ll
n_{\rm cr}$, Equation~(\ref{eq:6}) can be written as
\begin{equation}
\label{eq:3}
I^{e^-}_{\rm [CII] } = n_e N_{{\rm C}^+} \left [1.52\times10^{15}
n_{\rm cr} e^{\frac{91.21}{T_{\rm kin}}}\right ]^{-1}.
\end{equation}

The term $n_e N_{{\rm C}^+}$ can be written in terms of the fractional
abundance of ionized carbon, $X_{\rm C^+}=N_{\rm C^+}/N_{\rm H^+}$, as
$X_{\rm C^+} n_e N_{{\rm H}^+}=X_{\rm C^+} EM$.  \citet{Paradis2011}
estimated typical electron densities for regions emitting different
regimes of H$\alpha$ emission. For the range of H$\alpha$ intensities
in our sample, electron densities vary from 0.05 to 3.98\,cm$^{-3}$,
thus validating the assumption that $n_e \ll n_{\rm cr}$.
Equation~(\ref{eq:3}) can thus be rewritten in terms of the emission
measure as,

\begin{equation}
\label{eq:4}
I^{e^-}_{\rm [CII] } = X_{\rm C^+} EM \left [1.52\times10^{15}
n_{\rm cr} e^{\frac{91.21}{T_{\rm kin}}}\right ]^{-1},
\end{equation}
and can be used to estimate the contribution of ionized gas to the
observed [C\,{\sc ii}] emission from H$\alpha$ observations.

\begin{figure}
\centering
\includegraphics[width=0.45\textwidth,angle=0]{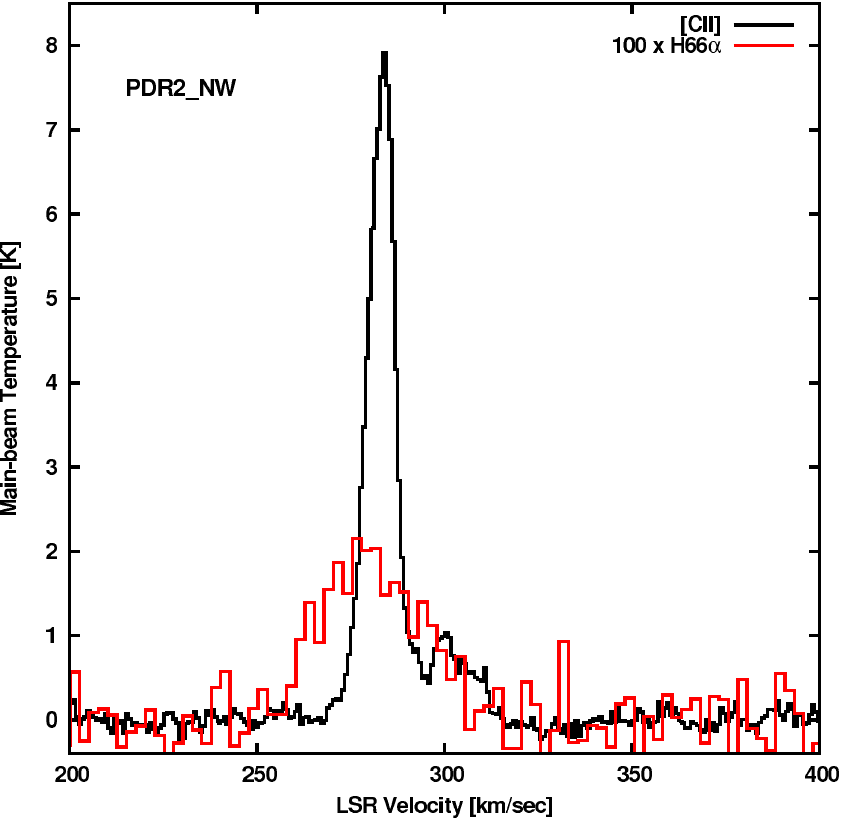}
\caption{Comparison between [C\,{\sc ii}] and the
  H66$\alpha$ radio recombination line in the PDR2\_NW position.}
\label{fig:RRL_obs}
\end{figure}

In Figure~\ref{fig:percentage_lmc}, we show the fraction of the
[C\,{\sc ii}] emission we estimate arises from ionized gas as a
function of the observed [C\,{\sc ii}] emission in the LMC and SMC.
 We assumed $T_{\rm kin}=8000$\,K for the kinetic
  temperature of the ionized gas.  In LOSs with multiple velocity
components, we assumed that each component contributes equally to the
derived [C\,{\sc ii}] emission from ionized gas, as suggested by the
H$66\alpha$ spectrum in Figure~\ref{fig:RRL_obs} that shows two
[C\,{\sc ii}] velocity components with varying peak intensities but
uniform recombination line emission.  We find that ionized gas tends
to contribute a small fraction of the [C\,{\sc ii}] emission, with
typical fractions around 19\% in the LMC and 5\% in the SMC.  These
contributions from ionized gas to the observed [C\,{\sc ii}] emission
are in agreement with those estimated using the unobscured [N\,{\sc
  ii}] fine structure lines by \citet{Chevance2016} and
\citet{Okada2015} in the 30\,Dor and N159 regions in the LMC,
respectively, and by \citet{Requena-Torres2016} in several star
forming regions in the SMC.  The derived contributions from ionized
gas to the observed [C\,{\sc ii}] emission in the LMC and SMC are also
consistent with those estimated in the Galactic plane
\citep{Pineda2013}.

We estimated the H$^+$ column density in our LOSs using $EM=n_e
N_{{\rm H}^+}$ and the volume densities of the ionized gas suggested
by \citet{Paradis2011} for different ranges of H$\alpha$
emission. Typical H$^+$ column densities in our sample are
$10^{20.8\pm0.35}$\,cm$^{-2}$ for the LMC and
$10^{20.5\pm0.14}$\,cm$^{-2}$ for the SMC.

Note that the H$\alpha$ emission used here might be affected by
extinction from dust grains making our estimate of the emission
measure from H$\alpha$ observations a lower limit.  We tested the
effect of dust extinction in the H$\alpha$ observations by assuming
that half of the visual extinction \citep[$A_{\rm H\alpha}=0.81A_{\rm
  V}$, e.g.][]{Viallefond1986,Parker1992} derived for each
line--of--sight (Section~\ref{sec:visu-extinct-determ}) is in front of
the H$\alpha$ sources.  We find that the H$\alpha$ intensities would
be underestimated by an average factor of 4.6 in the LMC and 1.4 in
the SMC. In the LMC the large factor is dominated by three warm and
dense PDRs, PDR4\_RIDGE, PDR3\_NW, and LMC\_1\_NW, which are bright in
[C\,{\sc ii}] and have the largest visual extinctions. Without these
PDR regions, we find that the H$\alpha$ emission would be affected by
a factor of 1.65 in the LMC. If a correction to the H$\alpha$
intensity for dust extinction is applied to the calculation of the
contribution of ionized gas to the observed [C\,{\sc ii}] emission,
the fraction would increase by a similar factor.  Note that for the
warm and dense PDRs mentioned above, applying a correction to the
contribution from ionized gas to the observed [C\,{\sc ii}] emission
would result in larger [C\,{\sc ii}] intensities than observed,
suggesting that less than $A_{\rm V}/2$ of extinction is in the
foreground of these sources.

We tested the reliability of H$\alpha$ as a tracer of ionized gas by
observing the unobscured H66$\alpha$ radio recombination line using
the DSS--43 NASA Deep Space Network Telescope in one of our
lines--of--sight (Figure~\ref{fig:RRL_obs}).  Following
\citet{Alves2015}, we converted the integrated intensity of the
H66$\alpha$ line to the emission measure, assuming $T_{\rm
  kin}=8000$\,K, to be 1.9$\times10^{4}$\,cm$^{-6}$\,pc. For the same
position we obtain $EM=7.8\times10^{3}$\,cm$^{-6}$\,pc from H$\alpha$,
a factor of $\sim$2.4 lower than that obtained from H66$\alpha$. The
position we observed in H66$\alpha$ corresponds to a warm and dense
photon dominated region likely associated with a large column of dust,
so we expect a smaller effect in more diffuse regions in our sample.

% Following
%e.g. \citet{Alves2015}, we converted the integrated intensity of the
%H67$\alpha$ line to the emission measure, assuming $T_{\rm
%  kin}=8000$\,K, to be 1.47$\times10^{4}$\,cm$^{-6}$\,pc. For the same
%position we obtain $EM=7.8\times10^{3}$\,cm$^{-6}$\,pc from H$\alpha$,
%a factor of $\sim$2 lower than that obtained from H67$\alpha$. The
%position we observed in H67$\alpha$ corresponds to a warm and dense
%photon dominated region likely associated with a large column of dust,
%so we expect a smaller effect in more diffuse regions in our sample.

\subsection{Atomic Gas}
\label{sec:atomic-gas}

We estimated the contribution from atomic gas to the observed [C\,{\sc
  ii}] emission in our sample in the Magellanic Clouds using the
H\,{\sc i} 21\,cm line observations. We calculated $N({\rm H^0})$ for
each LOS, assuming optically thin emission, using $N({\rm
  H^0})=1.82\times10^{18} I({\rm HI})$\,cm$^{-2}$, with $I({\rm HI})$
in units of ${\rm K\,km\,s^{-1}}$. We converted from $N({\rm H^0})$ to
$N({\rm C^+})$ using the carbon fractional abundances of the LMC and
SMC, assuming that all gas--phase carbon associated with atomic gas is
in the form of C$^+$.

We only considered H\,{\sc i} emission that is associated in velocity
space with [C\,{\sc ii}] emission, and to calculate $I({\rm HI})$ we
integrate the H\,{\sc i} spectrum over a velocity range defined by the
full width at half maximum (FWHM) of the [C\,{\sc ii}] lines.  Because
the [C\,{\sc ii}] emission is volume density sensitive for $n_{\rm
  H}<3000\,{\rm cm}^{-3}$, the critical density for collisions with H
\citep{Goldsmith2012}, we expect that the [C\,{\sc ii}] emission is
associated with cold, dense H\,{\sc i} (CNM;\citealt{Wolfire2003})
rather than diffuse, warm H\,{\sc i} (WNM; see discussion in
\citealt{Pineda2013}).)  We therefore consider the H\,{\sc i} emission
that is associated in velocity with [C\,{\sc ii}] to be CNM while the
remaining H\,{\sc i} emission is WNM. In the top panel of
Figure~\ref{fig:spectra}, we illustrate this decomposition by showing
the velocity range as a shaded region in which we assign H\,{\sc i} to
be CNM.  Note that due to the typical complexity of the H\,{\sc i}
line profiles, it is difficult to separate reliably components in
velocity for our entire data set.  

Our method for calculating the H\,{\sc i} column could be
overestimating the column density of cold atomic hydrogen associated
with [C\,{\sc ii}] if a broad WNM component overlaps in velocity with
the [C\,{\sc ii}] emission.  To quantify this effect, we attempted
  to fit multiple Gaussian components to the H\,{\sc i} spectra of
  SMC\_HI\_4 and SMC\_LIRS36 shown in Figure~\ref{fig:spectra}. For
  the Gaussian fitting, we fixed the LSR velocity of the [C\,{\sc ii}]
  components (see Table\,4) and left the peak main--beam temperature
  and line width as free parameters. We then subtracted all Gaussian
  components that are not associated with [C\,{\sc ii}] to the
  observed spectra, in order to isolate the Gaussian component
  associated with [C\,{\sc ii}]. In this component, we recalculated
  the integrated intensity by integrating over the velocity range
  defined by the FWHM of the [C\,{\sc ii}] line, as above. We find
  that we would typically overestimate the H\,{\sc i} intensity
  associated with CNM by a factor of about 2 in SMC\_HI\_4 and
  SMC\_LIRS36. Note, however, that complex H\,{\sc i} spectra such as
  those of SMC\_HI\_4 and SMC\_LIRS36 are typically rare in our
  sample, and mostly correspond to SMC lines--of--sight. In the SMC
  there are 5 LOSs with such complexity out of 10 detected in [C\,{\sc
    ii}]. In the LMC the H\,{\sc i} spectra is mostly similar to those
  of Diff1\_NW and PDR1\_NW as shown in Figure~\ref{fig:spectra}.

%#We therefore did not attempt a
%#%Gaussian decomposition of all the H\,{\sc i} data. 

To calculate the [C\,{\sc ii}] intensity from atomic gas using
Equation~(\ref{eq:6}), we need to assume a gas volume density and a
kinetic temperature. As discussed in Section~\ref{sec:uncertainties},
we adopted a kinetic temperature of $T_{\rm kin}=70$\,K in the case of
velocity components where H\,{\sc i} and [C\,{\sc ii}] are detected
but $^{13}$CO $J = 1\to0$ and/or $J=3\to2$ are not.  In these LOSs, we
were unable to determine the physical conditions of the C$^0$ or CO
layer.  In LOSs with kinetic temperatures of the C$^0$ and/or CO
layers derived in the excitation analysis (Section~\ref{sec:c-sc-i}),
and with [C\,{\sc ii}] and H\,{\sc i} detected, we still assume 70\,K
if the kinetic temperature of the C$^0$ layer is lower than 70\,K.
Otherwise, we assume the kinetic temperature of the C$^0$ layer.  The
volume density is then derived by combining the kinetic temperatures
with the thermal pressures of the diffuse ISM derived in
Section~\ref{sec:therm-press-diff}.

In Figure~\ref{fig:percentage_lmc}, we show the fraction of the
observed [C\,{\sc ii}] emission from atomic gas as a function of the
observed [C\,{\sc ii}]. The contribution from atomic gas to the
observed [C\,{\sc ii}] emission shows a large scatter with average
values of about 43\% for the LMC and 28\% for the SMC.  These average
contributions are consistent with the 20\% contribution derived in the
plane of the Milky Way by \citet{Pineda2013}.

\subsection{CO--dark H$_2$ gas and dense Photon Dominated Regions (PDRs)}
\label{sec:co-dark-h_2} When we subtract the contribution from the
ionized and atomic gas components to the observed total [C\,{\sc ii}]
emission, we typically find that there is residual [C\,{\sc ii}]
emission. This residual emission is produced in regions where carbon
is ionized but hydrogen is molecular.  The molecular hydrogen
component where carbon is mainly ionized, so that it is traced by
[C\,{\sc ii}] and not by CO, is what we refer to as CO--dark H$_2$
gas.  The CO--dark H$_2$ gas can either originate (1) from a diffuse
cloud that has enough dust shielding to maintain only a trace amount
of C$^0$ and CO, thus with most of the carbon in the form of C$^+$, or
(2) from the envelopes of a warm and dense molecular cloud in which
the inner parts have significant column densities of C$^0$ and CO. In
order to associate the observed [C\,{\sc ii}] velocity components to
either of these gas conditions, we use the detection limit of the
$^{13}$CO lines observed in our survey to identify clouds that have
enough shielding to maintain significant amounts of CO, and therefore
are likely associated warm and dense molecular clouds with ongoing
star formation.  We assigned the [C\,{\sc ii}] emission to warm and
dense PDRs in velocity components with enough CO column density for
the $^{13}$CO $J = 1 \to 0$ and/or $^{13}$CO $J = 3 \to 2$ lines to be
detected.  In LOSs where the $^{13}$CO lines are detected, we were
able to perform an excitation analysis
(Section~\ref{sec:kinet-temp-h_2}) that confirms our assumption of
warm and dense gas in these velocity components (see Table\,5).  In
LOSs that are detected in [C\,{\sc ii}] and H\,{\sc i}, but not in
$^{13}$CO $J= 1\to 0$ and/or $^{13}$CO $J = 3 \to 2$, we assumed that
the residual [C\,{\sc ii}] emission arises from diffuse CO--dark H$_2$
gas clouds.  In LOSs associated with PDRs, the [C\,{\sc ii}] emission
from this ISM component tends to be brighter than the emission arising
from other components, which is the result of the higher volume
densities and  temperatures of the gas.  The typical
contributions to the [C\,{\sc ii}] emission associated with PDRs are
62\% and 66\% of the observed emission in LOS in the LMC and SMC,
respectively.  The fraction of the [C\,{\sc ii}] emission arising from
CO--dark H$_2$ shows a large scatter, ranging from $\sim$10\% to
$\sim$80\%. A similar range of CO--dark H$_2$ fractions is found in
clouds in the Milky Way \citep{Langer2014a,Tang2016}.   This large
  scatter could be the result of clouds at different stages in the
  transition from diffuse to dense molecular gas, with clouds having
  varying fractions of CO--dark H$_2$ gas.

Our separation 
  between [C\,{\sc ii}] from diffuse CO--dark H$_2$ gas and from warm
  and dense PDRs is sensitive to our ability to detect the $^{13}$CO
  $J = 1 \to 0$ and/or $^{13}$CO $J = 3 \to 2$ lines.  For the thermal
  pressure of the diffuse ISM in the LMC and SMC derived in
  Section~\ref{sec:therm-press-diff} and a  kinetic temperature of the
  CO--dark H$_2$ of 49K (Section~\ref{sec:h-h_2-transition}), we
  obtain $n$(H$_2$)=694\,cm$^{-3}$ in the LMC and 2040\,cm$^{-3}$ in
  the SMC. For these conditions, the $^{13}$CO column density required for
  a $3\sigma$ detection of the $^{13}$CO $J=1 \to 0$ and $^{13}$CO $J = 3\to2$ is
  4$\times$10$^{14}$\,cm$^{-2}$ and 3$\times$10$^{15}$\,cm$^{-2}$
  for the LMC and 4.5$\times$10$^{14}$\,cm$^{-2}$ and
  1$\times$10$^{15}$\,cm$^{-2}$ of SMC, respectively.

\subsection{Uncertainties}
\label{sec:uncertainties}
 
%# Discuss Effect of H\,{\sc i} temperature. 
The main uncertainty in our decomposition originates from the
assumption of the temperature of atomic and molecular gas. The assumed
kinetic temperature is used in the derivation of thermal pressures
(Section~\ref{sec:therm-press-diff}) and in the calculation of the
contribution of atomic gas to the observed [C\,{\sc ii}] emission.
Observations of H\,{\sc i} absorption against background continuum
sources in the LMC and SMC suggest temperatures of the atomic gas as
low as 10\,K, with typical values of 30--40\,K \citep{Marx-Zimmer2000,
  Dickey2000}.  Note that H\,{\sc i} absorption features are stronger
the lower the kinetic temperature of the H\,{\sc i} gas, thus there
might be a systematic tendency to detect colder atomic gas in
absorption. At these low temperatures, most of the gas--phase carbon
is likely to be in the form of CO rather than in C$^+$, as suggested
by the detection of the $^{12}$CO $J=1\to0$ line in several H\,{\sc i}
absorption sources in the LMC \citep{Marx-Zimmer1999}.  While it is
possible that the H\,{\sc i} temperatures are lower in the Magellanic
clouds compared with our Galaxy, we consider it unlikely that [C\,{\sc
  ii}] will be detected for temperatures below 40\,K, as the [C\,{\sc
  ii}] intensity is significantly reduced for kinetic temperatures
below this value.  We therefore assume a temperature of 70\,K for the
atomic gas in our calculations, which is an intermediate value between
the temperature derived from H\,{\sc i} absorption in the Magellanic
Clouds and the 100\,K typically assumed in the Milky Way
\citep{Pineda2013}. For the temperature of the H$_2$ gas, we assume
that $T^{\rm H_2}_{\rm kin} = 0.7 T^{\rm H^0}_{\rm kin}$, which is
based on results in the Milky Way that suggest a temperature of 70\,K
for the CO--dark H$_2$ layer \citep{Goldsmith2013}, and which is in
agreement with PDR models calculations
\citep[e.g.][]{Goldsmith2016}. The temperature of the CO--dark H$_2$
gas corresponds to 49\,K for our assumed H\,{\sc i} temperature of
70K.

  We estimated the uncertainty associated with our H\,{\sc i} and
  H$_2$ kinetic temperature assumption by calculating the range in the
  [C\,{\sc ii}] intensities associated with the different ISM phases
  resulting from assuming $T^{\rm H^0}_{\rm kin} =40$\,K and $T^{\rm
    H^0}_{\rm kin} =100$\,K in the LMC and $T^{\rm H^0}_{\rm kin}
  =55$\,K and $T^{\rm H^0}_{\rm kin} =100$\,K in the SMC. We used
  $T^{\rm H^0}_{\rm kin} =55$\,K as a lower kinetic temperature limit
  in the SMC, as this is the lowest temperature where we could find
  solutions of the thermal pressure in the SMC (see
  Section~\ref{sec:therm-press-diff}).  The derived uncertainties are
  shown as error bars in Figure~\ref{fig:percentage_lmc}.

  Note that the assumed temperature of the CO--dark H$_2$ gas is lower
  than the H$_2$ rotational temperature in lines--of--sight compiled
  by \citet{Welty2012} in the LMC and SMC.  These H$_2$ temperatures
  range from 40 to 120\,K, with average values of 85\,K in both the
  LMC and SMC.  This range corresponds, for our assumption of $T^{\rm
    H_2}_{\rm kin} = 0.7 T^{\rm H^0}_{\rm kin}$, to a range in H$^0$
  kinetic temperature between 57 and 170\,K. However, the LOSs studied
  by \citet{Welty2012} have total hydrogen column densities that are
  much lower compared to the values derived in our survey (see
  Figure~\ref{fig:welty}), and therefore are diffuse and likely warmer
  LOSs.

% Discuss effect of beam filling factor.
   Another source of uncertainty originates from beam filling
    effects produced by the different angular resolution of our
    observations. The angular resolution of the data used in our
    [C\,{\sc ii}] decomposition varies from 12\arcsec\ for [C\,{\sc
      ii}] to 60\arcsec\ for H\,{\sc i}.  As described in
    Appendix~\ref{sec:beam-dilut-corr}, we derived correction factors
    using the 160\,$\mu$m map for [C\,{\sc ii}] and H\,{\sc i}, and the
    24\,$\mu$m continuum maps for the SHASSA observations. We estimate
    that the contribution from the different ISM phases to the
    observed [C\,{\sc ii}] emission would change typically by 10\%,
    14\%, 17\%, and 16\% for PDRs, CO--dark H$_2$ gas, H\,{\sc i} gas,
    and ionized gas in the LMC, respectively, and by 6\%, 4\%, 15\%,
    and 8\% for PDRs, CO--dark H$_2$ gas, H\,{\sc i} gas, and ionized
    gas in the SMC, respectively. In
    Figure~\ref{fig:contribution_dil}, we present the fraction of
    [C\,{\sc ii}] originating from different ISM phases as a function
    of the observed [C\,{\sc ii}] intensity in the case when the
    intensities are corrected by beam dilution effects.
    We choose not to include the variation in the intensities due to beam
    filling factors in Figure~\ref{fig:percentage_lmc} due to the
    uncertainty in whether the 24\,$\mu$m and 160\,$\mu$m emission
    represents the spatial distribution of the [C\,{\sc ii}]
    associated with ionized and neutral gas, respectively.

\section{Physical parameters}
\label{sec:physical-parameters}
\subsection{Determination of Visual Extinction}
\label{sec:visu-extinct-determ}

We used the {\it Herschel} HERITAGE \citep{Meixner2013} dust continuum
maps of the LMC and SMC to determine the dust column density, in terms
of the visual extinction $A_{\rm V}$, in our sample. We assume that
the dust spectral energy distribution (SED) in the LMC and SMC can be
described by the emission predicted from an optically thin  modified
black body at an equilibrium temperature $T_{\rm dust}$.  The dust
opacity at 160\,$\mu$m is given by
\begin{equation}
\label{eq:5}
\tau_{\rm 160\mu m}=\frac{I_{\rm 160\mu m}}{B_\nu (T_{\rm dust}, {\rm 160\mu m})},
\end{equation}
where $I_{\rm 160\mu m}$ is the dust continuum intensity at 160\,$\mu$m
and $B_\nu (T_{\rm dust}, {\rm 160\mu m})$ is the intensity of a black
body with a temperature $T_{\rm dust}$ at 160\,$\mu$m. We determined the
opacity at 160\,$\mu$m using the HERITAGE 160\,$\mu$m map together with
the dust temperatures fitted by \citet{Gordon2014} for a single
temperature blackbody modified by a broken power-law emissivity, their
preferred dust emission model. Both the 160\,$\mu$m and dust
temperature maps have a common resolution of 40\arcsec.  The average
($\pm$\,standard deviation) dust temperature is 22.8$\pm$4.1\,K in the
LMC and 23.1$\pm$2.5\,K in the SMC.

\citet{Lee2015} presented maps of the visual extinction in the LMC and
SMC derived from dust continuum emission. They converted the
160\,$\mu$m opacity to $A_{\rm V}$ using, $A_{\rm V}=2200\tau_{\rm
  160\mu m}$, which is the average of different methods to relate
these quantities based on data in the Milky Way. (We refer the reader
to their paper for more details on how this conversion factor is
determined.) The different methods described in \citet{Lee2015} show a
scatter from the adopted conversion factor between $A_{\rm V}$ and
$\tau_{\rm 160\mu m}$ of about 40\%, and we adopt this value as the
uncertainty in our determination of $A_{\rm V}$.

We compared the values of $A_{\rm V}$ derived here with those
presented by \citet{Lee2015} who derived the 160\,$\mu$m opacity using
the HERITAGE 160\,$\mu$m map and dust temperatures determined by
fitting a single temperature blackbody with an assumed $\beta=1.5$
wavelength dependence of the dust opacity. We find that our $A_{\rm
  V}$ values are in reasonable agreement with those derived by
\citet{Lee2015}, with differences typically within the assumed 40\%
uncertainty.

\subsection{Dust-to-Gas ratio}
\label{sec:dust-gas-ratio}

\begin{figure}[t]
\centering
\includegraphics[width=0.45\textwidth,angle=0]{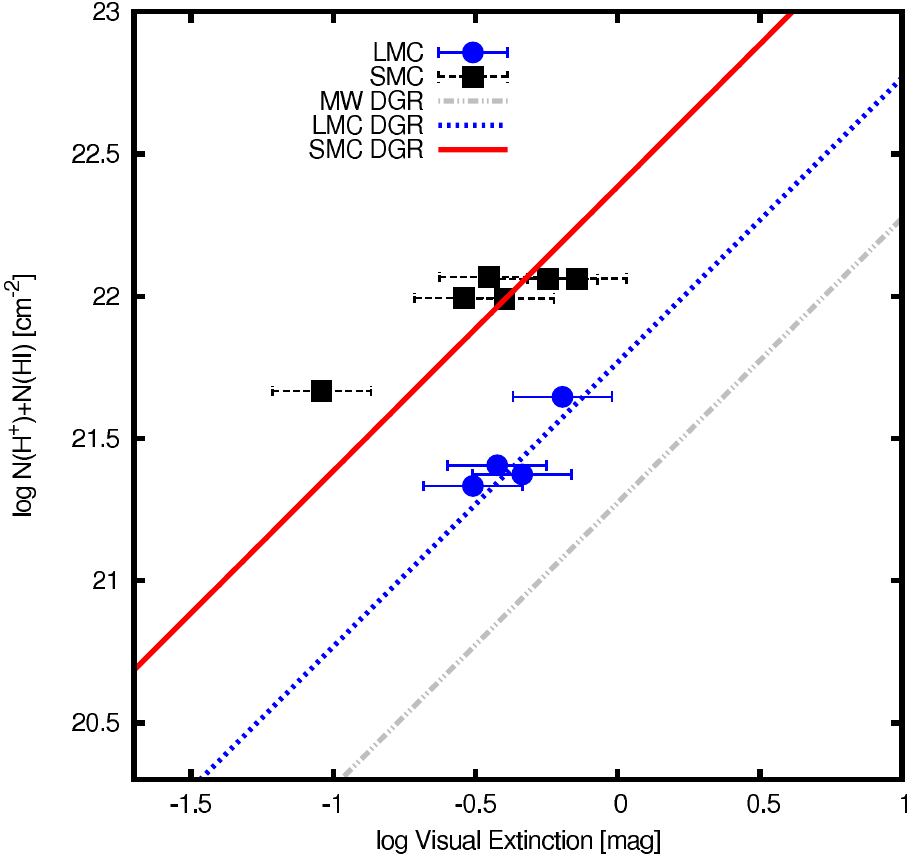}
\caption{The H$^+$+H$^0$ column density as a function the visual
  extinction associated with atomic gas for LOSs in the LMC and SMC
  where H\,{\sc i} and H$\alpha$ are the only spectral lines detected,
  therefore assumed to be diffuse gas. The straight lines represent
  the total hydrogen column density predicted for a given $A_{\rm V}$
  assuming $N$(H)/$A_{\rm
    V}$=$7.17\times10^{21}$\,cm$^{-2}$\,mag$^{-1}$ for the LMC,
  $1.68\times10^{22}$\,cm$^{-2}$\,mag$^{-1}$ for the SMC, and
  1.88$\times10^{21}$\,cm$^{-2}$\,mag$^{-1}$ for the Milky Way
  (Section~\ref{sec:dust-gas-ratio}).  }\label{fig:dgr}
\end{figure}

The dust--to--gas ratio (DGR) is a fundamental parameter relating the
quantities of gas and dust in the ISM of galaxies and the variation of
the DGR as a function of environment reflects the evolution of
galaxies \citep{Dwek1998}.  In the following, we test our
  determination of gas column densities and visual extinctions in our
  sample by checking whether these two quantities are related by the
  DGRs that are typically found in the LMC and SMC. We initially test
  diffuse LOSs in our sample, but we will extend this comparison to
  our entire sample in Section~\ref{sec:h-h_2-transition} to test our
  determination of H$_2$ column densities.

  The ratio of the total hydrogen column density, $N({\rm H})=N({\rm
    H^+})+N({\rm H^0})+2N({\rm H_2)}$, to color excess, $E$(B-V), in
  the ISM of the Milky Way is $N({\rm
    H})$/$E$(B-V)=$5.8\times10^{51}$\,cm$^{-2}$\,mag$^{-1}$
  \citep{Bohlin1978}.  Assuming a ratio of the total to selective
  extinction of $R_{\rm V}$=3.1, results in a relationship between the
  hydrogen column density and visual extinction of $N$(H)/$A_{\rm
    V}$=1.88$\times10^{21}$\,cm$^{-2}$\,${\rm mag}^{-1}$ in the Milky
  Way.  In the LMC, $N({\rm H})$/$E$(B-V) is observed to be
  $2\times10^{22}$\,cm$^{-2}$\,mag$^{-1}$ \citep{Koornneef1982,
    Fitzpatrick1985}, while in the SMC, $N({\rm H})$/$E$(B-V) ranges
  between $2.6\times10^{22}$\,cm$^{-2}$\,mag$^{-1}$ \citep{Martin1989}
  and $8.7\times10^{22}$\,cm$^{-2}$\,mag$^{-1}$
  \citep{Fitzpatrick1985b}. We adopt an intermediate value for the SMC
  of $6.6\times10^{22}$\,cm$^{-2}$\,mag$^{-1}$.  Assuming $R_{\rm
    V}$=3.4 in the LMC and $R_{\rm V}$=2.7 in the SMC
  \citep{Gordon2003},  results in $N({\rm H})/A_{\rm
    V}$=$5.56\times10^{21}$\,cm$^{-2}$\,mag$^{-1}$ for the LMC and
  $2.43\times10^{22}$\,cm$^{-2}$\,mag$^{-1}$ for the SMC.  Note that
  the value of $R_{\rm V}$ assumed here corresponds to that in diffuse
  regions, and it might be larger in denser regions
  \citep{Weingartner2001,Whittet2001}. Considering that a given line
  of sight might intersect both dense and diffuse regions,
  \citet{Whittet2001} estimated in the Milky Way an effective $R_{\rm
    V}$ along the line--of--sight that increases up to $\sim$4.0 for
  $A_{\rm V} \simeq$10\,mag. Thus, the $N({\rm H})/A_{\rm V}$ ratio
  can be lower by up to $\sim$30\% in denser LOSs.

  We tested our estimate of the visual extinction by comparing $A_{\rm
    V}$ with $N({\rm H}^+)$+$N({\rm H^0})$ in LOSs where [C\,{\sc ii}]
  was not detected. These lines--of--sight (5 in the LMC and 7 in the
  SMC) likely have little or no H$_2$ and/or have densities and
  temperatures that are insufficient for producing [C\,{\sc ii}]
  emission that will be detectable in the sensitivity limits of our
  survey. They are therefore well suited for our comparison between
  visual extinctions and the H$^{+}$+H$^0$ column densities as they
  are likely to have a negligible or no contribution from dust
  associated with H$_2$ to the observed $A_{\rm V}$.  The H$^+$ column
  densities for these LOSs were determined in Section
  \ref{sec:ionized-gas} and on average represent 5\% and 1.4\% of the
  $N({\rm H}^+)$+$N({\rm H^0})$ column densities in the LMC and SMC,
  respectively. The $A_{\rm V}$ and $N({\rm H^+})$ data points used
  here were derived from dust continuum and H$\alpha$ maps that were
  smoothed to the 60\arcsec\ resolution of the H\,{\sc i} data.  In
  Figure~\ref{fig:dgr}, we show $N({\rm H}^+)$+$N({\rm H^0})$ as a
  function of $A_{\rm V}$ for LOSs in our sample where the [C\,{\sc
    ii}] was not detected.  We also include in the plot straight lines
  that represent the predicted H$^+$+H$^0$ column density for a given
  value of $A_{\rm V}$ using the dust--to--gas ratios for the LMC,
  SMC, and Milky Way discussed above.  Given the uncertainties in the
  determination of $A_{\rm V}$, we consider that our derived values of
  $A_{\rm V}$ are consistent with the observed H\,{\sc i} column
  densities and the independently measured gas--to--dust ratios in the
  LMC and SMC.

\subsection{Thermal Pressures of the diffuse ISM in the Magellanic Clouds}
\label{sec:therm-press-diff}

\begin{figure}[t]
\centering
\includegraphics[width=0.45\textwidth,angle=0]{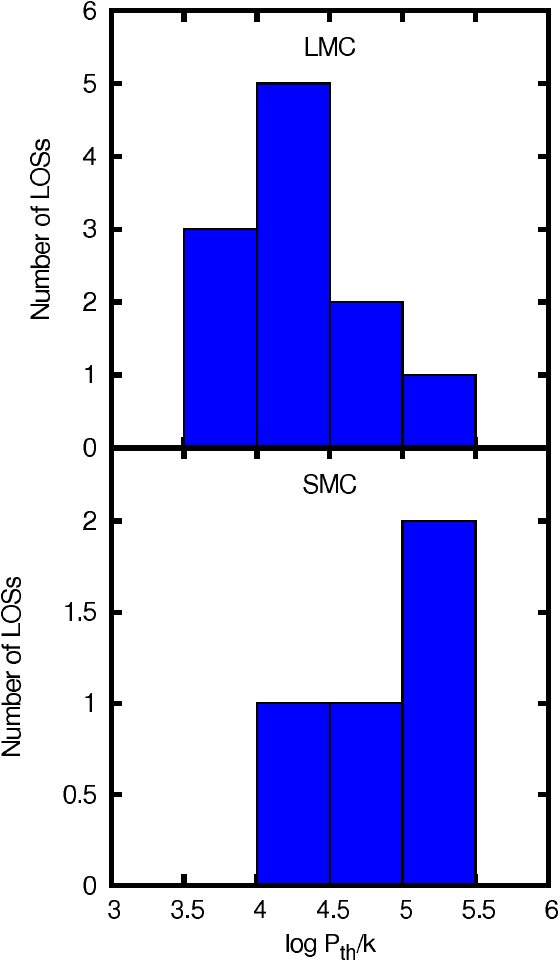}
\caption{Histograms of the thermal pressure of the diffuse
  interstellar medium of the LMC and SMC derived in
  Section~\ref{sec:therm-press-diff}.}\label{fig:pth}
\end{figure}

The thermal pressure of the diffuse ISM is an important parameter
which, despite being a small fraction of the total ISM pressure
\citep{Boulares1990}, plays a fundamental role in the phase transition
from warm and diffuse to cold and dense atomic gas
\citep{Pikelner1968,Field1969,Wolfire1995,Wolfire2003,Jenkins2011}. Thus, studying
the diffuse ISM thermal pressure is important for our understanding of
molecular cloud formation and the regulation of star formation in galaxies
\citep{Mckee1989,Blitz2006,Ostriker2010}.

In the Magellanic clouds the thermal pressure of the gas is expected
to be larger than the solar neighborhood because of larger FUV fields
and lower dust abundances, affecting the thermal balance (Wolfire et
al. 2017 in preparation; \citealt{Sandstrom2010};
\citealt{Welty2016}).  \citet{Bolatto2011} compared the relationship
between gas surface density and star formation rate in the SMC and
found that a thermal pressure $p_{\rm th}/k_{\rm
  B}=3\times10^4$\,K\,cm$^{-3}$ is required to find agreement between
observations and the theoretical predictions from
\citet{Ostriker2010}.

  An estimate of the thermal pressure of the diffuse ISM is also a
  requirement for the determination of the column density of the
  CO--dark H$_2$ gas using [C\,{\sc ii}] observations
  \citep{Pineda2013}. In the following we derive the thermal pressures
  of the diffuse ISM of the LMC and SMC by using LOSs where we detect
  H\,{\sc i} and [C\,{\sc ii}] but where no CO and/or [C\,{\sc i}] is
  detected. These LOSs (14 in the LMC and 6 in the SMC) are likely
  diffuse and the [C\,{\sc ii}] emission arises from both cold atomic
  and CO--dark H$_2$ gas.

  As discussed in Section~\ref{sec:origin-c-sc}, the observed [C\,{\sc
    ii}] intensity is the sum of the emission associated with the
  ionized, atomic, and molecular gas components.  For each LOS used
  here, we subtracted the [C\,{\sc ii}] emission associated with
  ionized gas derived in Section~\ref{sec:ionized-gas}.  The remaining
  [C\,{\sc ii}] intensity is thus described by six parameters (see
  Equation~\ref{eq:6}): the kinetic temperature ($T^{\rm H^0}_{\rm
    kin}$ and $T^{\rm H_2}_{\rm kin}$), volume density ($n_{\rm H^0}$
  and $n_{\rm H_2}$), and C$^{+}$ column densities ($N({\rm C}^+)_{\rm
    H^0}$ and $N({\rm C^+})_{\rm H_2})$ of gas associated with the
  H$^0$ and H$_2$ layers.

The total column density of C$^+$ along the line of sight is the sum
of the column of C$^+$ associated with the different collisional
partners,

\begin{equation}
N({\rm C}^+)_{\rm total}=N({\rm C}^+)_{\rm H^0}+N({\rm C}^+)_{\rm H_2}+N({\rm C}^+)_{\rm e^-}.
\end{equation}

We used the H\,{\sc i} 21\,cm observations to determine the column
density of atomic hydrogen that is associated in velocity with the
observed [C\,{\sc ii}] emission in Section~\ref{sec:atomic-gas} and
the H$^+$ column density associated with ionized gas in
Section~\ref{sec:ionized-gas}. The H$^0$ and H$^+$ column
densities are in turn converted to a C$^+$ column density by
multiplying the carbon fractional abundances of the LMC and SMC
discussed in Section~\ref{sec:origin-c-sc}.

To derive the column density of C$^+$ associated with H$_2$, we need
an estimate of the total hydrogen column density along the line of
sight. The visual extinction along the line--of--sight is the sum of
that from dust associated with molecular hydrogen, with CNM and WNM
atomic gas, and with ionized gas, so that
\begin{equation}
A^{\rm total}_{\rm V}=A^{{\rm H}_2}_{\rm V} + A^{\rm H^0, CNM}_{\rm V} + A^{\rm H^0, WNM}_{\rm V} + A^{\rm H^+}_{\rm V}. 
\end{equation}
We estimated $A^{\rm HI, CNM}_{\rm V}$ and $A^{\rm HI, WNM}_{\rm V}$
by converting the integrated intensities of H\,{\sc i} associated
(CNM) and not associated (WNM) to the observed [C\,{\sc ii}] emission
and applying the dust--to--gas ratios discussed in
Section~\ref{sec:dust-gas-ratio}. The same dust--to--gas ratio was
applied to the H$^+$ column densities to estimate visual extinction
associated with ionized gas.  The visual extinction associated with
H$_2$ is then estimated by subtracting that associated with H$^0$ and H$^+$
to the observed value. We then converted $A_{\rm V}^{{\rm H}_2}$ to a H$_2$
column density using a dust--to--gas ratio, and to $N({\rm C}^+)_{\rm
  H_2}$ using a fractional abundance of carbon.

With an estimate of $N({\rm C}^+)_{\rm H^0}$ and $N({\rm C}^+)_{\rm
  H_2}$, the remaining parameters are the kinetic temperatures and
volume densities of the molecular and cold atomic gas. We assumed that
for the diffuse lines--of--sight where only H\,{\sc i} and [C\,{\sc
  ii}] are detected there is thermal pressure equilibrium in the
interface between the cold H\,{\sc i} and the CO--dark H$_2$ layers,
\begin{equation}
p_{\rm th}/k_{\rm B}=n_{\rm H}T^{\rm H^0}_{\rm kin}=n_{\rm H_2}T^{\rm H_2}_{\rm kin}.
\end{equation}

Assuming a kinetic temperature for the H\,{\sc i} layer of 70\,K and
for the CO--dark H$_2$ layer of 49\,K (see discussion in
Section~\ref{sec:uncertainties}), we searched for a thermal pressure
that can reproduce the observed [C\,{\sc ii}]
intensity. Figure~\ref{fig:pth} shows histograms of the thermal
pressures derived in the LMC and SMC. The average ($\pm$ standard
deviation) thermal pressure is $p_{\rm th}/k_{\rm
  B}=10^{4.5\pm0.3}$\,K\,cm$^{-3}$ in the LMC and $p_{\rm th}/k_{\rm
  B}=10^{5.0\pm0.2}$\,K\,cm$^{-3}$ in the SMC.  We note that
components that are associated with the 30\,Doradus nebula in the LMC
tend to higher thermal pressures compared to those in other regions of
the LMC.  Without the LOSs associated with 30\,Doradus, we find an
average thermal pressure in the LMC of $p_{\rm th}/k_{\rm
  B}=10^{4.2\pm0.1}$\,K\,cm$^{-3}$ while LOSs associated with
30\,Doradus have $p_{\rm th}/k_{\rm B}=10^{4.9\pm0.2}$\,K\,cm$^{-3}$.
These LOSs appear to be influenced by the feedback effects of the R136
super star cluster \citep{Pellegrini2011}.  The average value for the SMC is in reasonable
agreement with that suggested by \citet{Bolatto2011}.

%70
%Pth + SD LMC no 30Dor
%14325 6065.22
%4.15609 0.0798584
%Pth + SD LMC  30Dor
%85333.3 69021.5
%4.93112 0.152558
%Pth + SD LMC  ALL
%33690.9 48229.9
%4.52751 0.270005
%Pth + SD SMC
%99225 79782
%4.99662 0.151654

We tested the sensitivity of the derived thermal pressures to the
assumed kinetic temperature of the gas.   For low temperatures,
  the thermal pressure solutions become uncertain, as higher volume
  densities are needed to reproduce the observed emission, and
  collisional deexcitation become important, reducing the dependence
  of the [C\,{\sc ii}] intensity on volume density
  \citep{Goldsmith2012}. In the LMC the thermal pressure increases by
  a factor of 1.2 between 100K and 70K and by a factor of 6 between
  100\,K and 40\,K. In the SMC, we could not find reliable solutions for
  temperatures lower than 55\,K. Between 100\,K and 70\,K, the thermal
  pressure increases by a factor of 1.5 and between 100\,K and 55\,K, by
  a factor of 3.5.

  We also studied the effects of beam dilution in our determination of
  thermal pressures in the LMC and SMC. For that purpose, we smoothed
  the dust continuum data used to calculate $A_{\rm V}$, and the
  H$\alpha$ data used to estimate the contribution from ionized gas to
  the [C\,{\sc ii}] emission, to the 60\arcsec\ resolution of the
  H\,{\sc i} data used to calculate $N({\rm H^0})$. We additionally
  corrected the observed [C\,{\sc ii}] emission with the beam dilution
  factor we estimated in Appendix~\ref{sec:beam-dilut-corr} to
  correspond to emission at 60\arcsec\ resolution. We used this data
  to recalculate the thermal pressures in our sub--sample. We find
  that applying a beam dilution factor to the data used in our
  calculation results in thermal pressures that are reduced by factors
  of $\sim$1.4 in both the LMC and SMC.

We note that the statistical significance of our determination of the
thermal pressure in the LMC and SMC is limited by the small number of
LOSs we were able to use. Using the technique presented here in future
large scale maps of the Magellanic Clouds in [C\,{\sc ii}] will
dramatically improve the significance of this result, and will allow
us to compare the distribution of thermal pressures in the LMC and SMC
with that in the Milky Way \citep{Jenkins2011}.

\begin{figure}[t]
\centering
\includegraphics[width=0.4\textwidth,angle=0]{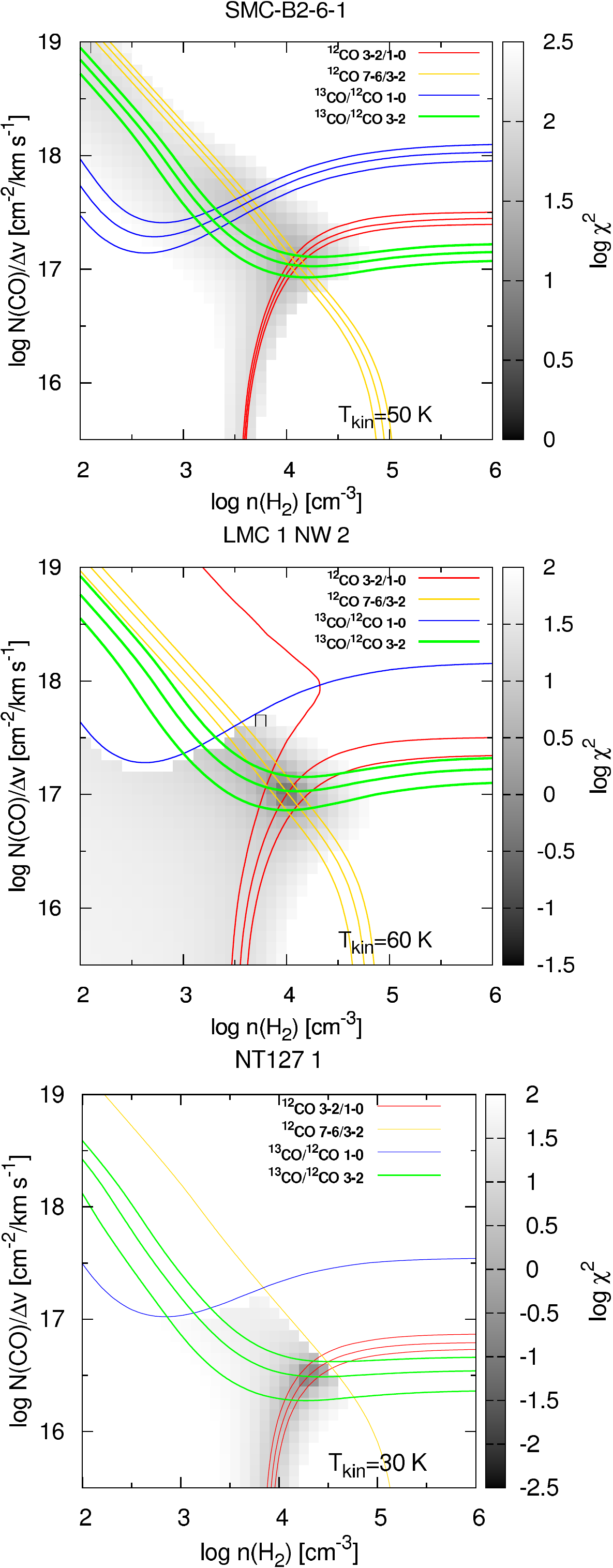}
\caption{Sample comparison between the observed CO and $^{13}$CO line
  ratios and the predictions from the RADEX model. The lines represent
  the observed line ratios $\pm$ the uncertainties in their
  determination.  The comparisons are shown at the kinetic temperature
  at which the minimum value of $\chi_{\rm total, min}^2$ is located.  In the middle
  and lower panels we show cases where the 3$\sigma$ upper limit of
  the $^{13}$CO $J=1\to0/^{12}$CO $J=1\to0$ and/or the $^{12}$CO
  $J=7\to6/J=3\to2$ ratios are used to discard solutions.}
\label{fig:co_excitation}
\end{figure}

\subsection{Kinetic temperatures and H$_2$ Column and Volume
  Densities}
\label{sec:kinet-temp-h_2}

\begin{deluxetable*}{lccccccccccccc}
\tabletypesize{\scriptsize} \centering \tablecolumns{12} \small
\tablewidth{0pt}
\tablecaption{Derived Parameters for LMC  and SMC Sample}
\tablenum{5}
\tablehead{
\colhead{LOS} &  \colhead{$N({\rm H^0})^1_{\rm CNM}$ } & \colhead{$T^{\rm C^+;2}_{\rm kin}$}  & \colhead{$n^{\rm C^+;3}_{\rm H_2}$} & \colhead{$N({\rm C^+})^4_{\rm H_2}$} &  \colhead{$T^{\rm C^0;2}_{\rm kin}$}  &  \colhead{$n^{\rm C^0;3}_{\rm H_2}$}   &   \colhead{$N({\rm C^0})^4$} &\colhead{$T^{\rm CO;2}_{\rm kin}$}   & \colhead{$n^{\rm CO;3}_{\rm H_2}$}    &   \colhead{$N({\rm CO})^{4}$}  & \colhead{filling$^{5}$} \\ 
\colhead{} &  \colhead{log } & \colhead{}  & \colhead{log} & \colhead{log} &  \colhead{}  &  \colhead{log}   &   \colhead{log} &\colhead{}   & \colhead{log}    &   \colhead{log}  & \colhead{factor} \\ 
\colhead{} &  \colhead{[cm$^{-2}$]} & [K] &   \colhead{[cm$^{-3}$]} & \colhead{[cm$^{-2}$]}&\colhead{[K]}  & \colhead{[cm$^{-3}$]} &  \colhead{[cm$^{-2}$]} &\colhead{[K]}  & \colhead{[cm$^{-3}$]} &  \colhead{[cm$^{-2}$]} &\colhead{}}\\
\startdata
\cutinhead{Large Magellanic Cloud}
Diff2\_SE\_1 & 20.5$^6$ & 49 & 2.8 & $17.1^{+0.2}_{-0.1}$ & -- & -- & -- & -- & -- & -- & -- \\
Diff7\_NW\_1 & 20.7$^6$ & 49 & 2.8 & 17.1$\pm$0.1 & -- & -- & -- & -- & -- & -- & -- \\
LMC11\_Ridge\_1 & 20.8$^6$ & 40$\pm$10 & $3.4^{+0.0^6}_{-0.3}$ & 17.5$\pm$0.3 & 40$\pm$10 & $3.9^{+0.2}_{-0.6}$ & $15.9^{+0.3}_{-0.7}$ & 20$\pm$10 & 4.5$\pm$0.1 & $15.9^{+0.3}_{-0.7}$ & $0.10^{+0.04}_{-0.32}$ \\
LMC12\_SE\_1 & 20.6$^6$ & 30$\pm$10 & 3.7$\pm$0.0$^6$ & $17.5^{+0.3}_{-0.6}$ & 30$\pm$10 & 4.4$\pm$0.1 & $15.3^{+0.2}_{-0.3}$ & 40$\pm$10 & 4.4$\pm$0.1 & $16.0^{+0.2}_{-0.3}$ & $0.10^{+0.02}_{-0.05}$ \\
LMC12\_SE\_2 & 20.7$^6$ & 60$\pm$10 & 3.1$\pm$0.0$^6$ & $17.1^{+0.1}_{-0.2}$ & 60$\pm$10 & 3.5$\pm$0.1 & $16.2^{+0.2}_{-0.4}$ & 30$\pm$10 & $4.6^{+0.3}_{-0.1}$ & $16.3^{+0.2}_{-0.4}$ & $0.34^{+0.09}_{-0.32}$ \\
LMC\_1\_NW\_1 & 20.7$^6$ & 120$\pm$20 & 2.6$\pm$0.0$^6$ & 17.2$\pm$0.1 & 120$\pm$20 & 2.7$\pm$0.1 & $16.3^{+0.2}_{-0.4}$ & $50^{+10}_{-20}$ & $6.0^{+1.9}_{-0.1}$ & $16.2^{+0.4}_{-0.5}$ & $0.01^{+0.00^7}_{-0.02}$ \\
LMC\_1\_NW\_2 & 20.8$^6$ & 160$\pm$10 & 2.6$\pm$0.0$^6$ & 16.9$\pm$0.1 & 160$\pm$10 & 2.9$\pm$0.1 & $16.1^{+0.2}_{-0.3}$ & 60$\pm$10 & 4.0$\pm$0.1 & $15.8^{+0.2}_{-0.3}$ & $0.02^{+0.00^7}_{-0.01}$ \\
LMC\_1\_NW\_3 & 21.2$^6$ & 49 & 2.8 & 17.6$\pm$0.0$^6$ & -- & -- & -- & -- & -- & -- & -- \\
LMC2\_NW\_1 & 21.1$^6$ & 49 & 2.8 & 17.2$\pm$0.1 & -- & -- & -- & -- & -- & -- & -- \\
LMC3\_NW\_1 & 20.7$^6$ & 49 & 2.8 & $16.7^{+0.3}_{-0.2}$ & -- & -- & -- & -- & -- & -- & -- \\
LMC4\_RIDGE\_1 & 21.1$^6$ & 49 & 2.8 & 17.4$\pm$0.1 & -- & -- & -- & -- & -- & -- & -- \\
LMC5\_SE\_1 & 20.1$^6$ & $170^{+10}_{-20}$ & 2.8 & 17.3$\pm$0.0$^6$ & $170^{+10}_{-20}$ & 3.3$\pm$0.1 & $16.0^{+0.2}_{-0.4}$ & 30$\pm$10 & $4.6^{+0.2}_{-0.1}$ & $15.9^{+0.2}_{-0.4}$ & $0.09^{+0.02}_{-0.07}$ \\
LMC7\_RIDGE\_1 & 21.1$^6$ & 49 & 2.8 & 17.4$\pm$0.1 & -- & -- & -- & -- & -- & -- & -- \\
LMC9\_NE\_1 & 20.8$^6$ & 49 & 2.8 & $16.7^{+0.3}_{-0.2}$ & -- & -- & -- & -- & -- & -- & -- \\
NT127\_1 & 20.1$^6$ & 30$\pm$10 & 2.6$\pm$0.0$^6$ & $17.2^{+0.5}_{-0.7}$ & 30$\pm$10 & 2.1$\pm$0.1 & $16.7^{+0.3}_{-0.4}$ & 30$\pm$10 & 4.3$\pm$0.1 & $16.0^{+0.3}_{-0.4}$ & $0.16^{+0.05}_{-0.15}$ \\
NT77\_1 & 20.0$^6$ & $190^{+20}_{-10}$ & 2.7$\pm$0.0$^6$ & $17.3^{+0.0^6}_{-0.1}$ & $190^{+20}_{-10}$ & 3.2$\pm$0.1 & $16.1^{+0.6}_{-0.2}$ & 50$\pm$20 & $4.3^{+0.6}_{-0.4}$ & $16.9^{+1.1}_{-0.6}$ & $0.74^{+0.49}_{-0.26}$ \\
PDR1\_NW\_1 & 21.1$^6$ & 30$\pm$10 & 3.7$\pm$0.0$^6$ & $17.8^{+0.3}_{-0.7}$ & 30$\pm$10 & $4.4^{+0.2}_{-0.1}$ & $16.6^{+0.3}_{-0.2}$ & 30$\pm$10 & $4.4^{+0.2}_{-0.1}$ & $16.8^{+0.7}_{-0.2}$ & $0.79^{+0.24}_{-0.21}$ \\
PDR2\_NW\_2 & 20.4$^6$ & 40$\pm$10 & 4.4$\pm$-0.0 & $17.9^{+0.2}_{-0.3}$ & 40$\pm$10 & 5.9$\pm$0.1 & 16.1$\pm$0.2 & 40$\pm$10 & 5.9$\pm$0.1 & 16.5$\pm$0.2 & $0.11^{+0.02}_{-0.03}$ \\
PDR2\_NW\_3 & 20.6$^6$ & 49 & 2.8 & 18.0$\pm$0.0$^6$ & -- & -- & -- & -- & -- & -- & -- \\
PDR2\_NW\_4 & 20.5$^6$ & 49 & 2.8 & 17.6$\pm$0.0$^6$ & -- & -- & -- & -- & -- & -- & -- \\
PDR3\_NE\_1 & 21.1$^6$ & 150$\pm$10 & 2.3$\pm$0.0$^6$ & 18.4$\pm$0.1 & 150$\pm$10 & 2.3$\pm$0.1 & $16.2^{+0.2}_{-0.3}$ & $60^{+10}_{-20}$ & $4.2^{+0.2}_{-0.3}$ & $16.8^{+0.2}_{-0.3}$ & $0.09^{+0.02}_{-0.04}$ \\
PDR3\_NE\_2 & 20.9$^6$ & 170$\pm$10 & 3.4$\pm$0.0$^6$ & 17.5$\pm$0.0$^6$ & 170$\pm$10 & 4.5$\pm$0.1 & $16.6^{+0.2}_{-0.3}$ & 40$\pm$10 & 4.5$\pm$0.4 & $17.3^{+0.2}_{-0.3}$ & $0.38^{+0.08}_{-0.28}$ \\
PDR3\_NE\_3 & 21.4$^6$ & 49 & 2.8 & 18.3$\pm$0.0$^6$ & -- & -- & -- & -- & -- & -- & -- \\
PDR4\_RIDGE\_1 & 21.0$^6$ & 80$\pm$20 & 2.7$\pm$0.0$^6$ & $17.5^{+0.1}_{-0.2}$ & 80$\pm$20 & 2.8$\pm$0.1 & $16.7^{+0.3}_{-0.5}$ & 30$\pm$10 & $4.6^{+0.7}_{-0.4}$ & $17.2^{+0.4}_{-0.6}$ & $0.40^{+0.16}_{-0.60}$ \\
SK-66D35\_2 & 21.0$^6$ & 49 & 2.8 & 17.1$\pm$0.1 & -- & -- & -- & -- & -- & -- & -- \\
SK-68D140\_1 & 21.1$^6$ & 49 & 2.8 & 17.2$\pm$0.1 & -- & -- & -- & -- & -- & -- & -- \\
SK-68D155\_1 & 20.8$^6$ & 49 & 2.8 & 17.6$\pm$0.0$^6$ & -- & -- & -- & -- & -- & -- & -- \\
SK-69D228\_1 & 20.8$^6$ & 49 & 2.8 & 17.1$\pm$0.1 & -- & -- & -- & -- & -- & -- & -- \\
\cutinhead{Small Magellanic Cloud}
AzV18\_1 & 21.5$^6$ & 49 & 3.3 & 17.1$\pm$0.1 & -- & -- & -- & -- & -- & -- & -- \\
AzV18\_2 & 21.1$^6$ & 49 & 3.3 & 16.9$\pm$0.1 & -- & -- & -- & -- & -- & -- & -- \\
SMC\_B2\_6\_1 & 21.1$^6$ & 160$\pm$10 & 2.9$\pm$0.0$^6 $& 17.4$\pm$0.1 & 160$\pm$10 & 3.1$\pm$0.1 & $16.1^{+0.3}_{-0.5}$ & -- & $4.1^{+0.3}_{-0.2}$ & $16.4^{+0.4}_{-0.6}$ & $0.06^{+0.03}_{-0.07}$ \\
SMC\_B2\_6\_2 & 21.4$^6$ & 49 & 3.3 & 17.1$\pm$0.1 & -- & -- & -- & -- & -- & -- & -- \\
SMC\_HI\_4\_1 & 21.3$^6$ & 49 & 3.3 & 17.1$\pm$0.0$^6 $& -- & -- & -- & -- & -- & -- & -- \\
SMC\_HI\_4\_2 & 21.0$^6$ & 49 & 3.3 & 16.7$\pm$0.1 & -- & -- & -- & -- & -- & -- & -- \\
SMC\_LIRS36\_1 & 20.8$^6$ & 60$\pm$10 & 3.5$\pm$0.0$^6 $& 17.5$\pm$0.1 & 60$\pm$10 & 3.8$\pm$0.1 & $16.4^{+0.4}_{-0.1}$ & -- & $4.3^{+0.4}_{-0.2}$ & $16.9^{+0.8}_{-0.3}$ & $0.89^{+0.42}_{-0.11}$ \\
SMC\_LIRS49\_1 & 20.9$^6$ & 49 & 3.3 & 17.2$\pm$0.0$^6 $& -- & -- & -- & -- & -- & -- & -- \\
SMC\_LIRS49\_2 & 21.0$^6$ & 50$\pm$10 & 3.1$\pm$0.0$^6 $& $17.8^{+0.1}_{-0.2}$ & 50$\pm$10 & 3.0$\pm$0.1 & 16.5$\pm$0.4 & -- & 4.4$\pm$0.3 & $16.8^{+0.8}_{-0.6}$ & $0.51^{+0.26}_{-0.49}$ \\
SMC\_LIRS49\_3 & 21.3$^6$ & 50$\pm$10 & $3.8^{+0.1}_{-0.0^6}$ & $16.8^{+0.1}_{-0.2}$ & -- & -- & -- & -- & $4.3^{+0.3}_{-0.1}$ & $16.3^{+0.4}_{-0.5}$ & $0.15^{+0.03}_{-0.24}$ \\
SMC\_LIRS49\_4 & 20.5$^6$ & 49 & 3.3 & 16.6$\pm$0.0$^6 $& -- & -- & -- & -- & -- & -- & -- \\
SMC\_NE\_1a\_1 & 21.0$^6$ & 110$\pm$10 & 3.3 & 16.7$\pm$0.1 & 110$\pm$10 & 3.6$\pm$0.1 & $15.8^{+0.1}_{-0.2}$ & -- & $3.9^{+0.3}_{-0.2}$ & $16.6^{+0.3}_{-0.2}$ & $0.12^{+0.00^7}_{-0.05}$ \\
SMC\_NE\_3c\_1 & 20.5$^6$ & 40$\pm$10 & 3.3$\pm$-0.0 & $16.6^{+0.2}_{-0.4}$ & 40$\pm$10 & 3.3$\pm$0.1 & $16.3^{+0.3}_{-1.0}$ & -- & $3.5^{+0.7}_{-0.1}$ & $16.5^{+0.4}_{-1.0}$ & $0.06^{+0.03}_{-0.42}$ \\
SMC\_NE\_3g\_1 & 21.1$^6$ & 190$\pm$10 & 3.0$\pm$0.0$^6 $& 16.6$\pm$0.1 & 190$\pm$10 & 3.3$\pm$0.1 & 15.9$\pm$0.3 & -- & 3.4$\pm$0.1 & 16.4$\pm$0.3 & 0.03$\pm$0.01 \\
SMC\_NE\_4a\_hi\_2 & 20.4$^6$ & 20$\pm$10 & $3.8^{+0.0^6}_{-0.1}$ & $17.5^{+0.7}_{-2.0}$ & 20$\pm$10 & 4.0$\pm$0.1 & $15.9^{+0.1}_{-0.2}$ & -- & 4.0$\pm$0.1 & $16.2^{+0.1}_{-0.2}$ & $0.10^{+0.00^7}_{-0.02}$ \\
SMC\_NE\_4a\_hi\_3 & 21.0$^6$ & 49 & 3.3 & $16.3^{+0.3}_{-0.2}$ & -- & -- & -- & -- & -- & -- & -- \\
SMC\_NE\_4c\_low\_1 & 21.2$^6$ & $30^{+10}_{-20}$ & 2.8$\pm$0.0$^6 $& $17.4^{+0.7}_{-0.8}$ & $30^{+10}_{-20}$ & $2.1^{+0.1}_{-0.2}$ & $16.4^{+0.5}_{-0.4}$ & -- & $2.1^{+0.1}_{-0.2}$ & $17.1^{+0.6}_{-0.5}$ & $0.15^{+0.09}_{-0.17}$ \\
\tablenotetext{1}{H$^0$ column density derived from the  integrated intensity of the H\,{\sc i} 21\,cm line over the velocity range defined by the FWHM of the [C\,{\sc ii}] velocity components (see Section~\ref{sec:atomic-gas}). }
\tablenotetext{2}{Kinetic temperature of the H$_2$ gas associated with C$^+$ (Section~\ref{sec:c-sc-i}), C$^0$ (Section~\ref{sec:ci_excitation}), and CO (Section~\ref{sec:co-emission}) . }
\tablenotetext{3}{Volume density of the H$_2$ gas associated with C$^+$ (Section~\ref{sec:c-sc-i}), C$^0$ (Section~\ref{sec:ci_excitation}), and CO (Section~\ref{sec:co-emission}) . }
\tablenotetext{4}{C$^+$ (Section~\ref{sec:c-sc-i}), C$^0$ (Section~\ref{sec:ci_excitation}), and CO (Section~\ref{sec:co-emission}) column densities associated with H$_2$ gas.  }
\tablenotetext{5}{Filling factor of the [C\,{\sc i}] and CO--emitting layers derived from $^{13}$CO observations in Section~\ref{sec:kinet-temp-h_2}. }
\tablenotetext{6}{The uncertainty is below 0.05. }
\tablenotetext{7}{The uncertainty is below 0.005.}
\end{deluxetable*}

We constrained the properties of the CO and [C\,{\sc i}]--emitting gas
by comparing line intensity ratios and absolute intensities of
[C\,{\sc i}], $^{12}$CO, and $^{13}$CO with the results of a radiative
transfer code.  We used the RADEX non--LTE radiative transfer code
\citep{vanderTak2007}, using the uniform sphere approximation, to
calculate line intensities as a function of the kinetic temperature,
H$_2$ volume density, and the column density per velocity interval
$N$/$\Delta$v of the species of interest. The collisional rate
coefficients were taken from the Leiden Atomic and Molecular Database
\citep[LAMDA;][]{Schoeier2005}.

 In the PDRs located at the borders of molecular clouds, the
  innermost region is at higher density, deeper in the cloud, and the
  outermost region is at lower density. Thus, the physical
conditions of the C$^0$ layer are not necessarily the same as those at
the CO layer. Since we will be modelling spheres with uniform volume
density and kinetic temperature, we analyzed the [C\,{\sc i}] and CO
data separately. We nevertheless assume that the [C\,{\sc i}] and CO
emission originate from the same regions with identical filling
factors. This assumption is justified considering that the CO and
[C\,{\sc i}] emission arise from the FUV illuminated surfaces of
clouds that are unresolved at the scale of our observations. For
optically thin clumps (or for clumps with moderate opacity), the
observed--to--model integrated intensity ratio gives the ratio of the
beam--averaged column density to the clump intrinsic column density,
assuming that the total integrated intensity is proportional to the
number of clumps, i.e. no velocity crowding is present. We therefore
determined the beam filling factor by comparing the observed absolute
intensity of an optically thin line ( $^{13}$CO $J=3\to2$ or $^{13}$CO
$J=1\to0$) with a constrained model.

We compared the observed line intensity ratios with the predictions
from a grid of RADEX models. The model grids predict line ratios for
the kinetic temperatures ranging from 10\,K to 200\,K in steps of
10\,K as a function of the CO or C$^0$ column densities per velocity
interval, $N_{\rm CO}/\Delta {\rm v}$ (or $N_{\rm C^0}/\Delta {\rm
  v}$), and H$_2$ volume density, $n_{\rm H_2}$. The coverage in
$N_{\rm CO}/\Delta {\rm v}$ and $N_{\rm C^0}/\Delta {\rm v}$ is
$10^{15}-10^{19}$\,cm$^{-2}$\,(km\,s$^{-1}$)$^{-1}$ and in $n_{\rm
  H_2}$ is $10^{2}-10^{6}$\, cm$^{-3}$, both in steps of 0.1\,dex.  In
our calculations, we initially assumed $\Delta {\rm
  v}=$1\,km\,s$^{-1}$, but later converted the constrained $N_{\rm
  C^0}/\Delta {\rm v}$ and $N_{\rm CO}/\Delta {\rm v}$ to total C$^0$
and CO column density by multiplying by the line FWHM resulting from
the Gaussian fit in each velocity component.  For velocity components
where the lines in the numerator of the line ratios are not detected
but those in the denominator are, we use the 3$\sigma$ upper limits of
the lines in the numerator to calculate line ratios which are used to
discard solutions in the RADEX grids
(Figure~\ref{fig:co_excitation}). In Table\,5, we summarize the
physical parameters derived in the LMC and SMC. We only list velocity
components where the [C\,{\sc ii}] line is detected.

  We quantified how a RADEX model $M$($T_{\rm kin}$, $N_{\rm
    C^0}/\Delta {\rm v}$, $n_{\rm H_2}$) reproduces a given line ratio
  $R$ with uncertainty $\sigma_R$ by calculating $\chi_R^2=\sum_i
  (R-M_i)^2/\sigma_R^2$ over the entire model grid. We later added the
  resulting $\chi_R^2$ for all line ratios available in a given
  velocity component to determine a total $\chi_{\rm total}^2$. We
  then identified the best matching model by searching for the minimum
  $\chi_{\rm total}^2$ in the model grid, $\chi_{\rm total,min}^2$.
  To determine the uncertainties in the derived parameters, we defined
  a region in the parameter space where models have a $\chi_{\rm
    total}^2$ that is within three times the minimum $\chi_{\rm
    total,min}^2$.  We then determined the range of possible values of
  a given parameter within this region when the other two are fixed.
  For well constrained models, $\chi_{\rm total,min}^2$ can be very
  small and the adjacent models in the parameter space often have
  $\chi_{\rm total}^2$ that are higher than $3\chi_{\rm
    total,min}^2$. In this case the accuracy in which a model
  parameter can be determined is given by the step size of the model
  grid.

 We studied the effect of beam dilution in the line ratios used
  our analysis in Appendix~\ref{sec:beam-dilut-corr}. Note that we do
  not correct the line ratios for beam dilution effects in our
  analysis, due to the uncertainties of whether dust continuum
  emission traces the distribution of gas.  In Table\,6, we present
  the effect of beam dilution in the line ratios that are calculated
  using observations at different angular resolutions. We find that
  these line ratios typically show a 10\% variation.  This small
  variation suggest that the observed structures are relatively
  extended at the resolution of our observations.  In Table\,7, we
  summarize the impact of the variation in the line ratios due to beam
  dilution in the derived physical properties of the C$^0$ and CO
  layers. The kinetic temperature, column density, and filling factor
  of the C$^0$ and CO layers are typically affected by factors between
  1.1 to 1.6. The largest impact of the variation in the line ratios
  is in the derived H$_2$ volume density with variations between 1.5
  and 4.  Note that the largest variation we see, in the resulting
  H$_2$ volume density of CO gas in the LMC, is dominated by three
  sources, PDR4\_RIDGE\_1, PDR3\_NW, and NT77, which vary by factors
  of 25, 8, and 8, respectively. Without these three sources we find
  typical variations of a factor 1.6 for $n^{\rm CO}_{\rm H_2}$ in the LMC.

\begin{deluxetable}{lccc}
\tabletypesize{\scriptsize} \centering \tablecolumns{4} \small
\tablewidth{0pt}
\tablecaption{Beam dilution effect on line ratios}
\tablenum{6}
\tablehead{
\colhead{Ratio$^1$} &  \colhead{$\Theta_a/\Theta_b^2$} & \colhead{ LMC$^3$} & \colhead{SMC$^3$} \\
\colhead{} & \colhead{}   & \colhead{}   & \colhead{}}  \\
\startdata
 $^{12}$CO $J = 7 \to 6$/$J = 3 \to 2$ & 26.5\arcsec/17.5\arcsec &  1.07   & 1.08 \\
 $^{12}$CO $J = 3 \to 2$/$J = 1 \to 0$ & 17.5\arcsec/33\arcsec &   1.11 & 1.11 \\

 [C\,{\sc i}] ${\rm ^3P}_2\to{\rm ^3P}_1/{\rm ^3P}_1\to{\rm ^3P}_0$  & 26.5\arcsec/44\arcsec & 1.09 & 1.05 \\
 \tablenotetext{1}{Line ratio involving spectral lines observed at different angular resolutions. }
 \tablenotetext{2}{Angular resolution of the spectra line in the
   numerator and denominator of the line ratio. }
 \tablenotetext{3}{Average variation in a line ratio due to beam
   dilution effects.}
\end{deluxetable}

\begin{deluxetable}{lcc}
  \tabletypesize{\scriptsize} \centering \tablecolumns{3} \small
  \tablewidth{0pt} \tablecaption{Impact of beam dilution effects on
    derived physical parameters} \tablenum{7} \tablehead{
    \colhead{Physical parameter} &  \colhead{ LMC$^1$} & \colhead{SMC$^1$} \\
    \colhead{} &  \colhead{}   & \colhead{}}  \\
  \startdata
  $N({\rm C^0})$  &  1.2   & 1.5 \\[4pt]
  $N({\rm CO})$  &  1.4   & 1.1 \\[4pt]
  $n^{\rm C^0}_{\rm H_2}$   & 2.2    & 3.2  \\[4pt]
 $n^{\rm CO}_{\rm H_2}$   &  4   & 1.5 \\[4pt]
  $T^{\rm C^0}_{\rm kin}$ & 1.2  & 1.3 \\[4pt]
 $T^{\rm CO}_{\rm kin}$ & 1.5 &  1.6 \\[4pt]
  filling factor & 1.4 & 1.2\\
  \tablenotetext{1}{Average variation on the physical parameter
    resulting from beam dilution effects.}
\end{deluxetable}
%
%
%LMC Tkin CO:  1.23889
%SMC Tkin CO:  1.26111
%LMC NCO:  1.39194
%SMC NCO:  1.05811
%LMC nh2 CO:  4.04206
%SMC nh2 CO:  1.47983
%#########################################################################
%LMC FF:  1.37479
%SMC FF:  1.20494
%LMC [CI]:  0.441833 0.230284 0.416249
%SMC [CI]:  0.473611 0.527155 0.586799
%LMC Tkin CI:  1.46399
%SMC Tkin CI:  1.59205
%LMC NCI:  1.23261
%SMC NCI:  1.53025
%LMC nh2 CI:  2.22481
%SMC nh2 CI:  3.19942
%#########################################################################

\subsubsection{CO--emitting layer}
\label{sec:co-emission}

We start by deriving the excitation conditions of the CO--emitting
cloud layer. Depending on availability, we compared the observed
$^{12}$CO $J=7\to6/J=3\to2$, $^{12}$CO $J=3\to2/1\to0$, $^{13}$CO
$J=3\to2/^{12}$CO $J=3\to2$, and $^{13}$CO $J=1\to0/^{12}$CO $J=1\to0$
intensity ratios with the predictions from a grid of RADEX models. In
Figure~\ref{fig:co_excitation} we show an example of the comparison
between the model grid and the observed line ratios. We assumed a
$^{12}$CO/$^{13}$CO abundance ratio of 49 derived in the N113 region
in the LMC by \citet{Wang2009}.

 Note that there are several velocity components in the LMC and
  SMC where $^{12}$CO $J=1\to0$ was detected but where $^{13}$CO $J =
  1 \to 0$ and/or $^{13}$CO $J = 3\to2$ were not detected. We were
  thus unable to derive the physical conditions of the CO--emitting
  layer in these LOSs. These LOSs are likely diffuse (see
  Section~\ref{sec:co-dark-h_2}), as $^{12}$CO $J=1\to0$ is known to
  be detectable even in low column density regions, where $^{13}$CO $J
  = 1\to0$ is not detected, and CO is not the dominant form of
  gas--phase carbon \citep{Pineda2010a}.

We find that single component models often provide a reasonable fit to
the observed line ratios in our sample. There are however some
indications by the $^{12}$CO/$^{13}$CO ratios of a colder gas
component.  When a best fitting model is identified, we compare the
observed $^{13}$CO $J=3\to2$ intensities, assuming this line to be
optically thin, with those predicted by the model to estimate the beam
filling factor. If the $^{13}$CO $J=3\to2$ is not detected, we use
instead the $^{13}$CO $J=1\to0$ when available.  We derived physical
conditions of 23 velocity components in 20 LOSs where at least two
pairs of line ratios could be calculated. The average value and
standard deviation of the physical conditions in the sample are
$43\pm19$\,K in the LMC and $50\pm20$\,K in the SMC for the kinetic
temperature, $10^{16.7 \pm 0.5}$\,cm$^{-2}$ in the LMC and $10^{16.6
  \pm 0.3}$\,cm$^{-2}$ in the SMC for the beam--averaged CO column
density, $10^{5.1\pm0.9}$\,cm$^{-3}$ in the LMC and
$10^{4.0\pm0.3}$\,cm$^{-3}$ in the SMC for the H$_2$ volume density,
and ($0.2\pm0.2$) in the LMC and ($0.20\pm0.3$) in the SMC for the
filling factor.

% From /home/jpineda/MagClouds_HSO/RADEX/table_conditions/extract_co_stats.sh
%Filling factor + SD
%0.203052 0.230454
%Tkin + SD
%51.7391 27.1313
%NCO + SD
%17.9529 0.807007
%nh2 + SD
%4.85999 1.2147

%#The beam filling factor varies from....

\subsubsection{[C\,{\sc i}]--emitting layer}
\label{sec:ci_excitation}

\begin{figure}[t]
\centering
\includegraphics[width=0.45\textwidth,angle=0]{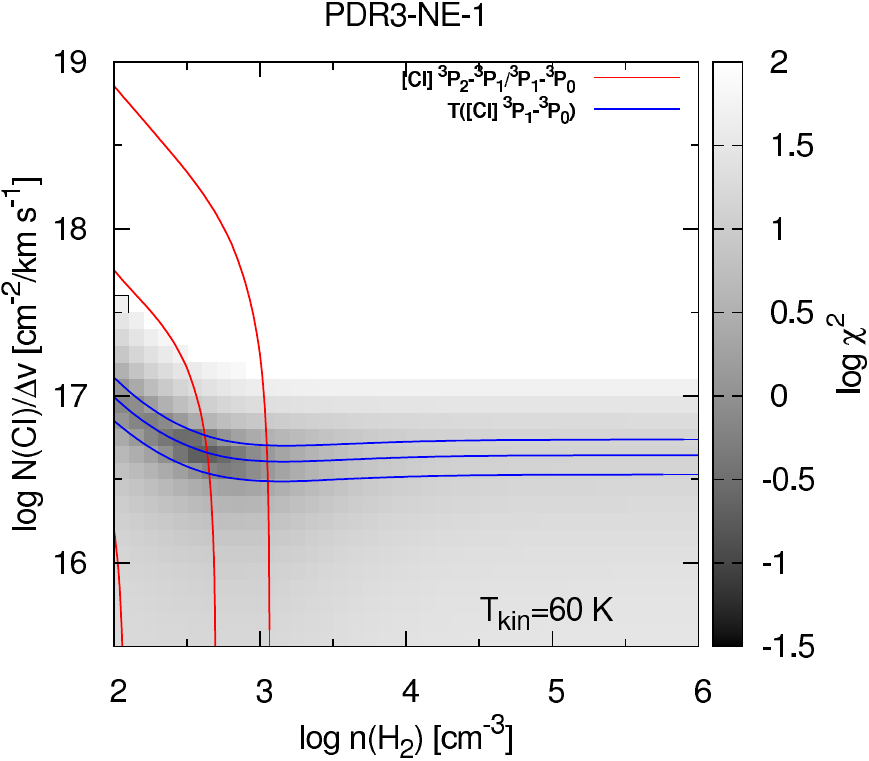}
\caption{Sample comparison between the [C\,{\sc i}] ${\rm
    ^3P}_2\to{\rm ^3P}_1/{\rm ^3P}_1\to{\rm ^3P}_0$ line intensity
  ratio and the peak intensity of the [C\,{\sc i}] ${\rm ^3P}_1\to{\rm
    ^3P}_0$ line with the grid of RADEX models.  The comparison is
  shown at the kinetic temperature at which the minimum value of
  $\chi^2$ is found. }
\label{fig:ci_excitation}
\end{figure}

We constrain the volume and column density and kinetic temperature of
the [C\,{\sc i}]--emitting layer comparing the observed [C\,{\sc i}]
${\rm ^3P}_2\to{\rm ^3P}_1/{\rm ^3P}_1\to{\rm ^3P}_0$ line intensity
ratio and the peak intensity of the [C\,{\sc i}] ${\rm ^3P}_1\to{\rm
  ^3P}_0$ line with the grid of RADEX models. We show a sample
comparison between observations and models in
Figure~\ref{fig:ci_excitation}.

For kinetic temperatures above $\sim50$\,K and C$^0$ column densities per
velocity interval below
$\sim10^{18}$ cm$^{-2}$\,(K\,km\,s$^{-1}$)$^{-1}$, a typical range of
solutions found in the [C\,{\sc i}]--emitting layer, the [C\,{\sc i}]
${\rm ^3P}_2\to{\rm ^3P}_1/^3{\rm P}_1\to{\rm ^3P}_0$ ratio provides a
good constraint on the volume density of the [C\,{\sc i}]--emitting gas. The
peak intensity of the [C\,{\sc i}] ${\rm ^3P}_1\to{\rm ^3P}_0$,
corrected by the filling factor derived in the CO--emitting layer,
depends mostly on the C$^0$ column density per velocity interval, and
thus constrain N(${\rm C^0}$)/$\Delta$v. The multiplication by a
filling factor is appropriate because, in the range of kinetic
temperatures and H$_2$ densities inferred in our analysis, the
[C\,{\sc i}] ${\rm ^3P}_1-{\rm ^3P}_0$ line is likely optically thin.

While the [C\,{\sc i}] lines typically provide good constraints on the
column and volume density of the [C\,{\sc i}]--emitting gas, the
kinetic temperature is not well constrained.  We therefore further
constrain the solutions by requiring that the [C\,{\sc i}]--emitting
region must have a lower or equal volume density and higher or equal
kinetic temperature than the CO--emitting layer.  This assumption is
justified as we expect that [C\,{\sc i}] and mid--$J$ CO lines are
emitted from adjacent layers in a PDR--like structure\footnote{For an
  example of a typical temperature distribution and abundance
  structure of the main carbon species in PDRs, see e.g. Figures 7 and
  9 in \citet{Tielens1985}.}, where the neutral carbon is located in a
somewhat warmer region with equal or lower volume density compared to
the one at which CO is located, and that the cloud's volume density
profile is smooth.

There are some velocity components where only the [C\,{\sc i}] ${\rm
  ^3P}_1\to{\rm ^3P}_0$ line is detected, and therefore we can only
calculate a 3$\sigma$ upper limit to the [C\,{\sc i}] ${\rm
  ^3P}_2\to{\rm ^3P}_1/^3{\rm P}_1\to{\rm ^3P}_0$ ratio. In these
cases we assume that the kinetic temperature and volume density at the
[C\,{\sc i}]--emitting layer are the same as those at the
CO--emitting layer and we use the [C\,{\sc i}] ${\rm ^3P}_1\to{\rm
  ^3P}_0$, corrected with its corresponding filling factor, to
constrain the C$^0$ column density per velocity interval.

We derived physical conditions of 22 velocity components in 20
LOSs. The average value and standard deviation of the physical
conditions in the [C\,{\sc i}]--emitting layer are $91\pm60$\,K in the
LMC and $73\pm62$ in the SMC for the kinetic temperature, $10^{16.4
  \pm 0.4}$\,cm$^{-2}$ in the LMC and $10^{16.2\pm 0.3}$\,cm$^{-2}$ in
  the SMC for the beam--averaged C$^0$ column density, and
  $10^{4.8\pm1.4}$\,cm$^{-3}$ in the LMC and
  $10^{3.5\pm0.5}$\,cm$^{-3}$ in the SMC for the H$_2$ volume density.

\subsubsection{[C\,{\sc ii}]--emitting layer}
\label{sec:c-sc-i}

As mentioned above, the observed [C\,{\sc ii}] emission is the result
of the combined emission from molecular, atomic, and ionized gas. In
Section~\ref{sec:ionized-gas} and \ref{sec:atomic-gas} we estimated
the physical conditions of the ionized and atomic gas associated with
C$^+$ to determine their contribution to the observed [C\,{\sc ii}]
emission. In the following, we describe the derivation of the physical
conditions of the [C\,{\sc ii}]--emitting gas associated with
molecular gas.

We estimated the volume density of the [C\,{\sc ii}]--emitting layer
associated with molecular hydrogen assuming a thermal pressure and a
kinetic temperature. In Section~\ref{sec:therm-press-diff}, we
estimated the typical thermal pressure of the diffuse gas of
$3.4\times10^{4}$\,K\,cm$^{-3}$ and $1\times10^{5}$\,K\,cm$^{-3}$, in
the LMC and SMC, respectively.  In LOS where H\,{\sc i} and [C\,{\sc
  ii}] are detected, but neither [C\,{\sc i}], CO, nor $^{13}$CO
emission are detected, we assume the derived diffuse ISM pressure and
a kinetic temperature of 49\,K (Section~\ref{sec:uncertainties}) to
derive a typical volume density of 694 and 2040\,cm$^{-3}$ for the LMC
and SMC, respectively. We also apply this criteria to LOSs where weak
CO emission is detected but no $^{13}$CO was
detected.

\begin{figure}[t]
\centering
\includegraphics[width=0.45\textwidth,angle=0]{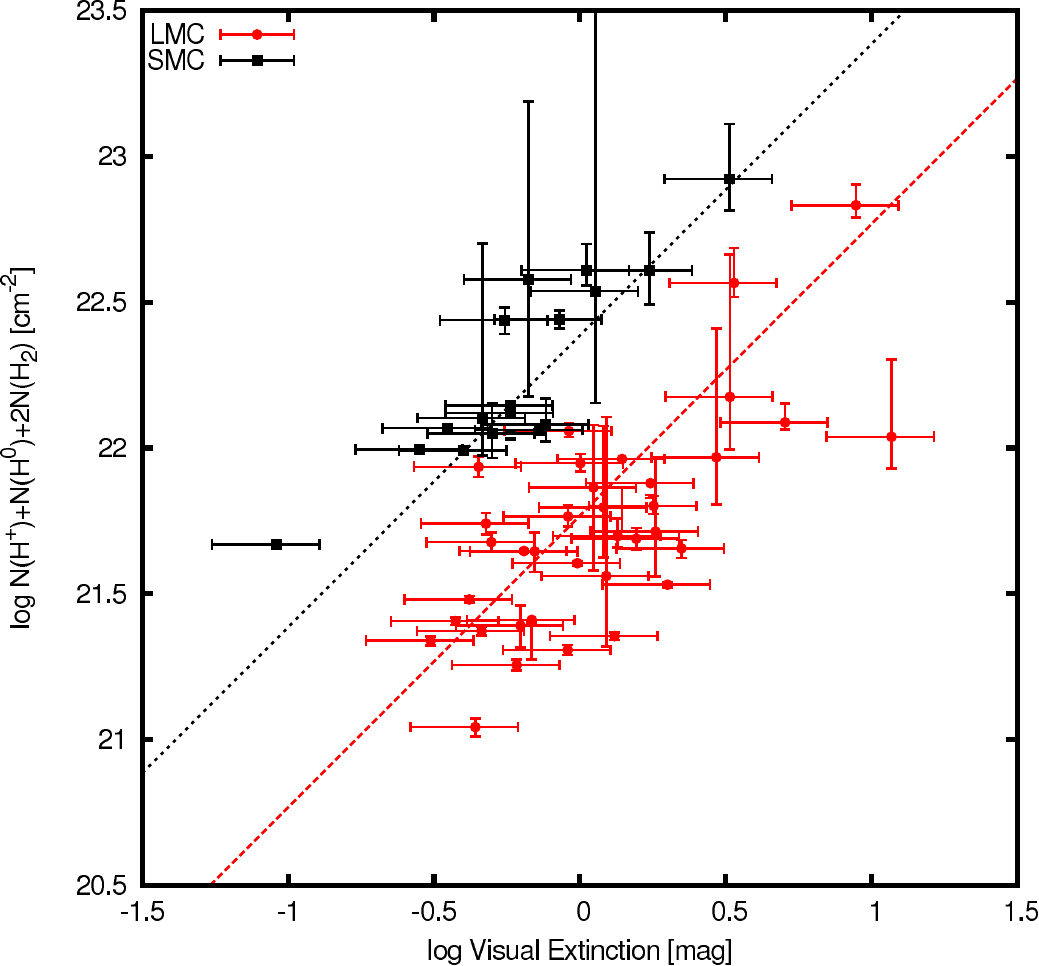}
\caption{Comparison between the $N$(H$^+$)+$N({\rm H^0})$+2$N$(H$_2)$
  hydrogen column densities derived in this study and the visual
  extinction derived from dust continuum observations.  The straight
  lines represent the total hydrogen column density predicted for a
  given $A_{\rm V}$ assuming $N$(H)/$A_{\rm
    V}$=$7.17\times10^{21}$\,cm$^{-2}$\,mag$^{-1}$ for the LMC and
  $1.68\times10^{22}$\,cm$^{-2}$\,mag$^{-1}$ for the SMC
  (Section~\ref{sec:dust-gas-ratio}).  }
\label{fig:total_H}
\end{figure}

\begin{figure*}[t]
\centering
\includegraphics[width=0.8\textwidth,angle=0]{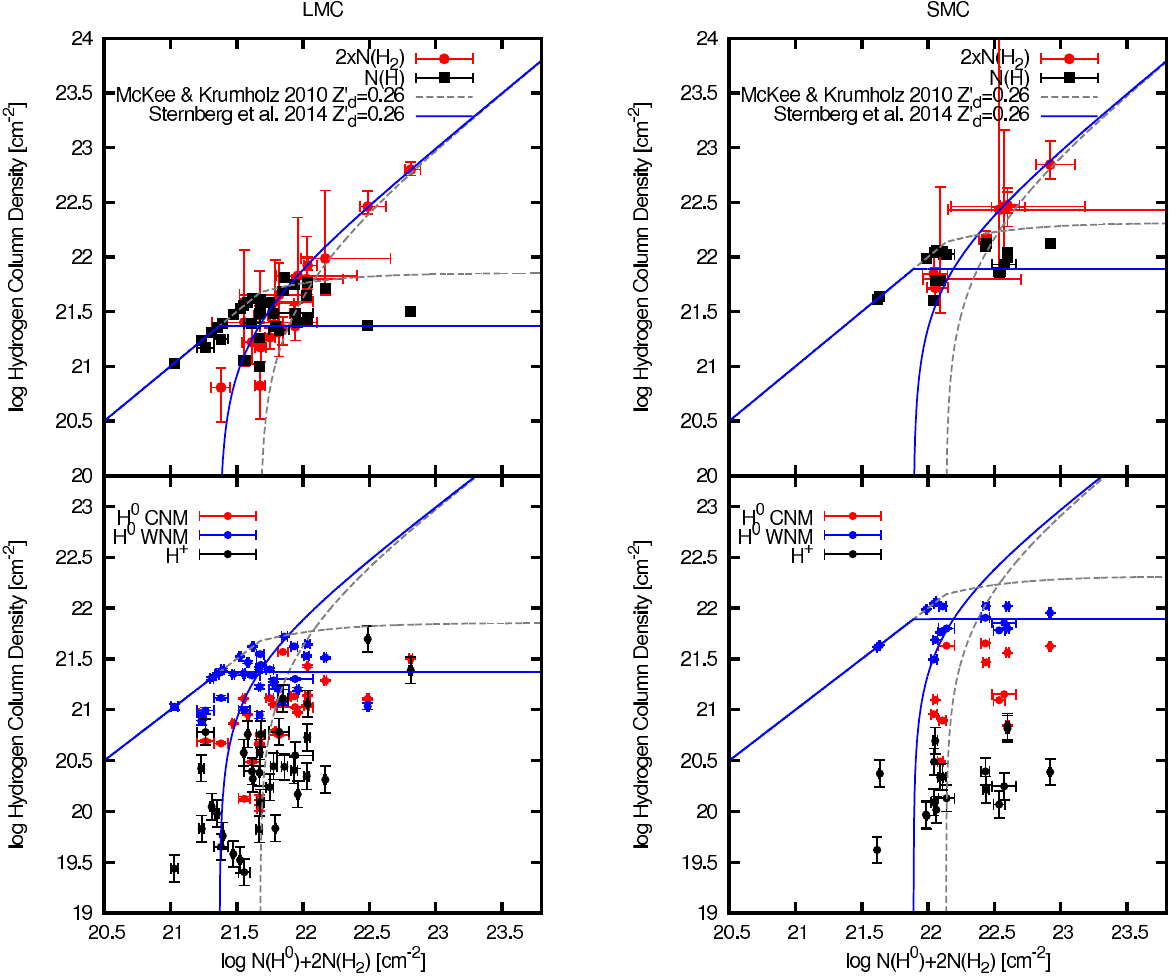}
\caption{({\it upper panels}) The column densities of H$^0$ and H$_2$
  as a function of the total hydrogen column density, $N({\rm
    H^0})$+2$N$(H$_2)$, for   LOSs in the  LMC ({\it left}) and SMC ({\it
    right}).  ({\it lower panels}) The column densities of H$^+$ and
  H$^0$ decomposed in CNM and WNM gas as a function of the total
  hydrogen column density for  LOSs in the LMC ({\it left}) and SMC ({\it
    right}). In all panels the dashed--dotted line indicates the
  predictions for $N({\rm H^0})$, and the dashed line for $N({\rm
    H}_2)$, from the model of \citet{Sternberg2014} and
  \citet{McKeeKrumholz2010} for $Z_{\rm d}^\prime=0.26$ and
  $\chi_0$=1.3 for the LMC and $Z_{\rm d}^\prime=0.1$ and $\chi_0$=1.9
  for the SMC (see Section~\ref{sec:h-h_2-transition}).  }
\label{fig:htoh2smc}
\end{figure*}

For velocity components where we were able to determine the physical
conditions of the [C\,{\sc i}]--emitting or $^{12}$CO--emitting layer,
we use a thermal pressure for the [C\,{\sc ii}] emitting layer that is
the geometrical mean between the diffuse ISM pressure derived in
Section~\ref{sec:therm-press-diff} and that of the [C\,{\sc
  i}]--emitting or that of the $^{12}$CO--emitting layer, in case we
were not able to determine the conditions at the [C\,{\sc
  i}]--emitting layer.  For the kinetic temperature we assume 49\,K in
case the temperature of the denser molecular gas is lower than or
equal to this temperature. We adopt the kinetic temperature of the
denser molecular gas in the case it is larger than 49\,K.  With these
two assumptions, we derive typical volume densities (and standard
deviations) of $10^{3.1\pm0.5}$\,cm$^{-3}$ for the LMC and
$10^{3.2\pm0.2}$\,cm$^{-3}$ for the SMC.

With the estimated volume density and kinetic temperatures of the
[C\,{\sc ii}]--emitting layer, we use Equation~(\ref{eq:6}) to derive
the C$^{+}$ column density for each velocity component. Typical values
of the beam--averaged column densities are
$10^{17.4\pm0.3}$\,cm$^{-2}$ and $10^{17.0\pm0.2}$\,cm$^{-2}$ for the
LMC and SMC, respectively.

\begin{figure*}[t]
\centering
\includegraphics[width=0.8\textwidth,angle=0]{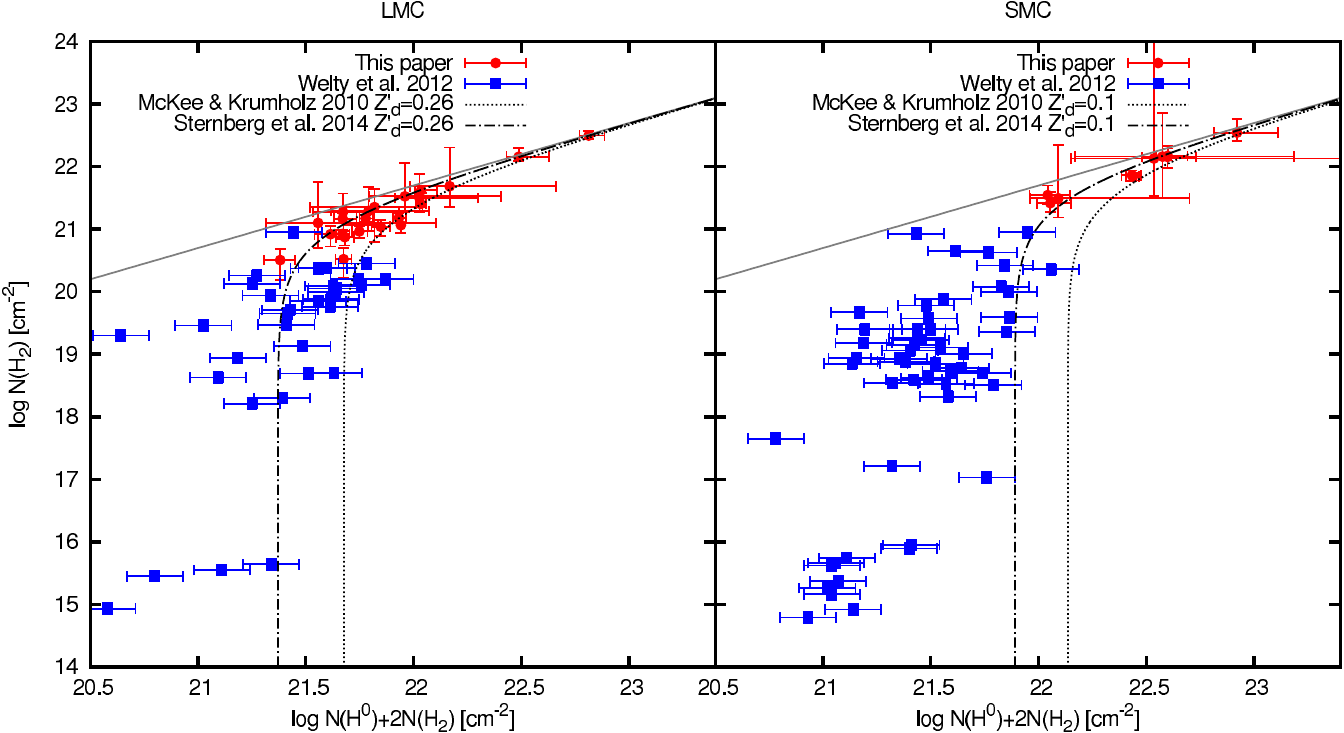}
\caption{The column density of molecular hydrogen as a function of the
  total hydrogen column density, $N({\rm H^0})$+2$N$(H$_2)$, for the
  values derived here and those observed in UV absorption for H$_2$
  and in optical absorption for H$^0$ compiled by
  \citet{Welty2012}. The straight line represents a molecular fraction
  of $f({\rm H}_2)$=1. We assume uncertainties of 10\% for $N({\rm
    H^0})$ and of 20\% for $N({\rm H}_2)$ in the \citet{Welty2012}
  data. We also show the predictions from the models of
  \citet{Sternberg2014} and \citet{McKeeKrumholz2010} for $Z_{\rm
    d}^\prime=0.26$ and $\chi_0$=1.3 for the LMC and $Z_{\rm
    d}^\prime=0.1$ and $\chi_0$=1.9 for the SMC (see
  Section~\ref{sec:h-h_2-transition}). }
\label{fig:welty}
\end{figure*}

\section{Discussion}
\label{sec:discussion}

\subsection{The H to H$_2$ transition}
\label{sec:h-h_2-transition}

The abundance of molecular hydrogen in interstellar clouds is set by
the balance between H$_2$ formation on dust grains and
photo--destruction by the ambient FUV photons.  The gas transitions
from being H$^0$--dominated to H$_2$--dominated when the dust column
density is large enough to allow H$_2$ self shielding to halt
photo--destruction, leading to a rapid increase of the H$_2$ abundance
\citep[e.g.][]{Gnedin2009, Krumholz2009,Sternberg2014,Bialy2016}.  The
transition between atomic to molecular hydrogen marks the onset of the
formation of molecular clouds and therefore has important implications
in star formation and galaxy evolution.

The atomic gas component in clouds is often traced by observations of
the H\,{\sc i} 21\,cm line, while molecular gas is determined by
observing the $^{12}$CO $J=1\to0$ line which is converted to a H$_2$
column density using a $X_{\rm CO}$ conversion factor. CO
observations, however, trace mostly well--shielded regions and the
diffuse H$_2$ component (CO--dark H$_2$) is not well probed.  Diffuse
H$_2$ has been observed in absorption against stars in the
far-ultraviolet
\citep[e.g.][]{Tumlinson2002,Cartledge2005,Sheffer2008,Welty2012}, but
is limited to individual lines--of--sight and tends to trace low H$_2$
column densities, as lines--of--sight with higher molecular content
will absorb the starlight required for reliable FUSE
measurements. [C\,{\sc ii}] observations, which can potentially image
large regions in the Magellanic clouds, can be used to trace the
CO--dark H$_2$ component and to study how it relates to the atomic
gas.

We study the transition from atomic to molecular gas in the Magellanic
clouds by estimating the column density of H, $N({\rm H^0})$, and of
H$_2$, $N({\rm H_2})$ in our sample.  As discussed in
Section~\ref{sec:atomic-gas}, we estimated the column density of
atomic hydrogen using $N({\rm H^0})=1.82\times10^{18} I({\rm
  HI})$\,cm$^{-2}$, with $I({\rm HI})$ in units of ${\rm
  K\,km\,s^{-1}}$.  The typical uncertainty in the H$^0$ column
density is 10$^{19.8}$\,cm$^{-2}$. To calculate the H$_2$ column
density, we first calculate the total carbon column density, $N^{\rm
  total}$(C)$=N({\rm C}^+)_{\rm H_2}+N({\rm C}^0)+N({\rm CO})$,
  with $N({\rm C}^+)_{\rm H_2}$ being the column density of C$^+$
  associated with H$_2$ (Section~\ref{sec:c-sc-i}). We converted from
$N^{\rm total}$(C) to $N({\rm H}_2)$ applying the gas--phase carbon
fractional abundances for the LMC and SMC discussed in
Section~\ref{sec:origin-c-sc}.  The estimated $N({\rm H_2})$ can
include well shielded gas, associated with C$^0$ and CO, as well as
the diffuse CO--dark H$_2$ gas, associated with C$^+$.  The
uncertainties in $N({\rm H_2}$) are the result of the propagation of
those from $N({\rm C}^+)$, $N({\rm C^0})$, $N({\rm CO})$. For $N({\rm
  C}^+)$ the uncertainty is given by the propagation of the errors in
the integrated intensity of the [C\,{\sc ii}] line and the errors on
the thermal pressures due to the uncertainties in the physical
conditions of the [C\,{\sc i}] or CO layer. The uncertainties of
$N({\rm C^0})$ and $N({\rm CO})$ are estimated as part of the
excitation analysis described in Section~\ref{sec:kinet-temp-h_2}.

 In order to test our determination of the H$^0$ and H$_2$ column
  densities, in Figure~\ref{fig:total_H}, we show a comparison between
  the $N$(H$^+$)+$N({\rm H^0})$+2$N$(H$_2)$ hydrogen column density
  derived from H$\alpha$, H\,{\sc i}, [C\,{\sc ii}], [C\,{\sc i}], and
  CO observations and the visual extinction derived from dust
  continuum (Section~\ref{sec:visu-extinct-determ}). We include the
  $N$(H$^+$) column density to account for the contribution from
  ionized gas to $A_{\rm V}$. Given the uncertainties in the
  assumptions used to derive all quantities involved, we find a
  reasonable correspondence between $N$(H$^+$)+$N({\rm
    H^0})$+2$N$(H$_2)$ and $A_{\rm V}$, with a scaling that
  corresponds to the dust--to--gas ratio of the LMC and SMC discussed
  in Section~\ref{sec:dust-gas-ratio}.  
% However, the data
%points in the SMC tend to have visual extinctions that are a factor of
%1.5 lower than what is expected for the dust--to--gas ratio of the
%SMC.

\begin{figure}[h]
\centering
\includegraphics[width=0.45\textwidth,angle=0]{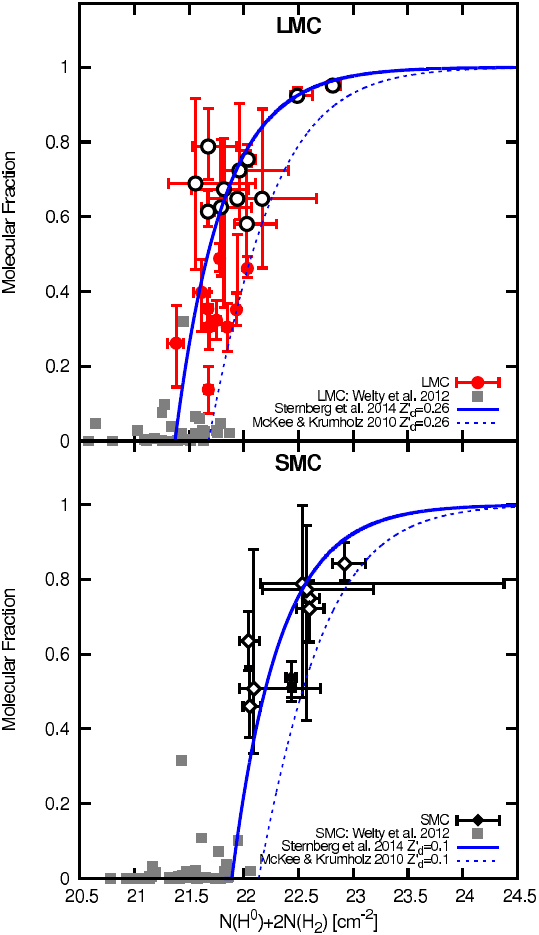}
\caption{Molecular fraction (Equation~\ref{eq:9}) as a function of the
  total hydrogen for the LMC and SMC. The lines are the predictions
  from the models of \citet{Sternberg2014} and
  \citet{McKeeKrumholz2010} for $Z_{\rm d}^\prime=0.26$ and
  $\chi_0$=1.3 for the LMC and $Z_{\rm d}^\prime=0.1$ and $\chi_0$=1.9
  for the SMC (see Section~\ref{sec:h-h_2-transition}).  The filled
  circles and squares correspond to lines--of--sight where we derived
  the H$_2$ column density from $N$(C$^+$) only while the open circles
  and squares represent LOSs where C$^0$ and CO also contribute to the
  total carbon column density.  We also include molecular fractions
  calculated from the sample of column densities observed in
  absorption in the FUV for H$_2$ and in the optical for H\,{\sc i}
  compiled by \citet{Welty2012}. For clarity we do not show the error
  bars in the \citet{Welty2012} data.  The uncertainties in the
  \citet{Welty2012} data are 30\% for the total hydrogen column
  density and 36\% for the molecular fraction based in the assumed
  uncertainties of 10\% for $N({\rm H^0})$ and of 20\% for $N({\rm
    H}_2)$. }
\label{fig:mol_fraction}
\end{figure}

\begin{figure}[t]
\centering
\includegraphics[width=0.48\textwidth,angle=0]{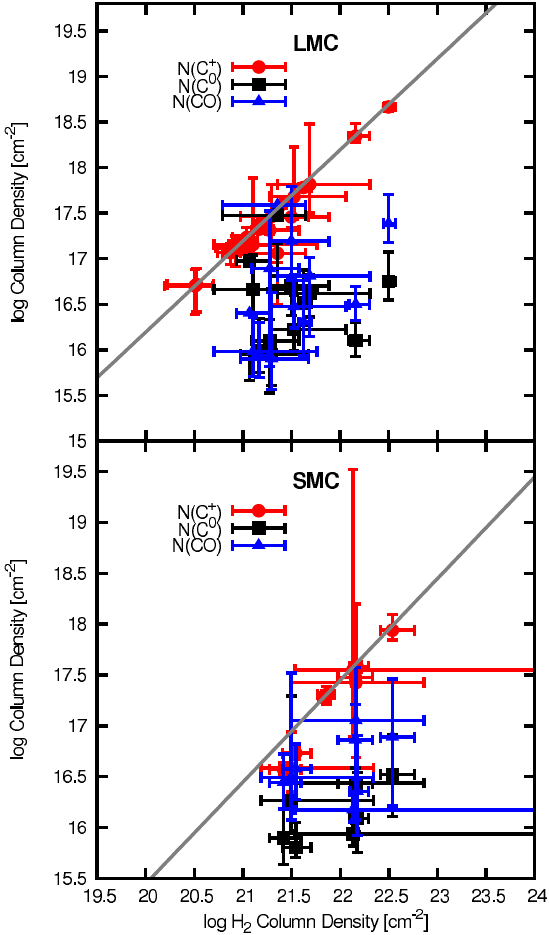}
\caption{The C$^{+}$, C$^{0}$ and CO column densities  associated
    with H$_2$ gas  as a function of the H$_2$ column density derived
  in our sample in the ({\it top}) LMC and ({\it bottom}) SMC. The
  straight grey lines represent the total gas--phase carbon column
  density, $N$(C$^+$)+$N$(C$^0$)+$N$(CO), that results by multiplying
  the H$_2$ column density by the [C]/[H$_2$] abundance ratio of
  1.56$\times10^{-4}$ for the LMC and 2.8$\times10^{-5}$ for the SMC
  (Section~\ref{sec:origin-c-sc}). }
\label{fig:lmc_colden}
\end{figure}

\begin{figure*}[t]
\centering
\includegraphics[width=0.8\textwidth,angle=0]{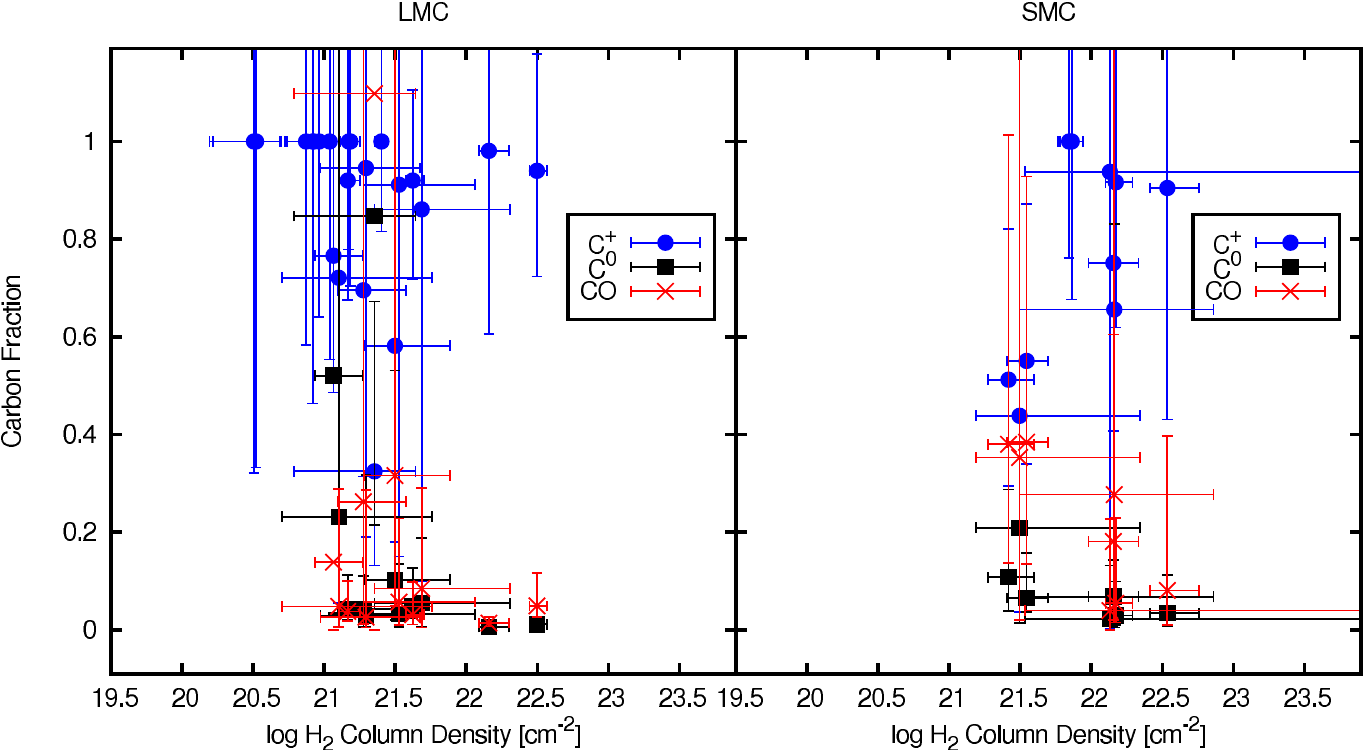}
\caption{The fraction of C$^+$, C$^0$, and CO with respect to the
  total carbon, C$^+$+ C$^0$+CO, column density  associated with H$_2$ gas as a function of the H$_2$ column
  density in the LMC ({\it left}) and SMC ({\it right}).}
\label{fig:lmc_fraction}
\end{figure*}

 Analytic theories for the atomic--to--molecular transition have been
 presented by \citet{Sternberg2014} and \citet{McKeeKrumholz2010} (see
 also \citealt{Krumholz2008} and \citealt{Krumholz2009}). Their work
 provides expressions for the molecular fraction as a function of
 total hydrogen column density (Eq. 97 for \citealt{Sternberg2014} and
 Eq. 93 for \citealt{McKeeKrumholz2010}) which can be used for
 comparison with our data.  Both models assume a value of the strength
 of the FUV radiation field, and that the gas metallicity and
 dust--to--gas ratio scale by the same factor, $Z_{\rm
   d}^\prime$. These two parameters determine the total H column
 density at which the transition from H$^0$ to H$_2$ takes place, with
 higher column densities when the metallicity is reduced and/or the
 FUV radiation field is increased.
%#Here is where you describe what are your assumptions.
  \citet{Hughes2010} studied the strength of the FUV field across the
  LMC using the dust temperature maps presented by
  \citet{Bernard2008}, finding an average value of $\chi_0$=1.3, in
  units of the local FUV radiation field calculated by
  \citet{Draine78}, which is similar to that found in the Solar
  neighborhood. In the SMC, however, the strength of the FUV field is
  found to be larger than that in the Solar neighborhood and the LMC
  \citep[e.g.][]{Lequeux1989}. \citet{Li2002} used IRAS, DIRBE, and
  ISO observations to constrain a dust model of the SMC, in which
  their best fit for the strength of the FUV field corresponds to
  $\chi_0$=1.9.  We therefore adopt $\chi_0$=1.3 and 1.9 for the LMC
  and SMC, respectively.  Note that because we are adopting a
    single value of the FUV radiation field in our model calculations,
    we do not expect an exact match between model and derived column
    densities, as the FUV field is likely to vary among our LOSs.
    Both \citet{Hughes2010} and \citet{Li2002} show variations of a
    factor $\sim2$ in the derived FUV radiation field in the LMC and
    SMC, respectively.  Because the model predicted column density at
    which the H$^0$ to H$_2$ transition takes place is proportional to
    the FUV radiation field, variations in this quantity among our
    sample should result in a factor $\sim2$ scatter around the model
    predicted column density. This predicted scatter is consistent
    with that seen in our comparison between model and observed column
    densities (see Figures~\ref{fig:htoh2smc}, \ref{fig:lmc_colden},
    and \ref{fig:lmc_fraction}).  For the metallicity scaling, we
  adopt the scaling on the dust--to--gas ratio rather than of the
  carbon abundance, which are Z$^\prime$=0.26 for the LMC and 0.1 for
  the SMC.  This assumption of using the dust--to--gas ratio is
  justified as the \citealt{Sternberg2014} and
  \citealt{McKeeKrumholz2010} models focus on describing the chemical
  structure of the cloud (the H$^0$ to H$_2$ transition) which depends
  on the absorption of photons, and on the formation rate of H$_2$, on
  dust grains. These models, however, do not consider the thermal
  structure of the gas, which depends on the abundance of gas
  coolants. Thus the most relevant parameter in these models is the
  dust--to--gas ratio rather than the gas--phase abundances of
  elements.

  As described in \citet{Sternberg2014}, see their Figure 14, their
  theory differs from that of \citet{McKeeKrumholz2010} in their
  consideration of shielding by dust associated with both H and H$_2$,
  instead of only by dust associated with H in
  \citet{McKeeKrumholz2010}, and their assumption of a larger dust
  absorption cross section
  ($\sigma_g$=1.9$\times10^{-21}$Z$^\prime$\,cm$^{-2}$ vs
  1.0$\times10^{-21}$Z$^\prime$\,cm$^{-2}$).  Additionally, they
  assumed a cloud geometry of a slab illuminated from both sides
  instead of the spherical geometry assumed by
  \citet{McKeeKrumholz2010}. They show that the differences between
  the two models are mostly explained by their consideration
  of absorption by dust associated with H$_2$ and the larger dust
  absorption cross section they assume.

  In the top panels of Figure~\ref{fig:htoh2smc}, we show the column
  densities of H$^0$ and H$_2$ we have derived as a function of the
  H$^0$+2H$_2$ column density.  We also show the H$^0$ and H$_2$
  column densities predicted by the \citet{McKeeKrumholz2010} and
  \citet{Sternberg2014}, as grey dashed and blue solid lines,
  respectively.  We find that the \citet{Sternberg2014} model is in
  good agreement with the derived H$^0$ and H$_2$ column
  densities. The \citet{McKeeKrumholz2010} model tends to predict
  lower H$_2$ column densities for a given value of $N_{\rm H}$.  Note
  that the predicted column densities by the \citet{Sternberg2014} and
  \citet{McKeeKrumholz2010} model are functions of the shielding
  column density, which is dominated by the cold neutral medium gas in
  the case of atomic gas. Thus, their predicted H$^0$ column densities
  correspond to that of the CNM only, while the derived H$^0$ column
  densities presented in the top panel of Figure~\ref{fig:htoh2smc}
  corresponds to the total H$^0$ along the LOS with includes the
  contribution from both the WNM and CNM.

  In the bottom panels of Figure~\ref{fig:htoh2smc}, we show a
  decomposition of the H$^0$ column density between WNM and CNM gas
  (see Section \ref{sec:atomic-gas}) and the contribution from H$^+$
  (Section~\ref{sec:ionized-gas}).  The \citet{Sternberg2014} and
  \citet{McKeeKrumholz2010} models tend to overestimate the CNM column
  densities. It is, however, possible that we are underestimating the
  CNM column density, as it can emit [C\,{\sc ii}] emission that is
  below the sensitivity of our observations.  Most of the H$^0$ in our
  sample is in the form of WNM.  We estimate that the CNM typically
  represents 28\% and 14\% of the total H$^0$ column density in the LMC
  and SMC, respectivelty.  The f CNM fractions in the LMC and SMC
  are in good agreement with those derived in H\,{\sc i} absorption
  studies by \citet{Dickey2000} using H\,{\sc i} absorption against
  background continuum sources (33\% for the LMC and 13\% for the SMC;
  \citealt{Dickey2000}).  Note that the CNM fractions derived by
  \citet{Dickey2000} are proportional to the assumed kinetic
  temperature of the gas which in their case is 55\,K.  While the 
    CNM fraction in the LMC is somewhat smaller than that found in
  Solar neighborhood (40\% ; \citealt{Heiles2003,Pineda2013}), this
  fraction is significantly smaller in the SMC.  \citet{Dickey2000}
  argued that the low CNM fraction in the SMC is a result of the lower
  abundance of gas coolants while the reduction of the photoelectric
  heating efficiency due to the lower dust--to--gas ratio is offset by
  the enhanced FUV radiation field in the SMC
  \citep{Lequeux1989,Li2002}, resulting in warmer atomic gas.  The
  H$^+$ component plays a small role in the derived column densities.

We find that the thresholds where H$_2$ formation becomes important
are $\sim$10$^{21.5}$\,cm$^{-2}$ for the LMC and
$\sim$10$^{22}$\,cm$^{-2}$ for the SMC. These correspond to 35$\,{\rm
  M}_{\odot}\,{\rm pc}^{-2}$ and 110$\,{\rm M}_{\odot}\,{\rm pc}^{-2}$
for the LMC and SMC, including the contribution from He,
respectively. These values are higher than the Milky Way value of
10\,${\rm M}_{\odot}\,{\rm pc}^{-2}$ \citep{Krumholz2008}, and reflect
the lower dust--to-gas ratio of the LMC and SMC that reduces the rate
of H$_2$ formation for a given total hydrogen column density. Thus a
larger hydrogen column is needed for H$_2$ self--shielding to balance
H$_2$ photo--destruction, allowing the gas to transition to a mostly
molecular phase in a low metallicity environment.

In Figure~\ref{fig:welty}, we show the H$_2$ column densities we
derive together with those compiled for the LMC and SMC by
\citet{Welty2012} as a function of the total hydrogen column
density. For clarity, we exclude upper limits in the \citet{Welty2012}
data. We also include the predictions of the model by
\citet{Sternberg2014} and \citet{McKeeKrumholz2010}\footnote{Note that
  \citet{Welty2012} also compared their data set with the
  \citet{McKeeKrumholz2010}. The predictions plotted in their
  Figure\,17 differ from those shown here, as they assumed a different
  metallicity/dust--to--gas ratio scaling factor of 0.5 and 0.2, for
  the LMC and SMC, respectively, and a FUV field of unity is
  implicitly assumed for both galaxies. }.  In the LMC, the
\citet{Sternberg2014} model is in good agreement with both the H$_2$
column densities derived here and most of those compiled by
\citet{Welty2012}. In the SMC, however, the \citet{Sternberg2014} and
\citet{McKeeKrumholz2010} models are unable to reproduce the data
points compiled by \citet{Welty2012} for $N({\rm
  H}_2)\leq10^{20}$\,cm$^{-3}$. A larger metallicity factor and/or a
lower FUV radiation field (e.g. Z$^\prime$=0.2 and $\chi_0=1$ for the
\citealt{Sternberg2014} model) would be required for the theoretical
models to reproduce the these data points.  We see that the column
densities derived from [C\,{\sc ii}], [C\,{\sc i}], and CO provide a
connection between the diffuse H$_2$ detected by FUV absorption and
the dense H$_2$ in well--shielded regions. This connection is better
illustrated in Figure~\ref{fig:mol_fraction}, where we show the
molecular fraction,
\begin{equation}
\label{eq:9}
f({\rm H}_2)=\frac{2N({\rm H}_2)}{N({\rm H^0})+2N({\rm H}_2)},
\end{equation}
calculated for both our sample and that of \citet{Welty2012}, as a
function of the total hydrogen column density. We again include the
predictions from the model of \citet{Sternberg2014} and
\citet{McKeeKrumholz2010}. We see that the FUV absorption observations
mostly trace $f({\rm H}_2)\le0.2$ while our determination using
[C\,{\sc ii}], [C\,{\sc i}], and CO observations traces the $0.1\le
f({\rm H}_2)\le1$ range.  In Figure~\ref{fig:mol_fraction}, we also
separated LOSs where we derived the H$_2$ column density from
$N$(C$^+$) only from those where C$^0$ and CO also contribute to the
total carbon column density.  We see that for molecular fractions
$f({\rm H}_2)\gtrsim0.45$, the C$^0$, $^{12}$CO, and $^{13}$CO column
densities are large enough for the detection of [C\,{\sc i}],
$^{12}$CO, and $^{13}$CO lines. The [C\,{\sc ii}] line alone traces a
larger range in the molecular fraction down to 0.1.  This result
suggests that [C\,{\sc ii}] observations, together with that of
[C\,{\sc i}] and CO, are important tools for the study of the
transition from atomic to molecular gas in the ISM of galaxies.

%What is the role of pressure here? Remember that the SMC LOSs have
%larger thermal pressure.

\subsection{The C$^+$ to C$^0$ to CO transition}
\label{sec:relat-betw-carb}

Another critical transition in interstellar clouds is that from
ionized to neutral carbon and to the main molecular form of carbon,
CO. The C$^+$/C$^0$/CO layer is the interface between the diffuse
CO--dark H$_2$ gas and the dense molecular gas where star formation
takes place. Therefore, it has an important role in the evolution of
molecular clouds and star formation. The formation of CO in the gas
phase is balanced by photo--destruction by FUV photons. The
C$^+$/C$^0$/CO transition can take place if there is sufficient dust
column density to efficiently shield the FUV photons that destroy
CO. In low metallicity environments, the lower dust--to--gas ratio
pushes the C$^+$/C$^0$/CO layer to higher hydrogen column
densities. This effect has important implications in the ability of
[C\,{\sc ii}], [C\,{\sc i}], and CO to trace the molecular hydrogen
mass in clouds.

In Figure \ref{fig:lmc_colden}, we show a comparison between the H$_2$
column density and the C$^{+}$, C$^{0}$, and CO column densities 
  associated with H$_2$ derived in our sample in the LMC and SMC.
The straight grey lines represent the total gas--phase carbon column
density, $N$(C$^+$)$_{\rm H_2}$+$N$(C$^0$)+$N$(CO), that results by
multiplying the H$_2$ column density by the [C]/[H$_2$] abundance
ratio of the LMC and SMC (Section~\ref{sec:origin-c-sc}).  
  Ionized carbon is the dominant form of gas--phase carbon that is
  associated with molecular gas in our sample.
 Both neutral carbon and CO represent a lower fraction of the total
carbon along the line--of--sight.  The low column densities of CO and
[C\,{\sc i}] in our sample suggest that a large fraction of the gas
column is associated with diffuse H$_2$ gas.

% We detect [C\,{\sc i}] and CO above $N({\rm
%   H}_2)\sim10^{21}$\,cm$^{-2}$ in the LMC and
% $\sim10^{21.5}$\,cm$^{-2}$ in the SMC. The difference in the
% thresholds for [C\,{\sc i}] and CO detection is plausibly influenced
% by the difference in metallicity and dust--to--gas ratio between the
% LMC and SMC. As we discussed in the previous section the detection
% thresholds for [C\,{\sc ii}] and CO correspond to $f({\rm H}_2)
% \simeq 0.45$ in both the LMC and SMC.

We were able to determine C$^0$ and CO column densities above $N({\rm
  H}_2)\sim10^{21}$\,cm$^{-2}$ in the LMC and
$\sim10^{21.5}$\,cm$^{-2}$ in the SMC. The difference in these
thresholds is likely influenced by the difference in metallicity and
dust--to--gas ratio between the LMC and SMC. As we discussed in the
previous section, the C$^0$ and CO column densities have a measurable
contribution to the total carbon abundance associated with H$_2$ for
$f({\rm H}_2) \simeq 0.45$ in both the LMC and SMC.

In Figure~\ref{fig:lmc_fraction}, we show the fraction of C$^+$,
C$^0$, and CO to the total gas--phase carbon (C$^+$+C$^0$+CO)
abundance  associated with H$_2$ gas as a function of $N({\rm
  H_2})$.  Ionized carbon is the dominant form of carbon in our
lines--of--sight in the LMC and SMC.  On average, C$^+$, C$^0$, and CO
represent 89\%, 9\%, and 10\% of the gas--phase carbon associated with
molecular gas in the LMC and 77\%, 6\%, and 17\% in the SMC,
respectively.

%  
%#LMC Average C+ Fraction 89.0602
%#LMC Average CI Fraction 5.68172
%LMC Average CO Fraction 7.61412
%############################
%SMC Average C+ Fraction 76.6606
%SMC Average CI Fraction 6.04411
%SMC Average CO Fraction 17.4899

\begin{figure*}[t]
\centering
\includegraphics[width=0.8\textwidth,angle=0]{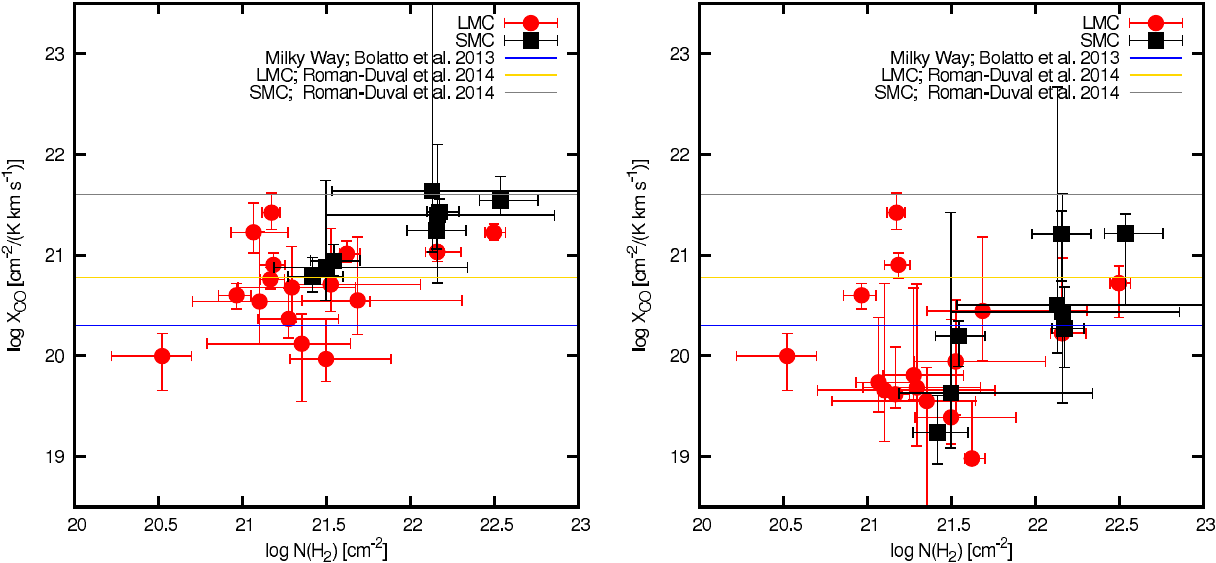}
\caption{({\it left}) The $X_{\rm CO}$ conversion factor as a function
  of the total H$_2$ column density derived in
  Section~\ref{sec:h-h_2-transition} in our LMC and SMC sample. The
  horizontal lines correspond to $X_{\rm
    CO}=2\times10^{20}$\,cm$^{-2}$(K\,km\,s$^{-1}$)$^{-1}$ compiled
  from several estimates in the Galaxy by \citet{Bolatto2013}, and
  $6\times10^{20}$\,cm$^{-2}$(K\,km\,s$^{-1}$)$^{-1}$ and
  $4\times10^{21}$\,cm$^{-2}$(K\,km\,s$^{-1}$)$^{-1}$, estimated by
  \citet{Roman-Duval2014} in the LMC and SMC, respectively. ({\it
    right}) The $X_{\rm CO}$ conversion factor as a function of the
  total H$_2$ column density corrected by the filling factor of the
  CO--emitting gas derived in our sample (see
  Section~\ref{sec:co-emission}).}
\label{fig:xco_ff}
\end{figure*}

\begin{figure}[t]
\centering
\includegraphics[width=0.48\textwidth,angle=0]{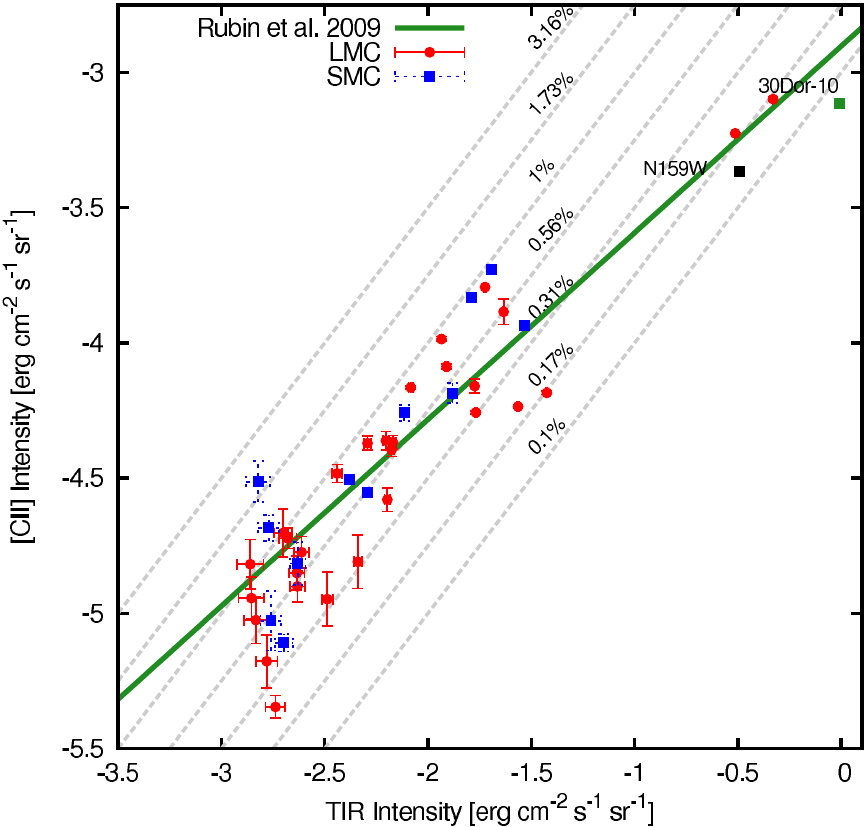}
\caption{The [C\,{\sc ii}] intensity as a function of the total
  infrared intensity in our LMC and SMC sample. We also include values
  for the 30\,Dor and N159W massive star forming regions in the LMC,
  with the [C\,{\sc ii}] data taken from \citealt{Boreiko91}. The
  dashed straight lines represent constant values of the photoelectric
  efficiency ranging from 0.1\% to 3.16\%, in steps of 0.25\,dex. The
  solid straight line corresponds to the fit of the [C\,{\sc ii}]--TIR
  relationship obtained by \citet{Rubin2009}. }
\label{fig:tir_vs_cii}
\end{figure}

\subsection{The CO--to--H$_2$ conversion factor}
\label{sec:c-sc-ii-2}

The mass of molecular clouds in the Milky Way and external galaxies is
often derived using observations of the $^{12}$CO $J=1\to0$ line
together with an empirically derived CO--to--H$_2$ conversion factor,
$X_{\rm CO}\equiv \frac{N({\rm H}_2)}{I_{\rm CO}}$ \citep[][and
references therein]{Bolatto2013}. The $X_{\rm CO}$ factor is
calibrated in the Milky Way to be
$2\times10^{20}$\,cm$^{-2}$(K\,km\,s$^{-1}$)$^{-1}$, but it is known
to vary with environmental conditions, in particular with metallicity,
which increases its value, as fainter CO emission per unit column
density is observed
\citep[e.g.][]{Rubio93,Muller2010,Hughes2010,Leroy2007,Roman-Duval2014}. In
low--metallicity environments, the reduced dust--to--gas ratio results
in an enhanced photo-destruction of CO, reducing the volume of H$_2$
that is traced by CO. However, the calibration in the Galaxy accounts
for regions where CO is not the dominant form of gas--phase carbon
\citep{Pineda2010a}.  With our estimate of $N$(H$_2$) together with
our $^{12}$CO $J=1\to0$ observations we can study the variation of
$X_{\rm CO}$ as a function of metallicity in the LMC and SMC.

In the left panel of Figure~\ref{fig:xco_ff} we show the $X_{\rm CO}$
conversion factor as a function of the total H$_2$ column density in
the LMC and SMC.   We calculated $X_{\rm CO}$ by dividing the
  H$_2$ column density (Section~\ref{sec:h-h_2-transition}) by the CO
  integrated intensity, both corresponding to their total value along
  the line--of--sight.  In the Figure, we show data points for all
  LOSs where CO was detected, including those where neither higher$-J$
  nor isotopic transitions were detected
  (Section~\ref{sec:co-emission}). In these LOSs, the H$_2$ column
  density was derived from $N({\rm C^+})$.   While showing a large
scatter, the values of $X_{\rm CO}$ in the LMC are consistent with,
and in the SMC somewhat lower than, those derived by
\citet{Roman-Duval2014} of
$6\times10^{20}$\,cm$^{-2}$(K\,km\,s$^{-1}$)$^{-1}$ and
$4\times10^{21}$\,cm$^{-2}$(K\,km\,s$^{-1}$)$^{-1}$ for the LMC and
SMC, respectively, which are similar to values commonly found in the
literature
\citep[e.g.][]{Rubio93,Pineda2009,Muller2010,Hughes2010,Leroy2007}.
We find a mean value of $X_{\rm CO}$ in the LMC of
$9.1\times10^{20}$\,cm$^{-2}$(K\,km\,s$^{-1}$)$^{-1}$ and in the SMC
of $2.2\times10^{21}$\,cm$^{-2}$(K\,km\,s$^{-1}$)$^{-1}$.  The
correlation between $X_{\rm CO}$ and $N({\rm H}_2)$ in
Figure~\ref{fig:xco_ff}, is a result of the intensity of CO having a
much smaller variation compared with the H$_2$ column density in our
sample.

As mentioned above, the reduced dust--to--gas ratio in the low
metallicity Magellanic Clouds results in an enhanced
photo--destruction of CO which results in CO tracing a smaller volume
of the H$_2$ gas. If this is the case, the $^{12}$CO $J=1\to0$
intensity is reduced by a filling factor, which in turn results in a
larger $X_{\rm CO}$ compared to a cloud with solar metallicity. We can
test this effect in our sample, as we derived the filling factor of
the CO--emitting gas as part of the excitation analysis in Section
\ref{sec:co-emission}.   For LOSs detected in $^{12}$CO $J =
  1\to0$ but where the H$_2$ column density was derived from $N$(C$^+$), we
  assumed a filling factor of unity.  In the right panel of
Figure~\ref{fig:xco_ff}, we show $X_{\rm CO}$ as a function of the
total H$_2$ column density in our sample, but this time we multiplied
$X_{\rm CO}$ by the filling factor derived in the excitation analysis,
effectively correcting the $I_{\rm CO}$ intensity for filling factor
effects.

While applying this correction increases the scatter in the observed
values, we see that the values of $X_{\rm CO}$ in the LMC and SMC are
closer to the value of $X_{\rm CO}$ in the Milky Way, where the
filling factor is closer to unity. The mean values of $X_{\rm CO}$ in
the LMC and SMC are reduced to
$2.9\times10^{20}$\,cm$^{-2}$(K\,km\,s$^{-1}$)$^{-1}$ and
$7.6\times10^{20}$\,cm$^{-2}$(K\,km\,s$^{-1}$)$^{-1}$, respectively,
which are closer to the value derived in the Milky Way of
$2\times10^{20}$\,cm$^{-2}$(K\,km\,s$^{-1}$)$^{-1}$
\citep{Bolatto2013}.  The reduced values of $X_{\rm CO}$ resulting
from applying a filling factor seem to confirm that the effect of
metallicity on the value of $X_{\rm CO}$ is to reduce the filling
factor of molecular gas traced by CO, resulting in a lower $^{12}$CO
$J=1\to0$ intensity which in turn enhances the $X_{\rm CO}$ factor.

\begin{figure*}[t]
\centering
\includegraphics[width=0.95\textwidth,angle=0]{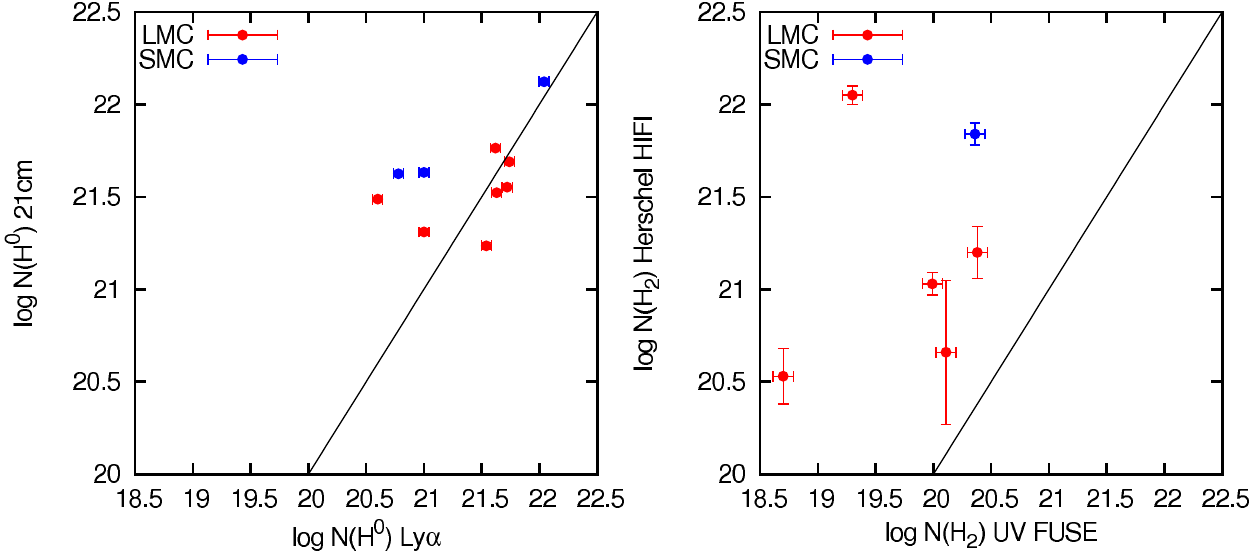}
\caption{({\it left}) Comparison of the H$^0$ column densities derived
  using 21\,cm observations and those derived using optical absorption
  of the Ly$\alpha$ line. ({\it right}) Comparison of the H$_2$ column
  densities derived in our analysis and those derived from UV
  absorption observations. The H$^0$ and H$_2$ column densities
  derived from absorption measurements were compiled by
  \citet{Welty2012}.}
\label{fig:fuse}
\end{figure*}

\subsection{Photoelectric heating and [C\,{\sc ii}] cooling}
\label{sec:relat-betw-c}

%#In atomic and diffuse molecular clouds the main heating mechanism is

Photoelectric (PE) heating is the main heating mechanism in atomic and
diffuse molecular clouds.  In this process the gas is heated by
energetic electrons that are expelled from dust grains after they
absorb FUV photons \citep{Spitzer1948}. The efficiency in which photo
electric heating works is dependent on the strength of the FUV
radiation field, the gas temperature, and the electron density, as
well as on grain charging, which depends on the type of dust grain
involved.

The efficiency of photoelectric heating can be studied by comparing
the total infrared intensity (tracing the energy absorbed by dust) and
the [C\,{\sc ii}] emission, which is the main cooling line in regions
where photoelectric heating dominates. Such comparison has been done
in the LMC at an angular resolution of 14.9\arcmin\ (217\,pc spatial
resolution) by \citet{Rubin2009}, resulting in PE heating efficiency
that varies by a factor of 1.4 between diffuse and bright
star--forming regions. They find that overall in the LMC, the [C\,{\sc
  ii}] constitutes 1.32\% of the LMC's far infrared luminosity.
Similar contributions from [C\,{\sc ii}] to the observed FIR
intensities were found by \citet{Israel2011} in several star forming
regions in the LMC and SMC with the KAO at a resolution of 55\arcsec\
(13.3\,pc for the LMC and 16.2\,pc for the SMC) and in a larger sample
of low--metallicity dwarf galaxies \citep{Cormier2015}.   The [C\,{\sc
  ii}]/FIR ratios found in the Magellanic clouds and other dwarf
galaxies are higher than the 0.1-0.2\% value found in the Milky Way
\citep{Wright1991}. This difference in the [C\,{\sc ii}]/FIR ratio has
been attributed to the lower dust--to--gas ratio in low metallicity
environments, which results in lower FIR intensity, and in an
increased volume of C$^+$--emitting regions.  In the following, we
compare our observations of [C\,{\sc ii}] in the LMC and SMC with that
of the total infrared emission derived from a combination of the {\it
  Spitzer} \citep{Meixner06} and {\it Herschel} \citep{Meixner2013}
dust continuum data.

\citet{Galametz2013} compared the resolved galaxy observations of
several {\it Herschel} and {\it Spitzer} bands with dust spectral
energy distribution (SED) models from \citet{Draine2007} to derive
empirical relationships between the specific intensity at different
bands and the total far infrared intensity (TIR) integrated between
3\,$\mu$m and 1100\,$\mu$m. In irregular galaxies (such as the LMC and
SMC), the total infrared is related to the far infrared intensity,
defined as the intensity integrated between 42\,$\mu$m and
122\,$\mu$m, as TIR/FIR$\approx 2$ \citep{Hunter2001}.

    We adopted the calibration for the TIR intensity (in units of
    erg\,cm$^{-2}$\,s$^{-1}$\,sr$^{-1}$) presented by
    \citet{Galametz2013} that uses the {\it Spitzer} 24\,$\mu$m and {\it
      Herschel} 100\,$\mu$m specific intensities (in units of
    erg\,cm$^{-2}$\,s$^{-1}$\,sr$^{-1}$\,Hz$^{-1}$),
 
\begin{equation}
I^{\rm TIR}=
2.421 \nu I_{\nu,24}
+
1.410 \nu I_{\nu,100},
\end{equation} 
as it ensures that the resulting TIR intensity can be smoothed to the
12\arcsec\ resolution of the {\it Herschel} [C\,{\sc ii}]
observations.   \citet{Galametz2013} found that this calibration
  is a good representation of the TIR resulting from full SED modeling
  of a large number of galaxies observed by {\it Herschel} and
  {\it Spitzer}.  At 12\arcsec\ resolution, our comparison between the
TIR and [C\,{\sc ii}] intensities is at 3\,pc and 3.5\,pc scales in
the LMC and SMC, respectively.  The resulting TIR maps have a typical
uncertainty of $2\times10^{-4}$ erg\,cm$^{-2}$\,s$^{-1}$\,sr$^{-1}$.

In Figure~\ref{fig:tir_vs_cii}, we show the [C\,{\sc ii}] intensity as
a function of the total infrared intensity estimated in the LMC and
SMC.  In the plot, we also include lines of constant [C\,{\sc ii}]/TIR
going from 0.1\% to 3.16\% in steps of 0.25\,dex, as well as the fit
of the [C\,{\sc ii}] and TIR relationship obtained by
\citet{Rubin2009}.   We find that [C\,{\sc ii}] intensities in
  both the LMC and SMC have a similar relationship with the total
  infrared emission despite their different metallicity and
  dust--to--gas ratio. On average [C\,{\sc ii}] emits 0.6\% and 0.8\%
  of the TIR intensity in the LMC and SMC, respectively. In terms of
  the FIR intensity, these averages correspond to 1.2\%, and 1.6\%,
  and thus they are in good agreement with the 1.32\% found by
  \citet{Rubin2009} over the whole LMC and the values found in star
  forming regions of the LMC and SMC by \citet{Israel2011}.

  Dense photon dominated regions in the LMC tend to have lower
  [C\,{\sc ii}]/TIR ratios, with [C\,{\sc ii}] emitting about 0.17\%
  of the TIR. A similar variation from diffuse to bright regions was
  suggested by \citet{Rubin2009}.  In Figure~\ref{fig:tir_vs_cii}, we also included
  data points that correspond to the 30\,Dor and N159W massive star
  forming regions in the LMC ([C\,{\sc ii}] data from
  \citealt{Boreiko91} at a 43\arcsec\ resolution), which together with
  the warm and dense PDRs in our sample, are in agreement with the
  lower [C\,{\sc ii}]/TIR ratios at high [C\,{\sc ii}] and TIR
  intensities suggested by the fit presented by \citet{Rubin2009}. The
  reduced [C\,{\sc ii}]/TIR ratio in dense PDRs can be understood as
  the combined effect of high volume densities and the high
  temperatures resulting from large FUV radiation fields. With
  increasing volume density and temperature, the excitation of the
  $^2{\rm P}_{3/2}$ level becomes a weak function of these quantities,
  and the [C\,{\sc ii}] intensity is only a function of column density
  \citep{Goldsmith2012}. The TIR intensity, which is a function of
  column density and dust temperature, is unaffected by the larger
  volume densities and FUV radiation fields.

\subsection{Comparison with FUSE LOSs}
\label{sec:comparison-with-fuse}

As mentioned in Section~\ref{sec:observations}, our sample contains 7
LOSs in the LMC and 3 in the SMC that coincide with stars that show UV
and optical absorption features that are used to determine the H$_2$ and H$^0$
column densities of their foreground gas
\citep[e.g.][]{Welty2012,Cartledge2005,Tumlinson2002}.  In this
sub--sample, we detected [C\,{\sc ii}] emission in 5 LOSs in the LMC
and in 1 in the SMC, with one LOS in the LMC also detected in [C\,{\sc
  i}] and CO.  Because the column densities determined in our analysis
correspond to the average column density within the beam of our
observations, and those from UV and optical absorption correspond to
column densities in a pencil beam, a comparison between these column
densities will be sensitive to substructure in the ISM of the LMC and
SMC. Another difference between these methods to determine column
densities is that the column densities in our analysis correspond to
the gas along the full sight--line through the LMC or SMC, while the
UV and optical absorption only corresponds the gas that is in the
foreground of the target stars. Therefore, a comparison between the
column densities determined with these different methods can also give
us some information on the structure of the ISM along the
line--of--sight in the LMC and SMC.

In Figure~\ref{fig:fuse}, we compare the H$^0$ column densities
derived using H\,{\sc i} 21\,cm observations ($N({\rm H^0})_{\rm
  21\,cm}$) and those derived using optical absorption of the
Ly$\alpha$ line ($N({\rm H^0})_{\rm Ly\alpha}$) as well as a
comparison between the H$_2$ column densities derived in our analysis
($N({\rm H}_2)_{\rm HIFI}$) and those derived from UV absorption
observations ($N({\rm H}_2)_{\rm UV}$). The H$^0$ and H$_2$ column
densities derived from absorption observations were compiled by
\citet{Welty2012}. We see that some LOSs are in reasonable agreement
between the column densities derived using H\,{\sc i} 21\,cm
observations and those derived using Ly$\alpha$ absorption, with some
having slightly larger $N({\rm H^0})_{\rm Ly\alpha}$ than that
expected from H\,{\sc i} 21\,cm observations. However, there are
several LOSs that have $N({\rm H^0})_{\rm Ly\alpha}$ that is
significantly lower than $N({\rm H^0})_{\rm 21\,cm}$. We also see that
the H$_2$ column densities in our sample are systematically larger
than those derived with FUSE.  A similar result is seen in the
comparison between atomic hydrogen column densities in a larger sample
presented by \citet{Welty2012} in the LMC and SMC. As explained by
these authors, $N({\rm H^0})_{\rm Ly\alpha}<$$N({\rm H^0})_{\rm
  21\,cm}$ could be a result of having the bulk of the atomic gas
behind the target stars. This effect can also be explained, as well as
the case when $N({\rm H^0})_{\rm Ly\alpha}>$$N({\rm H^0})_{\rm
  21\,cm}$, as a result of substructure within the 21\,cm beam where
the pencil beam of the optical absorption goes through a maximum or a
minimum in the gas distribution that have a much different column
density compared with the average within the beam. The first
explanation would require that H$_2$ is more extended along the LOSs
than H$^0$, which contradicts the spectral information that shows that
H\,{\sc i} is much more extended in velocity compared with [C\,{\sc
  ii}], [C\,{\sc i}], and CO. A more likely explanation for our data
is the existence of substructure within the beam of the
observations. The larger difference for the H$_2$ column densities
compared to that for the H$^0$ column densities could be a reflection
of the different structure of the atomic and molecular gas, with the
latter concentrated into smaller clumps.

\section{Conclusions}
\label{sec:conclusions}

In this paper we present a survey of the [C\,{\sc ii}], [C\,{\sc i}],
and CO emission observed with high--velocity resolution in the Large
and Small Magellanic clouds. The paper aims at characterizing the
transition from diffuse atomic to dense molecular clouds in the
low--metallicity environments of the Magellanic Clouds. Our sample was
selected based on the H\,{\sc i}, CO, and dust continuum emission and
includes regions in different stages in the transition from diffuse
regions to dense photon dominated regions associated with star
formation. Our results can be summarized as follows:

\begin{itemize}

\item We determined the contribution from different phases of the ISM
  to the observed [C\,{\sc ii}] emission in our sample. In LOSs
  associated with warm and dense PDRs, the [C\,{\sc ii}] emission from
  this ISM component tends to dominate.  The contribution from
  diffuse CO--dark H$_2$ to the observed [C\,{\sc ii}] shows a large scatter,
  ranging from $\sim$10\% to $\sim$80\%. This large scatter could be a
  reflection of clouds at different stages in the transition from
    diffuse to dense molecular gas, with different clouds having
    varying CO--dark H$_2$ fractions. We find that ionized and atomic
  gas have typically smaller contributions to the observed [C\,{\sc
    ii}] emission (Section~\ref{sec:origin-c-sc}).

\item Using lines--of--sight where only H\,{\sc i} and [C\,{\sc ii}]
  emission were detected, and using a derivation of the visual
  extinction from dust continuum emission, we determined a typical
  thermal pressure of the diffuse ISM to be $p_{\rm th}/k_{\rm
    B}=$3.4$\times$10$^{4}$\,K\,cm$^{-3}$ in the LMC and
  1$\times$10$^{5}$\,K\,cm$^{-3}$ in the SMC
  (Section~\ref{sec:therm-press-diff}).

  \item We used the [C\,{\sc i}] and CO observations to determine the
    column density, kinetic temperature, and volume density of the CO
    and [C\,{\sc i}]--emitting gas.  We find average values and
    standard deviations in the CO--emitting layer of $44\pm19$\,K in
    the LMC and $50\pm20$\,K in the SMC for the kinetic temperature,
    $10^{16.7 \pm 0.5}$\,cm$^{-2}$ in the LMC and $10^{16.6 \pm
      0.3}$\,cm$^{-2}$ in the SMC for the beam--averaged CO column
    density, $10^{5.1\pm0.9}$\,cm$^{-3}$ in the LMC and
    $10^{4.0\pm0.3}$\,cm$^{-3}$ in the SMC for the H$_2$ volume
    density, and ($0.2\pm0.2$) in the LMC and ($0.20\pm0.3$) in the
    SMC for the filling factor.  In the [C\,{\sc i}]--emitting layer
    we find $91\pm60$\,K in the LMC and $73\pm62$\,K in the SMC for the
    kinetic temperature, $10^{16.4 \pm 0.4}$\,cm$^{-2}$ in the LMC and
    $10^{16.2\pm 0.3}$\,cm$^{-2}$ in the SMC for the beam--averaged
    C$^0$ column density, and $10^{4.8\pm1.4}$\,cm$^{-3}$ in the LMC
    and $10^{3.5\pm0.5}$\,cm$^{-3}$ in the SMC for the H$_2$ volume
    density. This information combined with the [C\,{\sc ii}]
    observations allows us to calculate the column densities of
    CO--dark H$_2$ gas in our sample
    (Section~\ref{sec:kinet-temp-h_2}).

  \item We studied the transition between atomic and molecular gas in
    the LMC and SMC by comparing the H$_2$ column densities of
    CO--dark H$_2$ and CO--traced H$_2$ with the total hydrogen,
    $N({\rm H}^0)$+2$N$(H$_2$), column density in our sample.  We find
    reasonable agreement between our observations and theoretical
    models of the H$^0$ to H$_2$ transition
    (Section~\ref{sec:h-h_2-transition}).

  \item We found that most of the observed H\,{\sc i} is in the form
    of WNM.   We estimate that the CNM represents 28\% and 14\% of the total H$^0$
    column density in the LMC and SMC, respectively. The value in the
    LMC is similar to that in the Milky Way
    \citep[40\%;][]{Heiles2003,Pineda2013} while in the SMC the WNM
    represents a larger fraction (Section~\ref{sec:h-h_2-transition}).

\item We found that the H$_2$ column densities determined using
  [C\,{\sc ii}], [C\,{\sc i}], and CO observations in our survey trace
  molecular fractions between $0.1\le f({\rm H}_2)\le 1$ range. In
  contrast UV absorption observations mostly trace $f({\rm H}_2)\le
  0.2$.  The C$^0$ and CO column densities have a measurable
  contribution to the total gas--phase carbon column density for
  molecular fractions $f({\rm H}_2) \ge 0.45$, while [C\,{\sc ii}]
  alone traces a larger range in the molecular fraction down to 0.1
  (Section~\ref{sec:h-h_2-transition}).

\item Studying the C$^+$/C$^0$/CO transitions in our sample in
    the LMC and SMC reveals that most of the molecular gas in our
    sample is traced by [C\,{\sc ii}].  Both column densities of
    neutral carbon and CO represent a lower fraction of the total
    carbon associated with molecular gas along the line--of--sight.
    On average, C$^+$, C$^0$, and CO represent 89\%, 9\%, and 10\% of
    the gas--phase carbon in the LMC and 77\%, 6\%, and 17\% in the
    SMC, respectively.  The [C\,{\sc i}] and CO are detected above
    $N({\rm H}_2)\sim10^{21.5}$\,cm$^{-2}$ in the LMC and
    $\sim10^{21.8}$\,cm$^{-2}$ in the SMC. The difference in the
    thresholds for [C\,{\sc i}] and CO detection is possible a result
    of the metallicity  and dust--to--gas ratio difference between the
    LMC and SMC (Section~\ref{sec:relat-betw-carb}). 
%5
%5LMC Average C+ Fraction 89.0602
%5LMC Average CI Fraction 5.68172
%5LMC Average CO Fraction 7.61412
%5############################
%5SMC Average C+ Fraction 76.6606
%5SMC Average CI Fraction 6.04411
%5SMC Average CO Fraction 17.4899
%5

%XCO LMC
%9.10549
%XCO LMC corrected
%2.91441
%XCO SMC
%22.4061
%XCO SMC corrected
%7.60266

  \item We found a mean value of the $X_{\rm CO}$ conversion factor in
    the LMC of $9.1\times10^{20}$\,cm$^{-2}$(K\,km\,s$^{-1}$)$^{-1}$
    and in the SMC of
    $2.2\times10^{21}$\,cm$^{-2}$(K\,km\,s$^{-1}$)$^{-1}$, which are
    larger than the value of
    $2\times10^{20}$\,cm$^{-2}$(K\,km\,s$^{-1}$)$^{-1}$ in the Milky
    Way. When applying a filling factor correction to the CO emission
    we see that the values of $X_{\rm CO}$ in the LMC and SMC become
    closer to the value of $X_{\rm CO}$ in the Milky Way, where the
    filling factor is close to unity. This result suggests that the
    effect of metallicity in the value of $X_{\rm CO}$ is to reduce
    the filling factor of molecular gas traced by CO, which results in
    a lower $^{12}$CO $J=1\to0$ intensity which in turn results in an
    enhanced $X_{\rm CO}$ factor (Section~\ref{sec:c-sc-ii-2}).

  \item We found that most of the LOSs in our sample are consistent
    with a linear relationship between the [C\,{\sc ii}] and the total
    infrared emission (TIR). The [C\,{\sc ii}] emission represents
    about 1\% of the TIR in both the LMC and SMC despite their
    difference in metallicity and dust--to--gas ratio. This [C\,{\sc
      ii}]/TIR fraction is consistent with previous determination
    using the [C\,{\sc ii}] map of the LMC observed with the BICE
    balloon (Section~\ref{sec:relat-betw-c}).

\item We compared the H$^0$ and H$_2$ column densities derived in
  our analysis and those derived using optical and UV absorption in a
  sub-sample of our survey.  We find significant discrepancies between
  the column density observations which can be explained by
  substructure within the beam used for our observations
  (Section~\ref{sec:comparison-with-fuse}).

\end{itemize}

In conclusion, our results show that [C\,{\sc ii}], [C\,{\sc i}], and
CO observations are important tools for characterizing the transition
from atomic to molecular clouds in external galaxies. Future large
scale mapping of the LMC and SMC in these spectral lines with
current/future balloon, airborne (SOFIA), and space observatories will
make significant steps in our understanding of the evolution of the
ISM, star formation, and galaxy evolution.

\begin{acknowledgements}

  We would like to thank Drs. Julia Roman--Duval and Cheoljong Lee for
  providing their dust continuum data sets and Dr. Franck Le Petit for
  his help on comparing our H\,{\sc i} and H$_2$ data sets with
  theoretical predictions. We also thank an anonymous referee for a
  number of useful comments that significantly improved the
  manuscript.  This research was conducted at the Jet Propulsion
  Laboratory, California Institute of Technology under contract with
  the National Aeronautics and Space Administration.  We thank the
  staffs of the ESA and NASA {\it Herschel} Science Centers for their
  help.  \copyright\ 2016 California Institute of
  Technology. U.S. Government sponsorship acknowledged.

\end{acknowledgements}

\appendix
\section{Beam Dilution Correction}
\label{sec:beam-dilut-corr}

\subsection{Absolute Intensities}
\label{sec:absolute-intensities} 

\begin{figure*}[t]
\centering
\includegraphics[width=0.8\textwidth,angle=0]{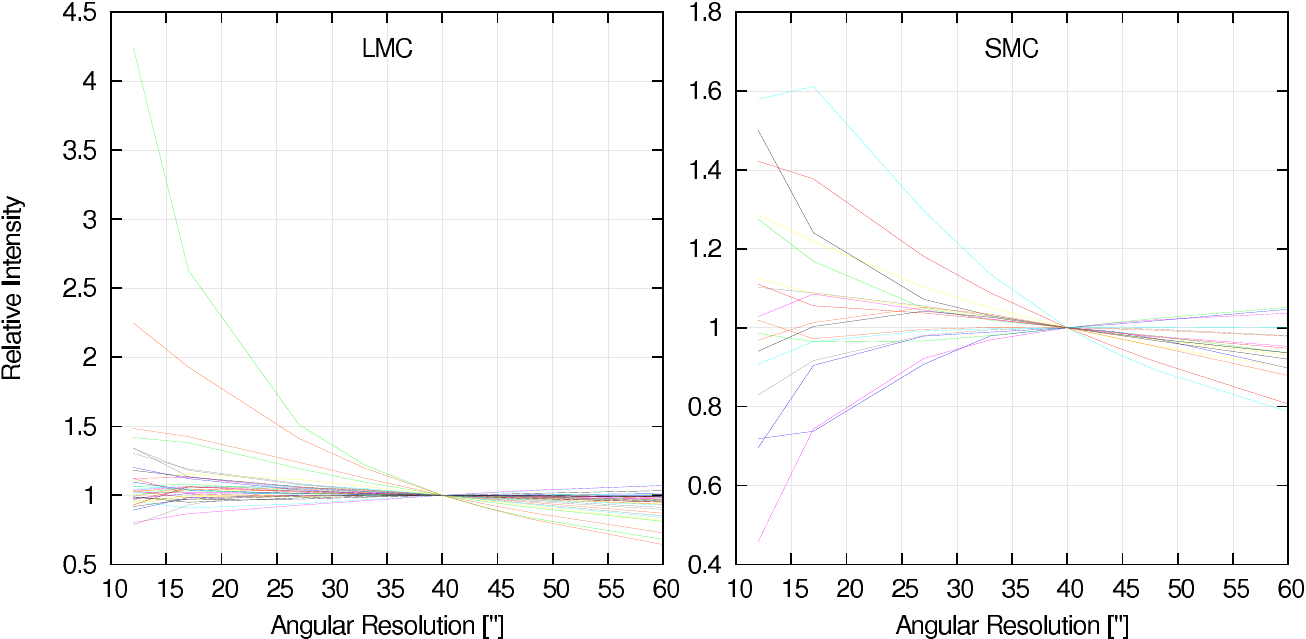}
\caption{The 160\,$\mu$m dust continuum intensity smoothed to
  different angular resolutions relative to that at 40\arcsec\ as a
  function of angular resolution for our sample in the LMC ({\it left
    panel}) and SMC ({\it right panel}).} \label{fig:resolution}
\end{figure*}

Our survey contains a set of multi--wavelength observations with
different angular resolutions. When possible our analysis was made
with data at matching resolutions, but there are cases in which we
have to rely on a combination of data with different angular
resolutions. In order to quantify how beam dilution affects the
results involving absolute intensities presented in this paper, we
smoothed the 12\arcsec\ angular resolution 160\,$\mu$m HERITAGE dust
continuum maps of the LMC and SMC to the different angular resolutions
of data set used in our analysis (12\arcsec, 17\arcsec, 27\arcsec,
33\arcsec, 44\arcsec, 48\arcsec, and 60\arcsec).  We assumed that the
distribution of dust continuum emission is similar to that of the
spectral lines used in our analysis. This assumption is motivated by
the agreement between dust continuum and [C\,{\sc ii}] maps in star
forming regions in the LMC (see e.g. \citealt{Okada2015} and
\citealt{Galametz2013} for N159).  We choose a reference angular
resolution of 40\arcsec\ which is an intermediate value between
12\arcsec\ and 60\arcsec. In Figure~\ref{fig:resolution}, we show the
variation of the absolute 160\,$\mu$m intensities for angular
resolutions between 12\arcsec and 60\arcsec\ with respect to that at
40\arcsec. Remarkably, most of our LOSs in the LMC and SMC show
variations within $\sim$40\% in the 12\arcsec\ to 60\arcsec\ range,
even though the beam area varies by a factor of 25. There are however
four exceptions, NT77, PDR3\_NE, PDR2\_NW, and LMC12\_NW where the
intensities vary by factors of 6, 3.5, 2, and 1.7 respectively. In the
SMC, AzV462, SMC\_NE\_1a, and SMC\_LIRS36, have the largest variations
with factors of 2.3, 2, and 1.8, respectively. Note that we did not
detect [C\,{\sc ii}] emission in AzV462.

 We also studied the variations of the SHASSA H$\alpha$
  intensities for varying angular resolution, by using the 24\,$\mu$m
  dust continuum maps of the LMC and SMC.  The 24\,$\mu$m continuum
  emission is a proxy for hot dust emission associated with H\,{\sc
    ii} regions.  The angular resolution of the 24\,$\mu$m images is
  6\arcsec. We convolved these images to 12\arcsec\ corresponding to
  [C\,{\sc ii}] observations and to 48\arcsec\ corresponding to that
  of the H$\alpha$ map.  We then computed the ratio of the 24\,$\mu$m
  intensities at 12\arcsec\ and 48\arcsec\ to determine a beam
  dilution correction factor for the H$\alpha$ data, which we use to
  quantify the impact of beam dilution in our results.  We find that
  the median flux variation between 12\arcsec\ and 48\arcsec\ is a
  factor of 1.15 in both in the LMC and SMC. The mean variations are
  factors of 1.6 in the LMC and of 1.7 in the SMC. The difference
  between the median and mean factors are mostly driven by three LOSs
  in both the LMC and SMC that show larger variations in their
  fluxes. In the LMC, they are SK-67D2, PDR3\_NE, and NT77, with
  fluxes varying by factors of 3.1, 3.4, and 8.3, respectively. (Note
  that SK-67D2 was not detected in [C\,{\sc ii}] in our survey.) In
  the SMC, the LOSs with the largest variations are SMC\_LIRS36,
  SMC\_NE\_1a, and SMC\_NE\_4c\_low, with fluxes varying by factors of
  3.6, 3.7, 5.2, respectively.

\begin{figure*}[t]
\centering
\includegraphics[width=0.8\textwidth,angle=0]{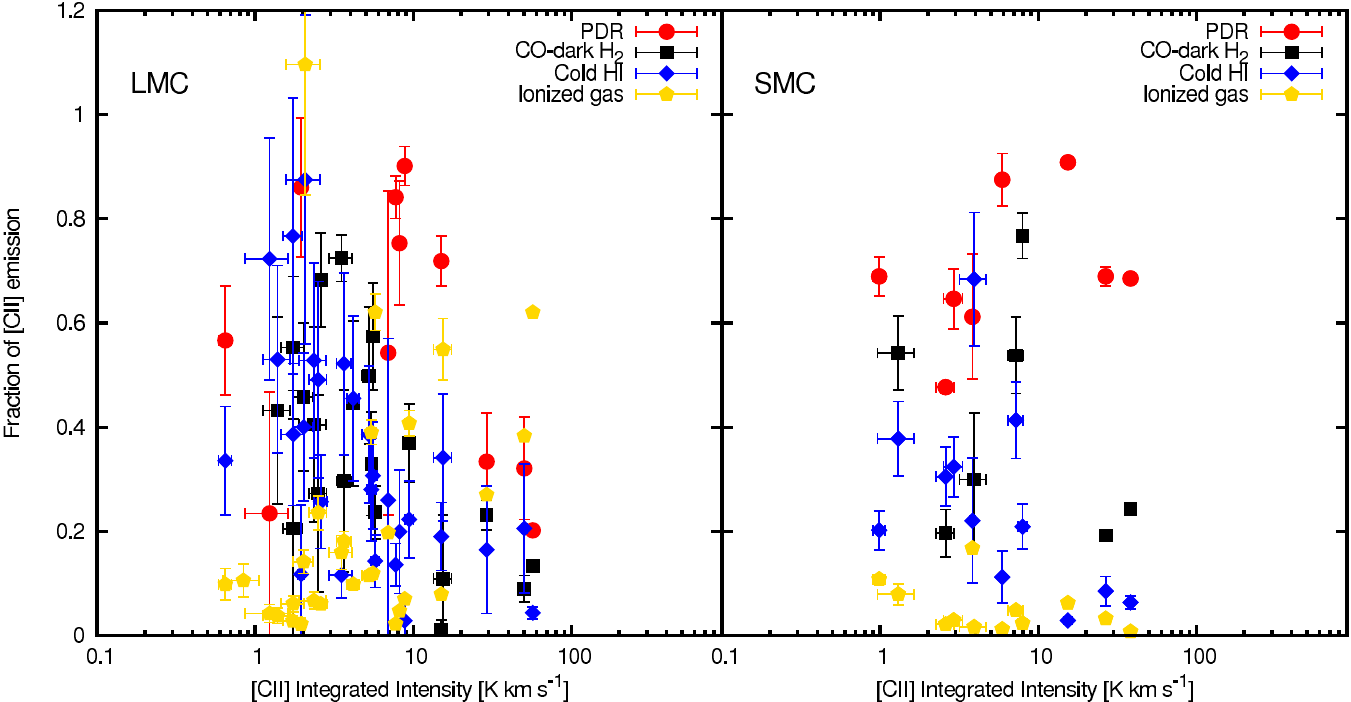}
\caption{The fraction of the [C\,{\sc ii}] emission that we estimate
  arises from ionized gas, cold atomic gas, CO--dark H$_2$, and photon
  dominated regions (PDRs) as a function of the observed [C\,{\sc ii}]
  emission in the LMC and SMC. All data points have been corrected by
  dilution effects as described in
  Appendix~\ref{sec:beam-dilut-corr}.}\label{fig:contribution_dil}
\end{figure*}

  In Figure~\ref{fig:contribution_dil}, we present the fraction of
  [C\,{\sc ii}] originating from different ISM phases as a function of
  the observed [C\,{\sc ii}] intensity in the case when the
  intensities are corrected by beam dilution effects. All quantities
  involved in the calculation of the contribution from different ISM
  phases to the observed [C\,{\sc ii}] emission have been corrected by
  beam filling effect to correspond to a common angular resolution of
  40\arcsec. In general, beam dilution effects have a small impact in
  the fraction of [C\,{\sc ii}] emission originating from the
  different phases of the ISM.

%Can you use the ratio of HSO vs BICE to calculate the f1,2?

\subsection{Line Ratios}
\label{sec:line-ratios}
Another instance where we use a combination of pointed observations at
different angular resolutions is in the excitation analysis in
Section~\ref{sec:kinet-temp-h_2}, where we use line ratios of CO and
[C\,{\sc i}] lines to estimate the physical conditions of the
CO and [C\,{\sc i}]--emitting gas.  In the following we describe a simple model that
we used to account for beam dilution effects in the line
ratios. Because we assume Gaussian sources, the model provides only a
first order correction to the observed line ratios, as the structure
of the ISM is likely to be more complex.

The ratio of the antenna temperatures of two lines, $T^*_1$ and
$T^*_2$, originating from a Gaussian source with FWHM size
$\Theta_{\rm s}$, when observed having the same FWHM beam size
$\Theta_{\rm f}$ centered in the source, is given by
\begin{equation}
  \frac{T^*_1}{T^*_{2}}=\frac{\frac{T_1 \Theta^2_{\rm s}}{\Theta^2_{\rm f}+\Theta^2_{\rm s}}}{\frac{T_2 \Theta^2_{\rm s}}{\Theta^2_{\rm f}+\Theta^2_{\rm s}}}=\frac{T_1}{T_2},
\end{equation}
where $T_1$ and $T_2$ are the intrinsic peak antenna temperature of the source. When the source is convolved with Gaussians having two different FWHM beam size $\Theta_{\rm a}$ and $\Theta_{\rm b}$, the integrated intensity ratio is given by
\begin{equation}
\frac{T^*_1}{T^*_{2}}=\frac{T_1}{T_2}\frac{\Theta^2_{\rm s}+\Theta^2_{\rm b}}{\Theta^2_{\rm s}+\Theta^2_{\rm a}},
\end{equation}
Thus, the intrinsic ratio is related to the observed  ratio as
\begin{equation}
\frac{T_1}{T_2}=\frac{T^*_1}{T^*_2}f_{1,2},
\end{equation}
where
\begin{equation}
f_{1,2}\equiv \frac{\Theta^2_{\rm s}+\Theta^2_{\rm a}}{\Theta^2_{\rm s}+\Theta^2_{\rm b}}.
\label{eq:2}
\end{equation}
The observed line ratio can therefore be corrected for beam dilution
effects if an estimate $f_{1,2}$ is available. We can estimate
$f_{1,2}$ by comparing the line or continuum emission of the same
source at different angular resolutions. In this case $T_1=T_2$ and
therefore the dilution factor is given by the inverse of the ratio of
the convolved peak intensities, $f_{1,2}=T^*_2/T^*_1$.

We used the 160\,$\mu$m HERITAGE dust continuum map with
12\arcsec\ angular resolution to estimate $f_{1,2}$ for the sources in
our survey.  We smoothed the 160\,$\mu$m map to the different angular
resolution of the spectral lines involved in our analysis (17\arcsec,
27\arcsec, 33\arcsec, and 44\arcsec) and calculated the dilution
factors of each line ratio using Equation~(\ref{eq:2}) by calculating
the ratio of the 160\,$\mu$m emission at the different angular
resolution pairs. We used the derived correction factors to study the
effect beam dilution in the line ratios used in our analysis.  Note
that we do not correct the line ratios for beam dilution effects in
our analysis, due to the uncertainties of whether dust continuum
emission traces the distribution of gas.  In Table\,6, we present the
effect of beam dilution in the line ratios that are calculated using
observations at different angular resolutions. We find that these line
ratios would typically vary by 10\%.  This small variation suggest
that the observed structures are relatively extended at the resolution
of our observations.

\bibliographystyle{aasjournal.bst}
%#$\bibliographystyle{/home/jpineda/astronat/apj/apj.bst}
\bibliography{papers}

%\bibliography{/root/latex/papers}
%\bibliographystyle{/root/astronat/apj/apj.bst}

\clearpage

\end{document}